\documentclass[iop]{emulateapj}

\usepackage{natbib,aas_macros,amsmath}
\usepackage{multirow,color}
\usepackage{natbib}

\newcommand{\carcsec}{$\!\!\arcsec$}
\newcommand{\m}[1]{\mathrm{#1}}
\newcommand{\kms}{\m{km\ s^{-1}}}
\newcommand{\Msun}{\ensuremath{M_{\odot}}}
\newcommand{\logMmin}{\m{log}M_\m{min}}
\newcommand{\sigmalogM}{\sigma_{\m{log}M}}
\newcommand{\redc}[1]{\textcolor{black}{#1}}

\begin{document}
\shortauthors{Harikane et al.}
\slugcomment{Accepted for publication in ApJ}

\email{hari@icrr.u-tokyo.ac.jp}
\author{
Yuichi Harikane\altaffilmark{1,2},
Masami Ouchi\altaffilmark{1,3},
Yoshiaki Ono\altaffilmark{1},
Surhud More\altaffilmark{3},
Shun Saito\altaffilmark{3},
Yen-Ting Lin\altaffilmark{4},
Jean Coupon\altaffilmark{5},
Kazuhiro Shimasaku\altaffilmark{6,7},
Takatoshi Shibuya\altaffilmark{1},
Paul A. Price\altaffilmark{8},
Lihwai Lin\altaffilmark{4},
Bau-Ching Hsieh\altaffilmark{4},
Masafumi Ishigaki\altaffilmark{1,2},
Yutaka Komiyama\altaffilmark{9},
John Silverman\altaffilmark{3},
Tadafumi Takata\altaffilmark{9},
Hiroko Tamazawa\altaffilmark{1,2},
and 
Jun Toshikawa\altaffilmark{9}
}

\altaffiltext{1}{
Institute for Cosmic Ray Research, The University of Tokyo, 5-1-5 Kashiwanoha, Kashiwa, Chiba 277-8582, Japan
}
\altaffiltext{2}{
Department of Physics, Graduate School of Science, The University of Tokyo, 7-3-1 Hongo, Bunkyo, Tokyo, 113-0033, Japan
}
\altaffiltext{3}{
Kavli Institute for the Physics and Mathematics of the Universe (Kavli IPMU, WPI), University of Tokyo, Kashiwa, Chiba 277-8583, Japan
}
\altaffiltext{4}{
Institute of Astronomy \& Astrophysics, Academia Sinica, Taipei 106, Taiwan (R.O.C.)
}
\altaffiltext{5}{
Astronomical Observatory of the University of Geneva, ch. d’Ecogia 16, 1290 Versoix, Switzerland
}
\altaffiltext{6}{
Department of Astronomy, Graduate School of Science, The University of Tokyo, Hongo, Bunkyo, Tokyo 113-0033, Japan
}
\altaffiltext{7}{
Research Center for the Early Universe, The University of Tokyo, Hongo, Tokyo 113-0033, Japan
}
\altaffiltext{8}{
Princeton University Observatory, Peyton Hall, Princeton, NJ 08544, USA
}
\altaffiltext{9}{
National Astronomical Observatory, Mitaka, Tokyo 181-8588, Japan
}
\altaffiltext{10}{
Center for Astronomy and Astrophysics, Department of Physics \& Astronomy, Shanghai Jiao Tong University, 800 Dongchuan Road, Shanghai, 200240, China
}

\shorttitle{Evolution of SHMR at $z=0-7$}

\title{
Evolution of Stellar-to-Halo Mass Ratio at $\lowercase{z}=0-7$ Identified by Clustering Analysis 
with the Hubble Legacy Imaging and Early Subaru/Hyper Suprime-Cam Survey Data
}

\begin{abstract}
We present clustering analysis results from 10,381 Lyman break galaxies (LBGs) at $z\sim 4-7$,
identified in the Hubble legacy deep imaging and 
new complimentary large-area Subaru/Hyper Suprime-Cam data.
We measure the angular correlation functions (ACFs) of these LBGs at $z\sim4$, $5$, $6$, and $7$, and fit these measurements using halo occupation distribution (HOD) models that provide an estimate of halo masses, $M_\m{h}\sim(1-20)\times10^{11}\ \Msun$.
Our $M_{\rm h}$ estimates agree with those obtained by previous clustering studies
in a UV-magnitude vs. $M_{\rm h}$ plane, and allow us to calculate stellar-to-halo mass ratios (SHMRs) of LBGs.
By comparison with the $z\sim 0$ SHMR, we identify evolution of the SHMR from 
$z\sim0$ to $z\sim4$, and $z\sim4$ to $z\sim7$ at the $>98\%$ confidence levels. 
The SHMR decreases by a factor of \redc{$\sim 2$} from $z\sim 0$ to $4$, and increases by a factor of \redc{$\sim 4$} from $z\sim 4$ to $7$ at the dark matter halo mass of $M_\m{h}\sim10^{11}\ \Msun$.
\redc{We compare our SHMRs with results of a hydrodynamic simulation and a semi-analytic model, and find that these theoretical studies do not predict the SHMR increase from $z\sim4$ to $7$.}
We obtain the baryon conversion efficiency (BCE) of LBGs at $z\sim 4$, and find that the BCE increases 
with increasing dark matter halo mass. 
Finally, we compare our clustering+HOD estimates with results from
abundance matching techniques, and conclude that
the $M_{\rm h}$ estimates of the clustering+HOD analyses agree
with those of the simple abundance matching
within a factor of 3, and that the agreement
improves when using more sophisticated abundance matching
techniques that include subhalos, incompleteness, and/or evolution in the star formation and stellar mass functions.
\end{abstract}


\section{Introduction}\label{ss_intro}
Dark matter halos play an important role in galaxy formation and evolution
in the framework of Lambda cold dark matter ($\Lambda$CDM) structure
formation models. Such halos can regulate processes such as gas cooling necessary for star formation.
Gas cooling is efficient in halos with masses of $10^{10}-10^{13}\ \Msun$, 
where the gas cooling time scale is shorter than the gas infall time scale \redc{\citep{1977MNRAS.179..541R,1993ApJS...88..253S,1993PhR...231..293S}}.
In low and high mass haloes, feedback from supernova (SN), radiation pressure, and active galactic nucleus (AGN) are thought to suppress star formation 
by thermal and kinetic energy input \citep[e.g.,][]{2005ApJ...618..569M,2009Natur.457..451D,2009MNRAS.396.2332K,2014ApJ...794..129H}.
The connection between galaxies and their dark matter halos is essential for understanding galaxy formation, 
and specifically the stellar-to-halo mass ratio (SHMR), which is defined as the ratio of a galaxy's stellar mass to its halo mass, is one of the key quantities.
The SHMR comprises the integrated efficiency of the past stellar mass assembly (i.e., star formation and mergers).
The SHMR has been theoretically investigated with the help of hydrodynamic simulations or semi-analytic models \redc{\citep[e.g.,][]{2014MNRAS.445..581H,2014ApJ...780..145T,2014ApJ...793...12B,2014PASJ...66...70O,2015arXiv151008463M,2015MNRAS.453.4337S}}.
From observational studies, the SHMR is measured by analyses of galaxy clustering, weak lensing, satellite kinematics, \redc{and rotation curves}
at low-redshift \redc{\citep[e.g.,][]{2006MNRAS.368..715M,2011MNRAS.410..210M,2012ApJ...744..159L,2015MNRAS.447..298H,2015arXiv150200313S,2015ApJ...799..130R,2015MNRAS.449.1352C,2015ApJ...807..152S,2015PASJ..tmp..266S}}.
These low-redshift studies find a SHMR with a peak at a dark matter halo mass of 
$M_\m{h}\sim10^{12}\ \Msun$, independent of redshift, and referred to as a pivot halo mass.
\citet{2012ApJ...744..159L} claim a redshift evolution of SHMRs (and pivot halo masses) from $z\sim0$ to $z\sim1$.
While some studies estimate the SHMR with clustering analysis at $z>1$ \citep[e.g.,][]{2010MNRAS.406..147F,2015A&A...576L...7D,2015MNRAS.449..901M,2015arXiv151105476H,2016MNRAS.tmp...55I}, it is difficult to investigate the evolution of SHMR at these high redshift due to poor statistics based on small galaxy samples available to date \citep[c.f.,][]{2015MNRAS.449..901M,2015arXiv151105476H}.

The abundance matching technique is another indirect probe of the SHMR.
The abundance matching technique connects galaxies to their host dark matter haloes by matching the cumulative stellar mass function (or the cumulative luminosity function) and the cumulative halo mass function.
Because this technique only requires one-point statistics that are easily measured, 
many recent studies apply this method from low-redshift to high redshift galaxies 
\citep{2013MNRAS.428.3121M,2013ApJ...770...57B,2015arXiv150400005F,2015arXiv150700999M,2015arXiv150702685T,2015arXiv150900482S}.
\citet{2013ApJ...770...57B} investigate the SHMR with abundance matching by using stellar mass functions
as well as specific star formation rates and cosmic star formation rate densities; they find that the SHMR evolves from $z=0$ to $z=8$.
While abundance matching is a useful and less expensive method 
to connect galaxies to their dark matter haloes, there are two major systematic uncertainties with respect to the application to high redshift galaxies.
One uncertainty is the star-formation duty cycle (DC) that is defined as the probability of a halo of given mass to host an observable star forming galaxy.
Most abundance matching studies of $z\gtrsim 4$ Lyman break galaxies (LBGs) use the UV luminosity function,
assuming their star-formation (i.e. UV-bright phase) DC is unity \citep[c.f.,][]{2013ApJ...770...57B}. 
However, star formation activity can be episodic.
Moreover, populations of passive and dusty star-forming galaxies are expected to exist that may be missed in LBG samples.
In fact, \citet{2009ApJ...695..368L} claim that the DC is $\sim0.3$ at $z\sim4$, 
and \cite{2001ApJ...558L..83O} indicate a halo mass-dependent DC based on  clustering analysis.
The other uncertainty, with respect to abundance matching, is the subhalo-galaxy relation.
While the majority of abundance matching studies include
subhalos \citep[subhalo abundance matching; e.g.,][]{2013MNRAS.428.3121M,2013ApJ...770...57B,2015arXiv150400005F,2015arXiv150700999M},
the subhalo-galaxy relation is poorly constrained.
For example, it is unclear which subhalo property best correlates with the stellar mass \citep[or luminosity;][]{2013ApJ...771...30R,2015arXiv150807012G}.
Preferably, one needs information independent from abundance to understand these systematics.
Because lensing analysis is not feasible for galaxies at $z\gtrsim 2$ due to the limited number of background galaxies and their lower image quality,
clustering analysis is a promising technique to test the abundance matching results and to extend our understanding of the connection between galaxies and dark matter halos to high redshift.

\redc{
The clustering analysis of the high redshift galaxies has been conducted with large survey data. 
\citet{2001ApJ...558L..83O,2004ApJ...611..685O,2005ApJ...635L.117O} obtained wide area data taken with the Subaru deep survey, and studied the clustering of LBGs at $z\sim4$ and $5$. 
As well, \citet{2009A&A...498..725H} estimated the angular correlation functions (ACFs) of LBGs at $z\sim3-5$ with high accuracy using the Canada-France-Hawaii Telescope Legacy Survey (CFHTLS) data. 
With the LBT Bo\"otes field survey data, \citet{2013ApJ...774...28B} studied the clustering properties of LBGs at $z\sim3$. 
Recently, the deep data of the Hubble Space Telescope legacy survey allowed us to study LBGs at $z\sim7$ \citep{2014ApJ...793...17B}.
Furthermore, \citet{2015MNRAS.454..205I} investigated the clustering properties of $z\sim2$ star forming galaxies using the wide area data taken by the United Kingdom Infra-Red Telescope (UKIRT), Subaru telescope, and CFHT.}

Recently a wide-field mosaic CCD camera, Hyper Suprime-Cam \citep[HSC;][]{2012SPIE.8446E..0ZM}, 
has been installed at the prime focus of the Subaru telescope \redc{\citep{2004PASJ...56..381I}}.
HSC has a field-of-view (FoV) of $1.75\ \m{deg}^2$ with a high sensitivity accomplished with 
the Subaru 8m primary mirror.
An HSC legacy survey under the Subaru Strategic Program (SSP; PI: S. Miyazaki) has been allocated 300 nights over 5 years, and
has been ongoing since March 2014.\footnote{http://www.naoj.org/Projects/HSC/surveyplan.html}
The HSC SSP has three survey layers of Wide, Deep, and Ultradeep
that will cover the sky areas of $1400$, $27$, and $3.5\ \m{deg^2}$ 
with the $5\sigma$ point-source limiting magnitudes of $r\simeq26\ \m{mag}$, $r\simeq27\ \m{mag}$, and $r\simeq28\ \m{mag}$, respectively.
It is expected that full HSC SSP data sets will provide us with $\sim2\times10^7$ LBGs at $z\gtrsim4$, 
which are $\sim400$ times larger than current samples identified in the deep fields of the CFHTLS \citep{2009A&A...498..725H}, and 
allow us to investigate statistical properties of LBGs down to $\sim 0.1 L^*$. 
Complementing these HSC SSP efforts, recent deep Hubble Space Telescope observations 
with Advanced Camera for Surveys (ACS) and Wide Field Camera 3 (WFC3) 
provide samples of $>10^4$ LBGs whose luminosities 
reach below $\sim 0.1 L^*$ \citep[e.g.,][]{2015ApJ...803...34B,2015ApJ...799...12I}.
In this study, we use the unique combined data sets of Subaru/HSC and Hubble/ACS+WFC3 
to investigate the galaxy-dark matter connection using LBGs over a wide luminosity range,
and investigate SHMRs at $z\gtrsim 4$, for the first time, using clustering analyses.

This paper is organized as follows.
We present the observational data sets of Subaru/HSC and Hubble/ACS+WFC3 in Section \ref{ss_dataset}.
We describe the photometry and sample selection of LBGs in Section \ref{ss_photosample}.
The clustering analysis is presented in Section \ref{ss_analysis}.
Sections \ref{ss_result_mass} and \ref{ss_result_SHMR} detail our results on the dark matter halo mass and SHMR, respectively.
We discuss the implications of the SHMR evolution and differences between our results and those from abundance matching in Section \ref{ss_discussion}.
Section \ref{ss_summary} summarizes our findings.
Throughout this paper we use the following cosmological model: 
$\Omega_\m{m}=0.3$, $\Omega_\Lambda=0.7$, $\Omega_\m{b}=0.045$, $H_0=70\ \m{\kms Mpc^{-1}}$, and $\sigma_8=0.8$.
We use $r_\m{200}$ that is the radius in which the mean enclosed density is 200 times higher than the mean cosmic density.
To define the halo mass, we use $M_\m{200}$ that is the total mass enclosed in $r_\m{200}$.
We assume a \citet{2003PASP..115..763C} initial mass function (IMF).
All magnitudes are in the AB system \redc{\citep{1983ApJ...266..713O}}.

\begin{deluxetable*}{ccccccccccccc}
\setlength{\tabcolsep}{0.35cm}
\tablecaption{Limiting Magnitudes of the Hubble Data}
\tablehead{ & \multicolumn{12}{c}{$5\sigma$ Limiting Magnitude}\\
\colhead{} & \colhead{Area} & \multicolumn{10}{c}{Hubble} & \colhead{CFHT/Subaru}\\
\colhead{Field} & \colhead{($\m{arcmin^2}$)} & \colhead{$B_{435}$} & \colhead{$V_{606}$} & \colhead{$i_{775}$} & \colhead{$I_{814}$} & \colhead{$z_{850}$} & \colhead{$Y_{105}$} & \colhead{$J_{125}$}& \colhead{$JH_{140}$} & \colhead{$H_{160}$} & \colhead{coadd\tablenotemark{a}} & \colhead{$r$}\\
\colhead{(1)}& \colhead{(2)}& \colhead{(3)}& \colhead{(4)} &  \colhead{(5)}& \colhead{(6)}& \colhead{(7)}& \colhead{(8)}& \colhead{(9)}& \colhead{(10)}& \colhead{(11)}& \colhead{(12)}& \colhead{(13)} }

\startdata

HUDF & 3.7 & 30.0 & 30.5 & 30.1 & 29.3 & 29.6 & 30.2 & 29.9 & 29.8 & 29.9 & 30.6 & \nodata \\
GOODS-N-Deep & 57.4 & 28.6 & 28.8 & 28.3 & 30.5 & 28.1 & 27.9 & 28.3 & \nodata & 28.1 & 28.6 & \nodata \\
GOODS-N-Wide & 58.2 & 28.6 & 28.7 & 28.2 & 29.9 & 28.0 & 27.7 & 27.6 & \nodata & 27.5 & 28.1 & \nodata \\
GOODS-S-Deep & 52.1 & 28.6 & 28.8 & 28.2 & 28.8 & 29.0 & 28.4 & 28.4 & \nodata & 28.3 & 29.0 & \nodata \\
GOODS-S-Wide & 30.4 & 28.6 & 28.8 & 28.2 & 28.4 & 28.0 & 27.7 & 27.8 & \nodata & 27.6 & 28.3 & \nodata \\
CANDELS-AEGIS & 174.9 & \nodata & 28.3  & \nodata & 27.8 & \nodata & \nodata & 27.6 & \nodata & 27.7 & 28.0 & 28.1 \\
CANDELS-COSMOS & 122.0& \nodata & 28.3 & \nodata & 28.0 & \nodata & \nodata & 27.6 & \nodata & 27.6 & 27.9 & 27.9/27.7 \\
CANDELS-UDS & 129.3 & \nodata & 28.2 & \nodata & 28.2 & \nodata & \nodata & 27.5 & \nodata & 27.6 & 27.9 & 28.2\\
HFF-Abell2744P & 3.1 & 28.8 & 29.1 & \nodata & 28.8 & \nodata & 29.0 & 28.8 & 28.8 & 28.8 & 29.3 & \nodata \\
HFF-MACS0416P & 3.8 & 28.6 & 28.9 & \nodata & 28.8 & \nodata & 29.3 & 29.1 & 29.1 & 29.0 & 29.5 & \nodata \\ \hline
PSF FWHM\tablenotemark{b} & & 0.\carcsec12 & 0.\carcsec11 & 0.\carcsec10 & 0.\carcsec11 & 0.\carcsec21 & 0.\carcsec20 & 0.\carcsec20 & 0.\carcsec20 & 0.\carcsec20 & 0.\carcsec21 & 0.\carcsec8
\enddata

\tablecomments{Columns: (1) Field. (2) Effective Area in $\m{arcmin^2}$. (3)-(12) Limiting magnitudes which correspond to $5 \sigma$ variations in the sky flux measured in a circular aperture of $0.\carcsec35$-diameter in PSF-matched images. (13) Limiting magnitude defined by a $5\sigma$ sky noise in a $1.\carcsec0$-diameter aperture. See \citet{2014ApJS..214...24S} for limiting magnitudes in other bands.}
\tablenotetext{a}{Coadd image of $Y_{105}J_{125}JH_{140}H_{160}$-bands. }
\tablenotetext{b}{Mean PSF FWHM values.}
\label{table_hst_image}
\end{deluxetable*}

\section{Observational Data Sets}\label{ss_dataset}
\subsection{Hubble Data}
\label{sec:hubble_data}
We use 10 deep optical-NIR imaging data sets 
of the Hubble Ultra Deep Field (HUDF), Great Observatories Origins Deep Survey (GOODS)-North-Deep, GOODS-North-Wide, GOODS-South-Deep, GOODS-South-Wide, 
Cosmic Assembly Near-infrared Deep Extragalactic Legacy Survey (CANDELS)-All-Wavelength Extended Groth Strip International Survey (AEGIS), CANDELS-Cosmological Evolution Survey (COSMOS), CANDELS-Ultra Deep Survey (UDS), Hubble Frontier Field (HFF)-Abell2744P, and HFF-MACS0416P
that are taken with ACS and WFC3 on the Hubble Space Telescope.
The total area of the Hubble data is $\sim600\ \m{arcmin^2}$.
We mask regions that are contaminated by the halos of bright stars or diffraction spikes by visual inspection,
and measure limiting magnitudes in a  $0.\carcsec35$-diameter circular aperture 
with {\sc sdfred} \citep{2002AJ....123...66Y,2004ApJ...611..660O},
after homogenizations of the point-spread functions (PSFs; see Section \ref{ss_PSF_photo} for more details).
The typical FWHMs of the PSFs of ACS and WFC3 images are $0.\carcsec1$ and $0.\carcsec2$, respectively.
The limiting magnitudes, PSF FWHMs, and effective areas of these images are summarized in Table \ref{table_hst_image}.

\subsubsection{HUDF}
HUDF has the deepest ACS and WFC3 imaging data, ever taken,
from the combination of the three surveys, HUDF \citep{2006AJ....132.1729B},
HUDF09 \citep[GO 11563; PI: G. Illingworth; e.g.,][]{2010ApJ...708L..69B},
and HUDF12 \citep[GO 12498; PI: R. Ellis; e.g.,][]{2013ApJ...763L...7E,2013ApJS..209....3K}.
We use the combined HUDF data set compiled by the eXtreme Deep Field (XDF) team\footnote{https://archive.stsci.edu/prepds/xdf/} \citep{2013ApJS..209....6I}.
The HUDF data consist of 9-band images of 
$B_{435}V_{606}i_{775}I_{814}z_{850}Y_{105}J_{125}JH_{140}H_{160}$, and 
cover $\sim4\ \m{arcmin^2}$ sky area. The $5 \sigma$ limiting magnitudes 
are $\sim30\ \m{mag}$ over these 9 bands. 

\subsubsection{GOODS-North and GOODS-South}
We use the data sets of the GOODS-North and GOODS-South fields,
available from the CANDELS and 3D-HST teams \citep{2012ApJS..200...13B,2014ApJS..214...24S} 
,\footnote{http://candels.ucolick.org/} \footnote{http://3dhst.research.yale.edu/Home.html}
that are obtained by CANDELS 
\citep[PIs: S. Faber and H. Ferguson;][]{2011ApJS..197...35G,2011ApJS..197...36K}. 
The GOODS fields are comprised of deep and wide survey data
whose $5 \sigma$ limiting magnitudes are typically $\sim28.5\ \m{mag}$ and $\sim27.5\ \m{mag}$, respectively.
About half of the GOODS-North and GOODS-South fields are deep survey areas, while
the remaining half of GOODS-North and quarter of the GOODS-South
are wide survey areas. The GOODS-North and GOODS-South data are observed with 
bands of $B_{435}V_{606}i_{775}I_{814}z_{850}Y_{105}J_{125}H_{160}$ 
with effective areas of $\sim120$ and $\sim90\ \m{arcmin^2}$, respectively.

\subsubsection{CANDELS-AEGIS, CANDELS-COSMOS, and CANDELS-UDS}
The largest area Hubble data sets in our study come from
CANDELS-AEGIS, CANDELS-COSMOS, and CANDELS-UDS imaging data 
\citep{2011ApJS..197...35G,2011ApJS..197...36K}
available from the 3D-HST \citep{2012ApJS..200...13B,2014ApJS..214...24S}.
These imaging regions are covered by ACS $V_{606}I_{814}$ and WFC3 $J_{125}H_{160}$ observations
with the typical $5\sigma$ limiting magnitude of $27.5\ \m{mag}$.
Ground-based optical images taken with the CFHT and Subaru telescope are also available for these fields.
We use the CFHT $ugr$ band images of the CANDELS-AEGIS field, the CFHT $ugr$ and Subaru $BVr$ band images 
of the CANDELS-COSMOS field, and the CFHT $u$ and Subaru $BVr$ band images of the CANDELS-UDS field.

\subsubsection{HFF-Pallarels}
Our study also includes imaging data from 
HFF \cite[PI J. Lotz; e.g.,][]{2015ApJ...799...12I,2015ApJ...804..103K}.
These data are parallel-field observations of Abell2744 and MACS0416 galaxy clusters
that are taken from the HFF team.\footnote{http://www.stsci.edu/hst/campaigns/frontier-fields/}
Because lensing effects such as magnification and survey volume distortion 
are negligibly weak in HFF parallel fields (see, e.g.,  \citealt{2015ApJ...799...12I}),
we regard these HFF parallel images as blank field data.
These two HFF parallel fields are observed with 
7 bands of $B_{435}V_{606}I_{814}Y_{105}J_{125}JH_{140}H_{160}$ 
over a total effective area of $\sim7\ \m{arcmin^2}$.
The typical $5\sigma$ limiting magnitude is $29.0\ \m{mag}$.

\subsection{Subaru Data}
Our study includes early data of the HSC SSP survey taken from March to November of 2014 (S14A\_0b). 
We use the HSC SSP Wide layer $griz$ data of the XMM field ($\m{R.A.}=2^\m{h}17^\m{m}00^\m{s}$, $\m{decl.}=-5{\arcdeg}12{\arcmin}00{\arcsec}$ [J2000]) and GAMA09h field ($\m{R.A.}=8^\m{h}47^\m{m}00^\m{s}$, $\m{decl.}=0{\arcdeg}45{\arcmin}00{\arcsec}$ [J2000]).\footnote{These are the central coordinates of the early HSC data that we use.}
While the HSC data is $\sim3-6$ magnitudes shallower than the Hubble data, the HSC data cover $\sim90$ times larger effective area: $8.3$ and $6.9\ \m{deg}^2$, in XMM and GAMA09h, respectively. As a result, the HSC data can provide clustering measurements at the bright end.
The HSC data are reduced by the HSC SSP collaboration with hscPipe (version 3.4.1) that is 
the HSC data reduction pipeline based on the Large Synoptic Survey Telescope (LSST) software pipeline 
\citep{2008arXiv0805.2366I,2010SPIE.7740E..15A}.
The HSC data reduction pipeline performs CCD-by-CCD
reduction and calibration for astrometry, warping, coadding, and
photometric zeropoint measurements.
The astrometric and photometric calibration are based on the data of 
Panoramic Survey Telescope and Rapid Response System (Pan-STARRS) 1 imaging survey 
\citep{2013ApJS..205...20M,2012ApJ...756..158S,2012ApJ...750...99T}.
We mask imaging regions contaminated with diffraction spikes and halos of bright stars using the mask extension outputs from the HSC data reduction pipeline and information of bright stars from our source catalogs (Section \ref{ss_subaru_source}) and the Sloan Digital Sky Survey (SDSS) DR12 \citep{2015ApJS..219...12A}.\footnote{http://www.sdss.org/dr12/}
We use the PSF outputs from the pipeline \citep{2011PASP..123..596J}, and typical PSF FWHMs are $0.\carcsec6-0.\carcsec9$.
The $5\sigma$ limiting magnitudes measured with {\sc sdfred} are $\sim25-26\ \m{mag}$
(Table \ref{table_hsc_image}).

\begin{deluxetable}{cccccc}
\setlength{\tabcolsep}{0.35cm}
\tablecaption{Limiting Magnitudes of the Subaru/HSC Data}
\tablehead{
\colhead{} & \colhead{Area} & \multicolumn{4}{c}{$5\sigma$ Limiting Magnitude}\\
\colhead{Field} & \colhead{($\m{arcmin^2}$)} & \colhead{$g$} & \colhead{$r$} & \colhead{$i$} & \colhead{$z$} \\
\colhead{(1)}& \colhead{(2)}& \colhead{(3)}& \colhead{(4)} &  \colhead{(5)}& \colhead{(6)} }
\startdata
HSC-XMM &30100 & 26.3 & 25.8 &  25.8 & 25.1\\
HSC-GAMA09h &24800 &  26.3 &25.7  &  25.3 &  25.0\\ \hline
PSF FWHM\tablenotemark{a} & & 0.\carcsec82 & 0.\carcsec85 & 0.\carcsec62 & 0.\carcsec67
\enddata

\tablecomments{Columns: (1) Field. (2) Effective Area in $\m{arcmin^2}$. (3)-(6) Limiting magnitudes defined by a $5\sigma$ sky noise in a PSF-$75\%$-flux-radius circular aperture in PSF-matched images.
}
\tablenotetext{a}{Mean PSF FWHM values.}
\label{table_hsc_image}
\end{deluxetable}

\section{Photometric Samples and LBG Selections}\label{ss_photosample}
\subsection{Hubble Samples}
\subsubsection{Multi-band Photometric Catalogs}\label{ss_PSF_photo}
We construct multi-band source catalogs from the Hubble data.
To measure object colors, we match the image PSFs to the WFC3 $H_{160}$-band images 
whose typical FWHM of the PSF is $\simeq 0.\carcsec2$, the largest of the Hubble multi-band images.
We use {\sc SWarp} \citep{2002ASPC..281..228B} to produce our detection images 
that are the co-added data of $Y_{105}$, $J_{125}$, $JH_{140}$ and $H_{160}$-band images.
The $5\sigma$ limiting magnitudes of the detection images are typically $\sim0.5\ \m{mag}$ 
deeper than those of the single-band images (Table \ref{table_hst_image}).

We perform source detection and photometry
with SExtractor \citep{1996A&AS..117..393B}.
We run SExtractor (version 2.8.6) in dual-image mode for each multi-band image with its detection image,
having the parameter set as follows:
${\tt DETECT\_MINAREA}=6$, ${\tt DETECT\_THRESH}=2.0$, 
${\tt ANALYSIS\_THRESH}=2.0$, ${\tt DEBLEND\_NTHRESH}=32$, and ${\tt DEBLEND\_MINCOUNT}=0.005$.
The total number of the objects detected is 130,655.
We measure the object colors with {\tt MAG\_APER} magnitudes defined in a $0.\carcsec35$-diameter circular aperture.
\redc{We use the {\tt MAG\_AUTO} measurements of SExtractor for total magnitudes.}
In the CANDELS-AEGIS, CANDELS-COSMOS, and CANDELS-UDS fields, we use the CFHT and Subaru imaging data
to reduce low-$z$ interlopers from high-$z$ galaxy samples.
Because we only need magnitude upper limits of high-$z$ galaxy candidates for this purpose,
we do not homogenize the PSFs of the CFHT and Subaru images. We obtain aperture magnitudes
of SExtractor {\tt MAG\_APER} with a $1.\carcsec0$-diameter circular aperture.
If a source is not detected either in a Hubble or CFHT/Subaru band, 
we replace the source flux with the $1\sigma$-upper limit flux.

\subsubsection{Lyman Break Galaxy Selection}\label{selection}
We select LBGs from our source catalogs using color information.
From the HUDF, GOODS-North, and GOODS-South source catalogs, 
we select LBGs at $z\sim4$, $5$, $6$, and $7$
with the following LBG color criteria as given in \citet{2015ApJ...803...34B}:

$z\sim4$ 
\begin{eqnarray}
B_{435}-V_{606} > 1, \label{eq_z4_1} \\
i_{775}-J_{125} < 1, \label{eq_z4_2} \\
B_{435}-V_{606} > 1.6(i_{775}-J_{125})+1. \label{eq_z4_3}
\end{eqnarray}

$z\sim5$
\begin{eqnarray}
\label{eq:lbg_hudf_z5}
V_{606}-i_{775} > 1.2, \label{eq_z5_1}\\
z_{850}-H_{160} < 1.3, \label{eq_z5_2}\\
V_{606}-i_{775} > 0.8(z_{850}-H_{160})+1.2. \label{eq_z5_3}
\label{eq:lbg_hudf_z5_e}
\end{eqnarray}

$z\sim6$
\begin{eqnarray}
\label{eq:lbg_hudf_z6}
i_{775}-z_{850} > 1.0, \label{eq_z6_1}\\
Y_{105}-H_{160} < 1.0, \label{eq_z6_2}\\
i_{775}-z_{850} > 0.777(Y_{105}-H_{160})+1.0. \label{eq_z6_3}
\label{eq:lbg_hudf_z6_e}
\end{eqnarray}

$z\sim7$
\begin{eqnarray}
z_{850}-Y_{105} > 0.7, \label{eq_z7_1}\\
J_{125}-H_{160} < 0.45, \label{eq_z7_2}\\
z_{850}-Y_{105} > 0.8(J_{125}-H_{160})+0.7. \label{eq_z7_3}
\end{eqnarray}

\begin{figure*}
 \begin{minipage}{0.48\hsize}
  \begin{center}
   \includegraphics[width=0.99\hsize,bb=10 10 420 420,clip]{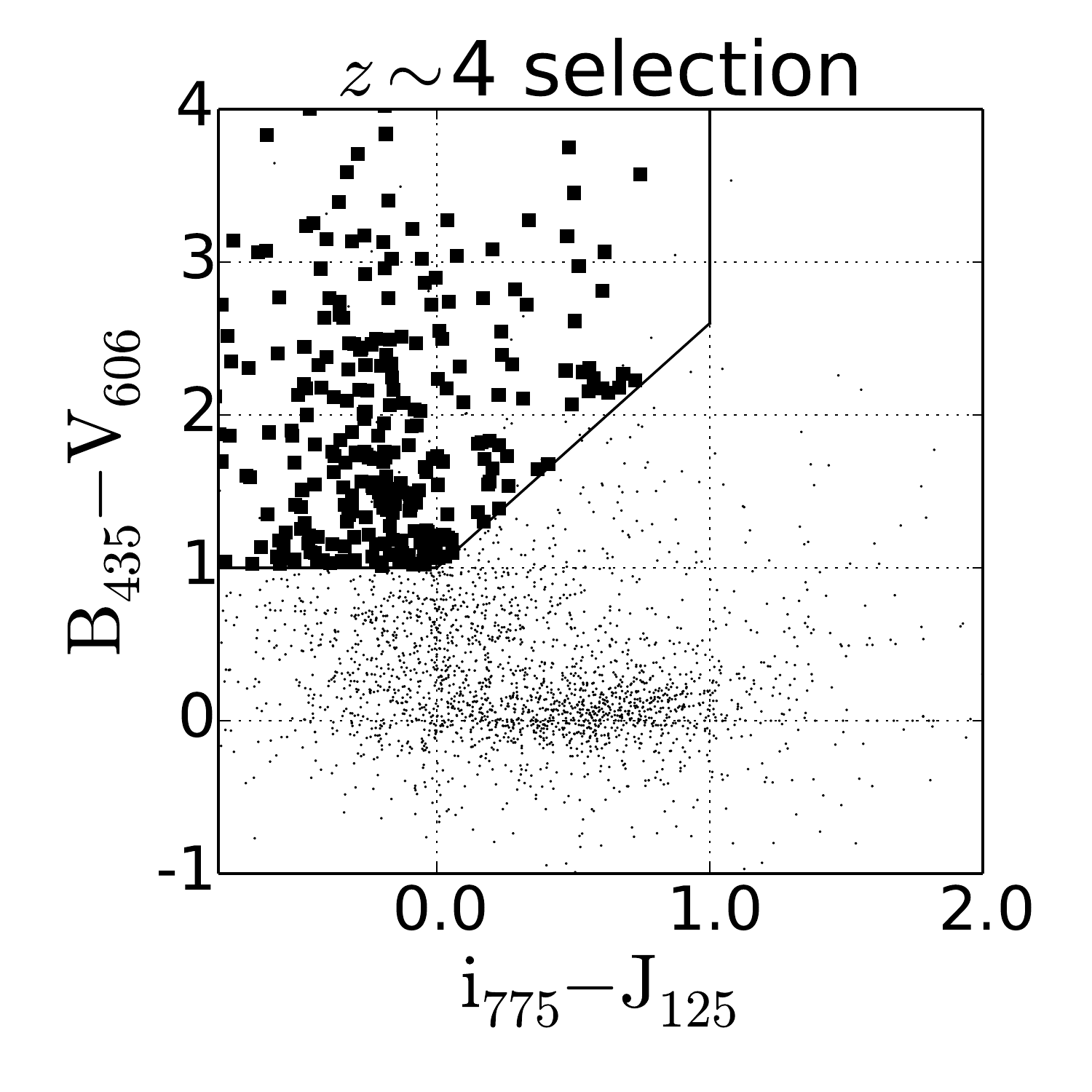}
  \end{center}
 \end{minipage}
 \begin{minipage}{0.48\hsize}
 \begin{center}
  \includegraphics[clip,bb=10 10 420 420,width=1\hsize]{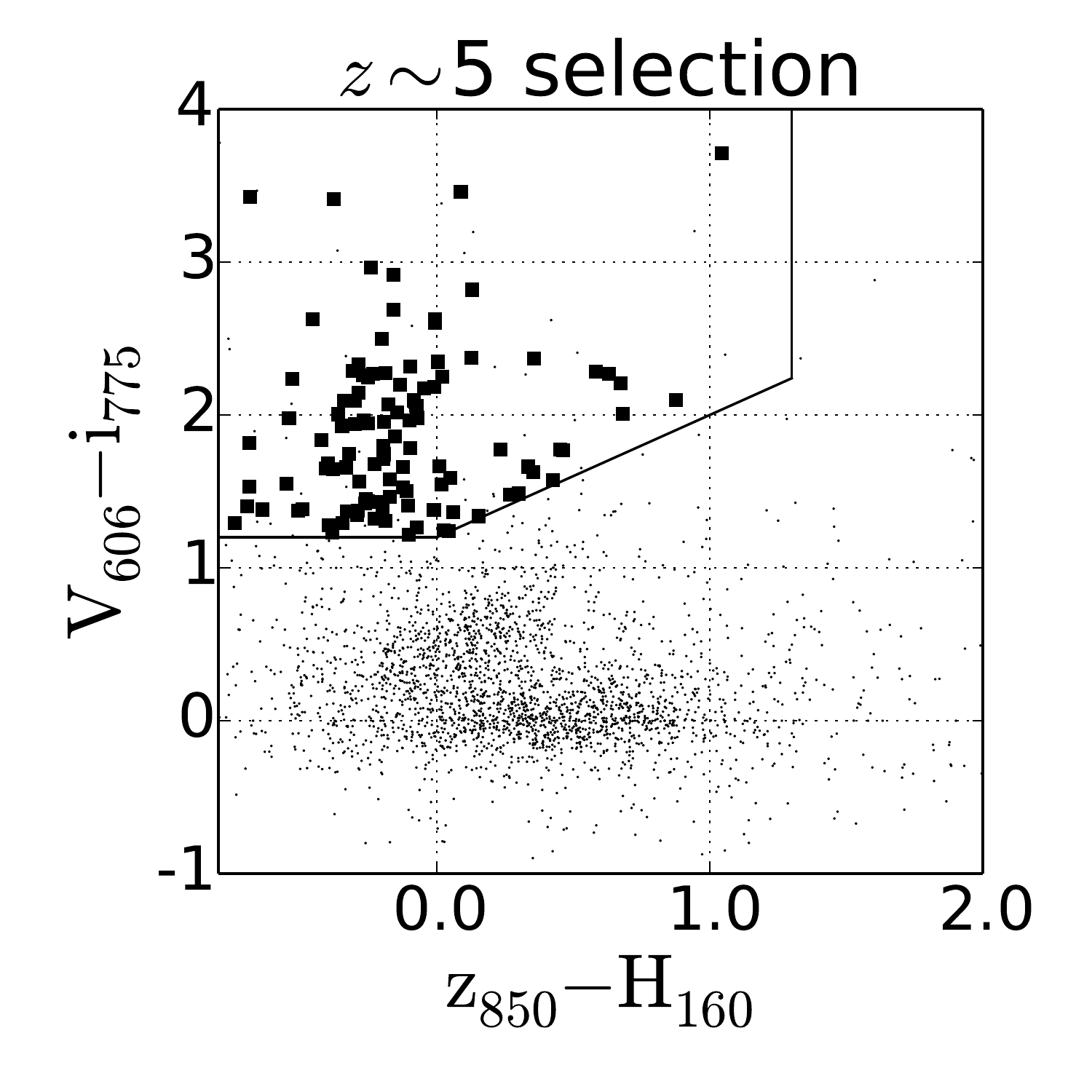}
 \end{center}
 \end{minipage}
  \\
  \begin{minipage}{0.48\hsize}
  \begin{center}
   \includegraphics[clip,bb=10 10 420 420,width=1\hsize]{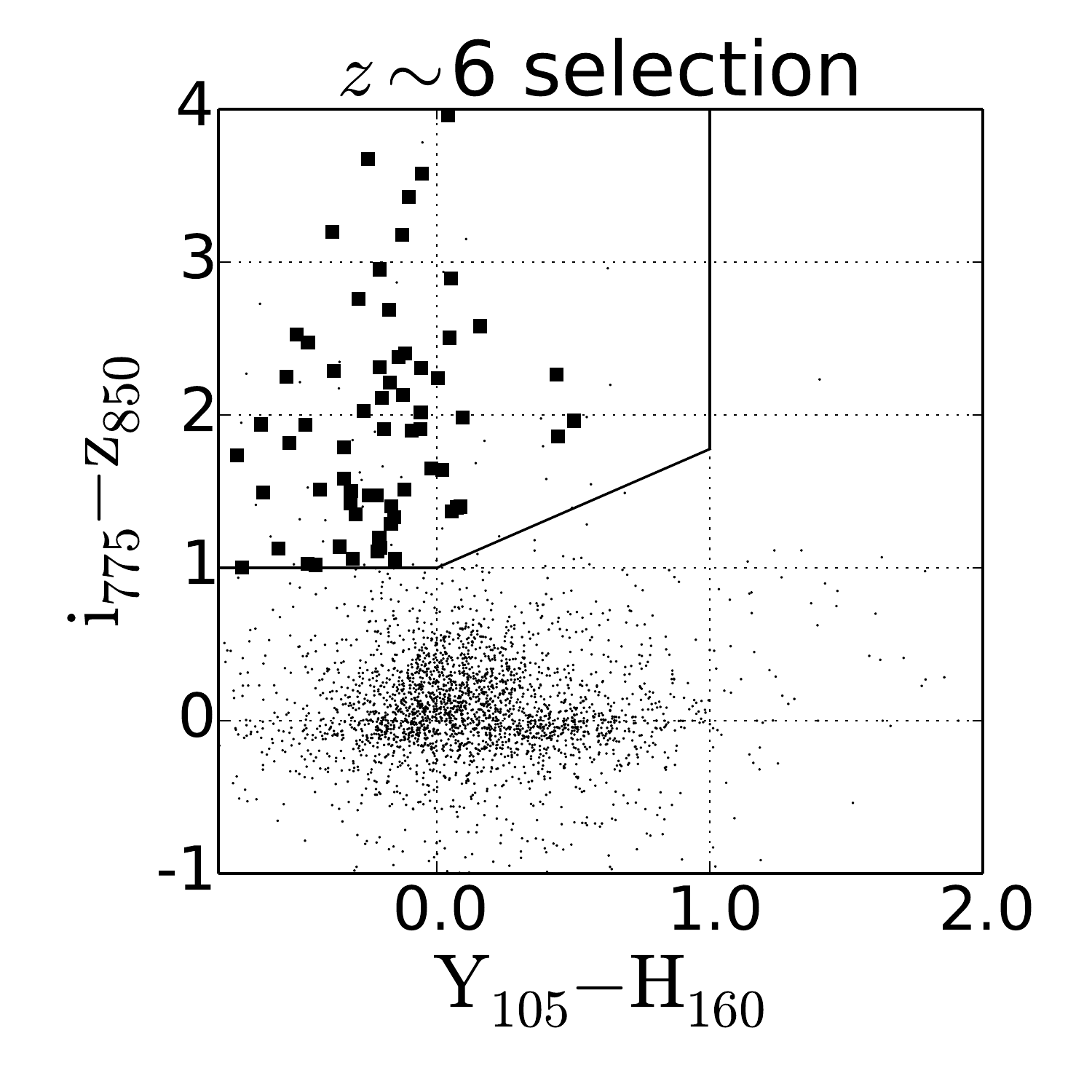}
  \end{center}
 \end{minipage}
 \begin{minipage}{0.48\hsize}
 \begin{center}
  \includegraphics[clip,bb=10 10 420 420,width=1\hsize]{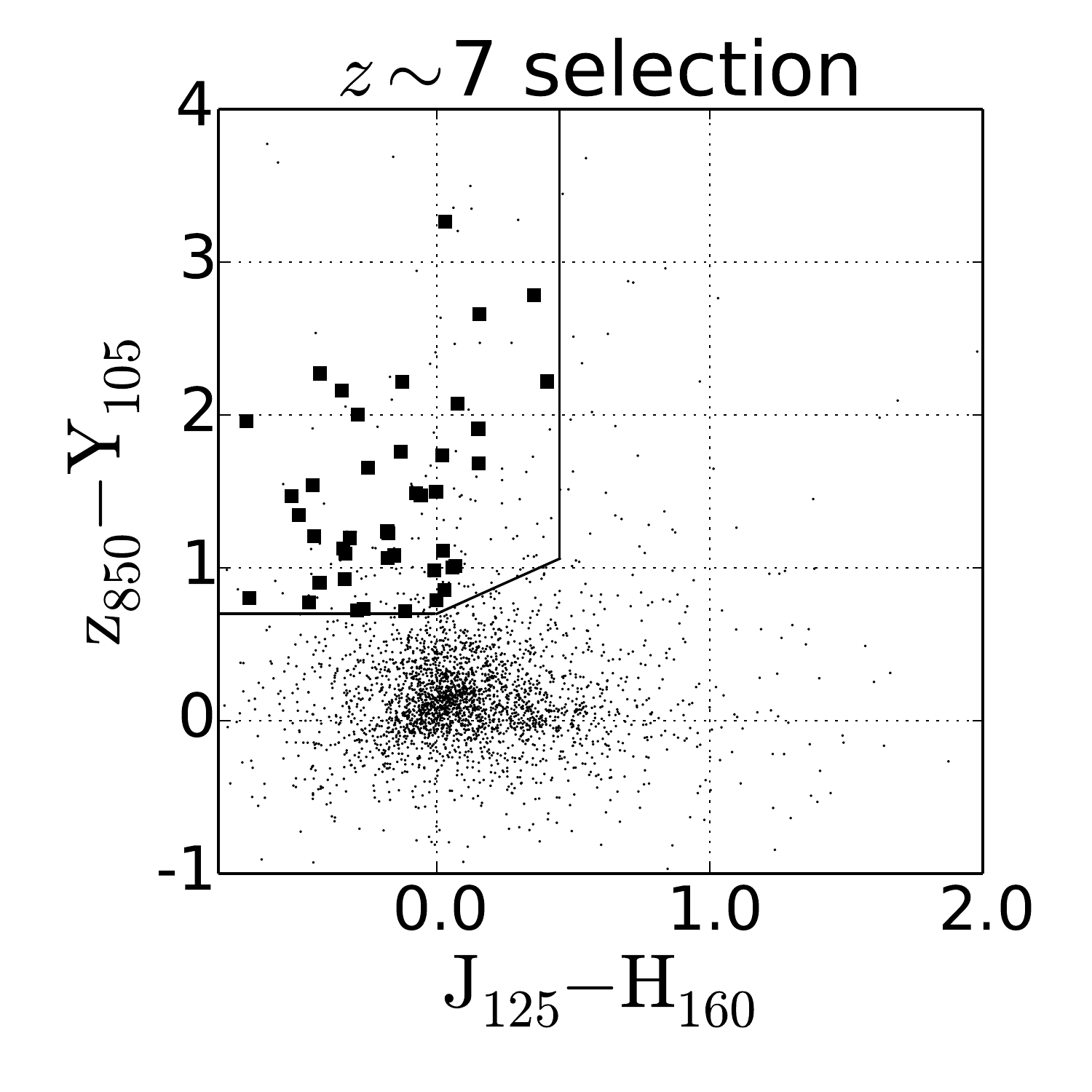}
 \end{center}
 \end{minipage}
 \caption{Two-color diagrams for selection of LBGs at $z\sim4$, $5$, $6$, and $7$ 
from the Hubble data.
 The color selection criteria are indicated with solid lines.
The black squares and dots denote colors of selected LBGs and 
other objects in the HUDF region, respectively.
 In addition to the color criteria indicated by the solid lines, 
we also enforce other criteria such as a non-detection in the images blueward of the Lyman break (see Section \ref{selection} for details).
 \label{fig_color_color}}
\end{figure*}

We select galaxies that have a Lyman break according to the criteria of Equations (\ref{eq_z4_1}), (\ref{eq_z5_1}), (\ref{eq_z6_1}), and (\ref{eq_z7_1}), and exclude intrinsically-red galaxies by the additional constraints of Equations (\ref{eq_z4_2}), (\ref{eq_z4_3}), (\ref{eq_z5_2}), (\ref{eq_z5_3}), (\ref{eq_z6_2}), (\ref{eq_z6_3}), (\ref{eq_z7_2}), and (\ref{eq_z7_3}).
Figure \ref{fig_color_color} presents
these color selection criteria, together with
all sources from the HUDF catalog.
These LBG color selection criteria are extensively tested by simulations, and used to study evolution of the UV luminosity functions \citep[e.g.,][]{2015ApJ...803...34B}.

In the five fields of CANDELS-AEGIS, CANDELS-COSMOS, CANDELS-UDS, HFF-Abell2744P, and HFF-MACS0416P, 
the number of the available multi-bands are smaller than those in HUDF and GOODS.
We use different color criteria, and select LBGs at $z\sim5$, $6$, and $7-8$ in these five fields \redc{\citep{2015ApJ...803...34B}} as follows:

$z\sim5$
\begin{eqnarray}
V_{606}-I_{814} > 1.3,\label{eq_z5a_1}\\
I_{814}-H_{160} < 1.25,\label{eq_z5a_2}\\
V_{606}-I_{814} > 0.72(I_{814}-H_{160})+1.3.\label{eq_z5a_3}
\end{eqnarray}

$z\sim6$
\begin{eqnarray}
I_{814}-J_{125} > 0.8,\label{eq_z6a_1}\\
J_{125}-H_{160} < 0.4,\label{eq_z6a_2}\\
I_{814}-J_{125} > 2(J_{125}-H_{160})+0.8.\label{eq_z6a_3}
\end{eqnarray}

$z\sim7-8$
\begin{eqnarray}
I_{814}-J_{125} > 2.2,\label{eq_z7a_1}\\
J_{125}-H_{160} < 0.4,\label{eq_z7a_2}\\
I_{814}-J_{125} > 2(J_{125}-H_{160})+2.2.\label{eq_z7a_3}
\end{eqnarray}
We select galaxies that have a Lyman break by the criteria given in Equations (\ref{eq_z5a_1}), (\ref{eq_z6a_1}), and (\ref{eq_z7a_1}), and exclude intrinsically-red galaxies by the criteria of Equations (\ref{eq_z5a_2}), (\ref{eq_z5a_3}), (\ref{eq_z6a_2}), (\ref{eq_z6a_3}), (\ref{eq_z7a_2}), and (\ref{eq_z7a_3}).

In addition, we also adopt the following four criteria
that are similar to those in \citet{2015ApJ...803...34B}.
First, to identify secure sources, we apply detection limits of
$>5\sigma$ and $>5.5 \sigma$ levels in the detection images in HUDF and the other fields, respectively.
Since the HUDF data are deep and clean, the detection criterion of HUDF is moderately loosened than the other fields.
Second, for reducing foreground interlopers, we remove sources 
with continuum detected at wavelengths shortward of the Lyman breaks of the target LBGs.
In all of the Hubble fields except CANDELS-AEGIS, CANDELS-COSMOS, and CANDELS-UDS,
we apply the criterion of $< 2\sigma$ no-detection in the $B_{435}$ band
for candidate LBGs at $z\sim5$ and $z\sim6$, if $B_{435}$ data are available.
Additionally, we require $< 2\sigma$ non-detection in $V_{606}$ band 
or $V_{606}-z_{850}>2.6$ for the $z\sim6$ LBG candidates.
For the $z\sim 7$ LBG candidates, we calculate an optical $\chi^2$ value
for each source with the $B_{435}V_{606}i_{775}$ flux measurements, if available, 
in the same manner as \citet{2011ApJ...737...90B}. The optical $\chi^2$ value
is defined by $\chi^2_{\m{opt}}=\Sigma_i \m{SGN}(f_i)(f_i/\sigma_i)^2$, where $f_i$, 
$\sigma_i$, and $\m{SGN}(f_i)$ are the flux in each band, its uncertainty, and its sign, respectively. 
We remove $z\sim 7$ LBG candidates whose $\chi^2_{\m{opt}}$ values are larger than 4.
For the rest of the fields, CANDELS-AEGIS, CANDELS-COSMOS, and CANDELS-UDS,
we
calculate $\chi^2_{\m{opt}}$ values using the ground-based data whose wavelength are shorter than the redshifted 
Lyman break for the target LBGs at $z\sim5$, $6$, and $7$.
We use a threshold value of 2, 3, or 4 that corresponds to the number of the ground-based bands of $< 3$, $3$, or $> 4$, respectively \citep{2015ApJ...803...34B}, 
and remove LBG candidates whose $\chi^2_{\m{opt}}$ value is larger than the threshold.
Third, to isolate LBGs from foreground Galactic stars, 
the LBG candidates should have an SExtractor stellarity parameter, {\tt CLASS\_STAR}, less than $0.9$ \citep{2009A&A...498..725H,2015ApJ...803...34B},
if the candidates are $1\ \m{magnitude}$ brighter than the detection limit.
Finally, to avoid multiple identifications of a source satisfying two sets of selection criteria at different redshifts,
we keep the source in a catalog of LBGs at a redshift higher than the other, and remove
the source from the low-$z$ catalog. For example, if a source meets the criteria of
Equations (\ref{eq:lbg_hudf_z5})-(\ref{eq:lbg_hudf_z5_e}) and (\ref{eq:lbg_hudf_z6})-(\ref{eq:lbg_hudf_z6_e}), the source is not included
in the LBG catalog of $z\sim 5$, but $z\sim 6$.
After adopting these criteria, the estimated contamination fractions by foreground galaxies are estimated to be
$f_\m{c}\sim2$, $5$, $7$, and $9\%$ for the $z\sim4$, $5$, $6$, and $7$ LBG samples, respectively, based on the Monte-Carlo simulations in \citet{2015ApJ...803...34B}.

\begin{figure*}
\begin{center}
  \includegraphics[clip,bb=80 180 2000 2700,width=1\hsize]{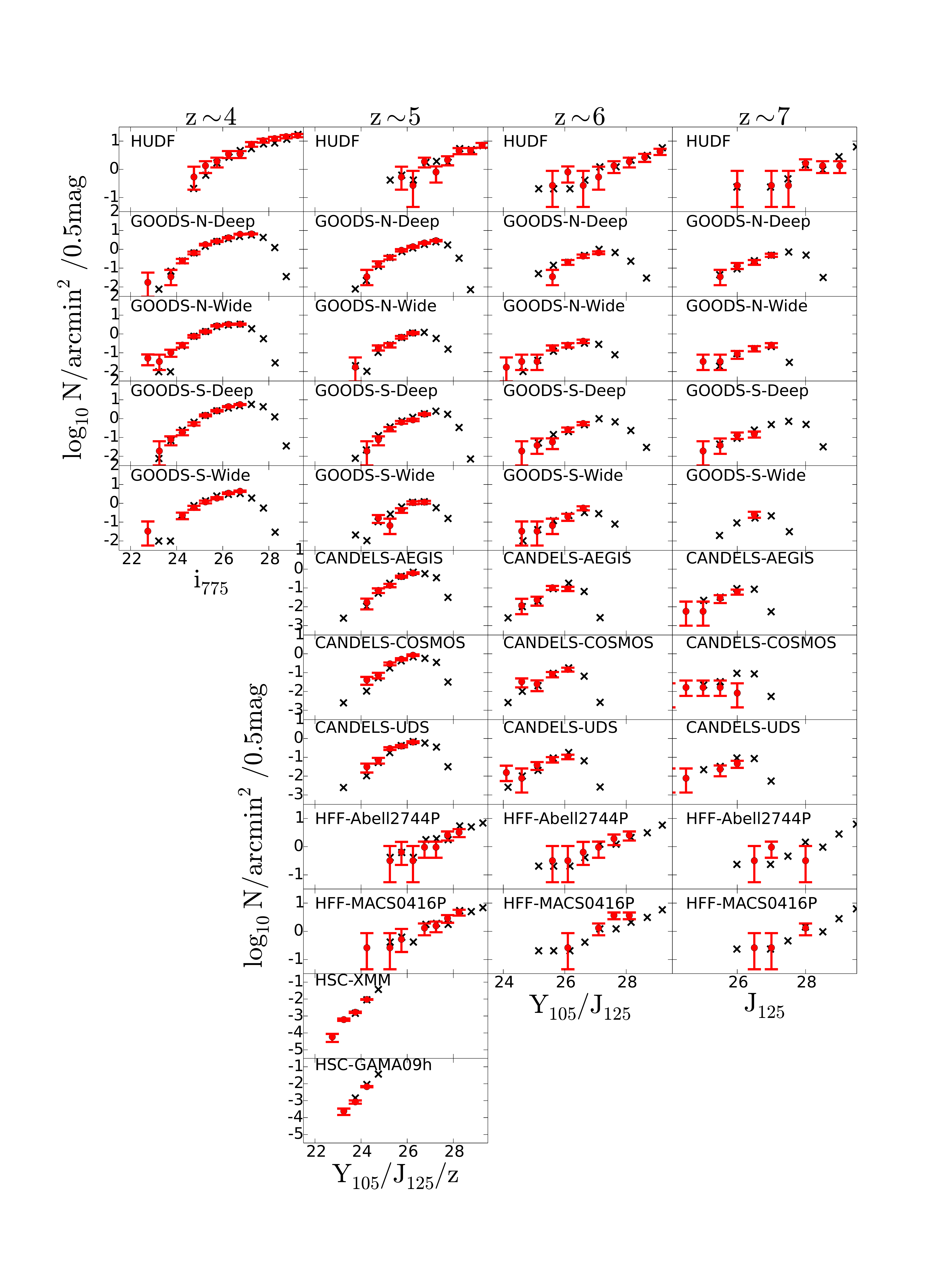}
\\
 \end{center}
 \caption{Surface number densities of LBGs at $z\sim4$, $5$, $6$, and $7$.
The red circles represent the surface number densities of our LBGs, while 
the black crosses denote the surface number densities of the LBGs presented in 
the literature \citep{2015ApJ...803...34B,2009A&A...498..725H}.
 The surface number densities of our LBGs are consistent with 
the previous results. 
We confirm that the errors of our surface number densities are comparable with \citet{2015ApJ...803...34B} in HUDF.
  \label{fig_number}}
\end{figure*}

We construct a total sample of $5185$, $2964$, $978$, and $524$ LBGs at $z\sim4$, $5$, $6$, and $7$, respectively, based on the Hubble data.
Table \ref{table_catalog_lbg} shows magnitudes of the LBGs.
For conservative estimates of the clustering signals, we use the LBGs whose aperture magnitudes in the rest-frame UV band, $m^\m{aper}_\m{UV}$, are brighter than the $5\sigma$ limiting magnitudes.
The rest-frame UV band is defined by the observed band whose central wavelength is nearest 
to the rest-frame wavelength of $1500\ \m{\AA}$ 
for the Hubble data as well as the Subaru data (Section \ref{sec:subaru_samples}).
Table \ref{table_number_lbg} summarizes the numbers of LBGs for each field. 
\redc{We compare our sample with the sample of \citet{2015ApJ...803...34B}, and find that our sample is consistent with that of \citet{2015ApJ...803...34B}. 
In the deep fields used in our comparison, more than $\sim80\%$ of the galaxies in our sample are included in the sample of \citet{2015ApJ...803...34B} at magnitudes brighter than the $10\sigma$ limiting magnitude. 
Similarly, more than $\sim70\%$ of the galaxies of the \citet{2015ApJ...803...34B} sample are included in our sample. 
The remaining $20-30\%$ galaxies are located near the border of the color selection window, and are missed due to photometric errors. 
We also compare the surface number densities of our LBGs with those of \citet{2015ApJ...803...34B} in Figure \ref{fig_number};}
we confirm that the surface number densities of our LBGs are consistent.
The mean redshifts of the $z\sim4$, $5$, $6$, and $7$ LBGs are $z_\m{c}=3.8$, $4.9$, $5.9$, and $6.8$, respectively,
and the redshift distributions are the same as those shown in Figures 1 and 19 of \citet{2015ApJ...803...34B}.

\begin{turnpage}
\begin{deluxetable*}{ccccccccccccc}
\tablecaption{Catalog of LBGs in the Hubble Data}
\tablehead{\colhead{Catalog ID} & \colhead{$\m{R.A.\ (J2000)}$} & \colhead{$\m{decl.\ (J2000)}$} & \colhead{$B_{435}$} & \colhead{$V_{606}$} & \colhead{$i_{775}$} & \colhead{$I_{814}$} & \colhead{$z_{850}$} & \colhead{$Y_{105}$} & \colhead{$J_{125}$}& \colhead{$JH_{140}$} & \colhead{$H_{160}$} & \colhead{coadd\tablenotemark{a}}  \\
 }
\startdata
z4\_gdsd\_7260 & 53.074684744 & -27.880245679 & $29.4\pm0.5$ & $28.0\pm0.1$ & $27.7\pm0.1$ & $27.9\pm0.1$ & $28.0\pm0.1$ & $27.8\pm0.1$ & $28.1\pm0.2$ & \nodata  & $27.6\pm0.1$ & $28.2\pm0.1$ \\
z4\_gdsd\_7269 & 53.075142909 & -27.880051154 & $29.1\pm0.4$ & $27.6\pm0.1$ & $27.3\pm0.1$ & $27.4\pm0.1$ & $27.3\pm0.0$ & $27.7\pm0.1$ & $27.6\pm0.1$ & \nodata  & $27.8\pm0.1$ & $27.9\pm0.1$ \\
z4\_gdsd\_7328 & 53.064744909 & -27.87986999 & $29.4\pm0.5$ & $27.6\pm0.1$ & $27.3\pm0.1$ & $27.3\pm0.1$ & $27.3\pm0.0$ & $27.4\pm0.1$ & $27.4\pm0.1$ & \nodata  & $27.7\pm0.1$ & $28.6\pm0.2$ \\
z4\_gdsd\_7433 & 53.065422101 & -27.879309789 & $<30.3$  & $29.4\pm0.4$ & $28.2\pm0.2$ & $28.2\pm0.1$ & $28.6\pm0.2$ & $28.5\pm0.2$ & $28.1\pm0.2$ & \nodata  & $28.0\pm0.2$ & $28.9\pm0.2$ \\
\nodata & \nodata & \nodata & \nodata & \nodata  & \nodata & \nodata & \nodata & \nodata & \nodata  & \nodata & \nodata & \nodata
\enddata

\tablecomments{All magnitudes listed are measured in $0.\carcsec35$-diameter circular apertures. 
Upper limits are $1\sigma$.
}
\tablenotetext{a}{Coadd image of $Y_{105}J_{125}JH_{140}H_{160}$-bands. \\ (The complete table is available in a machine-readable form in the online journal.) }
\label{table_catalog_lbg}
\end{deluxetable*}
\end{turnpage}

\begin{deluxetable*}{ccccccc}
\setlength{\tabcolsep}{0.35cm}
\tablecaption{Number of LBGs for Our Analysis}
\tablehead{
\colhead{} & \colhead{Area} & \colhead{} & \colhead{} & \colhead{} & \colhead{}& \colhead{}\\
\colhead{Field} & \colhead{($\m{arcmin^2}$)} & \colhead{$5\sigma$ depth} & \colhead{$z\sim4$} & \colhead{$z\sim5$} & \colhead{$z\sim6$} & \colhead{$z\sim7$}\\
\colhead{(1)}& \colhead{(2)}& \colhead{(3)}& \colhead{(4)} &  \colhead{(5)}& \colhead{(6)}& \colhead{(7)} }

\startdata
HUDF & 3.7 & 30.6 & 290 (348) & 48 (130) & 0 (86) & 0 (50) \\
GOODS-N-Deep & 57.4 & 28.6 & 1411 (1655) & 431 (630) & 43 (136) & 81 (113) \\
GOODS-N-Wide & 58.2 & 28.1 & 788 (800) & 193 (223) & 63 (69) & 27 (31) \\
GOODS-S-Deep & 52.7 & 29.0 & 1139 (1872) & 205 (696) & 142 (311) & 66 (203) \\
GOODS-S-Wide & 30.4 & 28.3 & 461 (510) & 92 (142) & 28 (51) & 13 (31) \\
CANDELS-AEGIS & 174.9 & 28.0 & \nodata & 304 (381) & 73 (101) & 0 (28) \\
CANDELS-COSMOS & 122.0 & 27.9 & \nodata & 314 (348) & 76 (80) & 0 (27) \\
CANDELS-UDS & 129.3 & 27.9 & \nodata & 268 (310) & 54 (65) & 0 (25) \\
HFF-Abell2744P & 3.1 & 29.3 & \nodata & 30 (37) & 0 (26) & 0 (7) \\
HFF-MACS0416P & 3.8 & 29.5 & \nodata & 56 (67) & 0 (53) & 0 (9) \\
HSC-XMM & 30100 & 25.1&  \nodata &451 (451)&   \nodata& \nodata\\
HSC-GAMA09h & 24800 & 25.0 &  \nodata & 279 (279)&  \nodata &  \nodata
 \\ \hline
$N_{\rm total}(z)$ & & & 4089 (5185) & 2671 (3694) & 585 (978) & 291 (524) \\
$N_\m{total}$ & & &  \multicolumn{4}{c}{7636 (10381)}
\enddata

\tablecomments{Columns: (1) Field. (2) Effective area in $\m{arcmin^2}$. (3) $5\sigma$ limiting magnitude in
the coadd image. (3)-(7) Number of the LBGs for our analysis at each redshift that are brighter than
the $5\sigma$ limiting magnitude in the rest-frame UV band whose central wavelength is nearest to rest-frame $1500${\AA}. The value in parentheses is the number of LBGs in the parent sample.}
\label{table_number_lbg}
\end{deluxetable*}

\subsection{Subaru Samples}
\label{sec:subaru_samples}
\subsubsection{Multi-band Photometric Catalogs}\label{ss_subaru_source}
We make HSC source catalogs from the reduced images
in the same manner as the Hubble source catalogs.
First, we homogenize the PSFs of the HSC images to $\sim0.\carcsec9$ in FWHM by convolving images with a Gaussian,
matching a PSF's $75\%$-flux circular radius that includes 75\% of a total flux for a PSF profile source.
We then run SExtractor (version 2.8.6) in dual-image mode
to detect sources in the detection image, and to
carry out photometry in the HSC images 
for {\tt MAG\_APER} in a circular aperture of 
the PSF's $75\%$-flux radius.
\redc{The total magnitude is estimated from the aperture magnitude with an aperture correction. 
The aperture correction is estimated to be $0.31\ \m{mag}$ under the assumption of the PSF profile.}
We use the parameter set of
${\tt DETECT\_MINAREA}=6$, ${\tt DETECT\_THRESH}=1.5$, 
${\tt ANALYSIS\_THRESH}=1.5$, ${\tt DEBLEND\_NTHRESH}=32$, and ${\tt DEBLEND\_MINCOUNT}=0.005$.

\subsubsection{Lyman Break Galaxy Selection}
\redc{We select LBGs at $z\sim5$ with the HSC data, because we can conduct secure selections with the $g$-band image.
We apply color criteria}
similar to those of the CFHT study \citep{2009A&A...498..725H}
that uses the photometric system almost identical to
the one of our HSC data.
The $z\sim5$ color selection criteria for the HSC sources are as follows:
\begin{eqnarray}
r-i > 1.2,\label{eq_s_1}\\
i-z < 0.7,\label{eq_s_2}\\
r-i> 1.5(i-z)+1.0.\label{eq_s_3}
\end{eqnarray}
We select galaxies that have a Lyman break by the criterion of Equation (\ref{eq_s_1}), and exclude intrinsically-red galaxies by the criteria of Equations (\ref{eq_s_2}) and (\ref{eq_s_3}).
These LBG color selection criteria are used for the study of
clustering evolution \citep[e.g.,][]{2009A&A...498..725H}.

In addition to the selection criteria above, we require sources 
to be detected at the $>5\sigma$ level in the $z$-band image, and to be undetected at the $<2\sigma$ level 
in the $g$-band image.
We also apply a criterion of SExtractor stellarity parameter, {\tt CLASS\_STAR}, of $<0.9$.
We obtain 730 LBGs at $z\sim5$.
The surface number densities of our HSC LBGs are presented in Figure \ref{fig_number},
which agree with the previous results of \citet{2009A&A...498..725H}.
More details of the data reduction and the LBG selection are presented in Y. Ono et al. (in preparation).

\section{Clustering Analysis}\label{ss_analysis}
\subsection{ACF}
We derive the ACFs, $\omega(\theta)$,
with our LBG samples. We calculate
the observed ACFs, $\omega_{\m{obs}}(\theta)$, 
using the estimator presented in \citet{1993ApJ...412...64L},
\begin{equation}
\omega_{\m{obs}}(\theta)=\frac{DD(\theta)-2DR(\theta)+RR(\theta)}{RR(\theta)},
\end{equation}
where $DD(\theta)$, $DR(\theta)$, and $RR(\theta)$ are numbers of galaxy-galaxy, galaxy-random, and 
random-random pairs normalized by the total number of pairs.
We create a random sample composed of 10,000 (100,000) sources for each Hubble (Subaru) field
with the geometrical shape same as the observational data including the mask positions.
The errors are estimated by the bootstrap technique of \citet{1986MNRAS.223P..21L} 
with 100 resamples replacing individual galaxies for each field.
It is known that this bootstrap technique tends to overestimate 
the errors of the correlation function \citep{1994MNRAS.267..927F,1992ApJ...392..452M}.
Although we cannot quantitatively evaluate this trend with our data
that are not large enough, the forthcoming data of the HSC survey
will enable us to investigate this effect.

Due to the finite size of our survey fields, the observed ACF is 
underestimated by a constant value known as the integral constraint, 
$IC$ \citep{1977ApJ...217..385G}. Including
the correction for the number of objects 
in the sample, $N$ \citep{1980lssu.book.....P},
the true ACF is given by
\begin{equation}
\omega(\theta)=\omega_{\m{obs}}(\theta)+IC+\frac{1}{N}.
\label{eq:acf}
\end{equation}
We estimate the integral constraint with 
\begin{equation}
IC=\frac{\Sigma_{i}RR(\theta_i)\omega_{\m{model}}(\theta_i)}{\Sigma_{i}RR(\theta_i)},
\label{eq:ic}
\end{equation}
where $\omega_{\m{model}}(\theta)$ is the best-fit model ACF, and $i$ refers the angular bin.

To test the dependence of the clustering strength on the luminosity of the galaxies, 
we make subsamples that are brighter than the threshold UV magnitudes, $m_{\rm UV, th}$,
that are listed in Table \ref{table_PL}. 
\redc{In each subsample, we obtain the best-estimate of the ACF that is the weighted mean of the ACFs of the different fields in an angular bin.}
Table \ref{table_PL} shows the numbers of LBGs in the subsamples.
Note that the numbers of the faint-magnitude subsamples are
smaller than those of bright-magnitude subsamples (e.g. $m^\m{aper}_\m{UV, th}=29.2-29.8$ and $27.2-28.2$).
This is because the faint-magnitude subsamples are only composed of 
LBGs in very deep data covering a small field (e.g. HUDF).
Using the UV luminosity functions of \citet{2015ApJ...803...34B},
we calculate the the number density of LBGs for each subsample and associated errors corrected for incompleteness.
We estimate the cosmic variance in the number densities using the bias values obtained in Section 4.2,
following the procedures in \citet{2004ApJ...600L.171S}. 
We include the uncertainty from cosmic variance in our estimate of the error on the number density.
The LBG number densities and the errors are presented in Table \ref{table_PL}.

We fit the ACFs with a simple power law model,
\begin{equation}
\omega(\theta)=A_{\omega}\theta^{-\beta}.
\label{eq:power_law_model}
\end{equation}
Because we obtain no meaningful constraints on $\beta$ for most of the subsamples, we fix the value of $\beta$ to $0.8$ 
that is used in previous clustering analyses 
\citep[e.g.,][]{2001ApJ...558L..83O,2004ApJ...611..685O,2010ApJ...723..869O,2003A&A...409..835F,2010MNRAS.406..147F}.
We use Equation (\ref{eq:power_law_model}) for $\omega_{\rm model}(\theta)$ to determine $IC$ (Equation \ref{eq:ic}),
and obtain the best-fit $A_{\omega}$ values with Equation (\ref{eq:acf}).

\begin{deluxetable*}{cccccccccccc}
\setlength{\tabcolsep}{0.35cm}
\tablecaption{Summary of the Clustering Measurements with the Power Law Model}
\tablehead{
\colhead{$z_c$} & \colhead{$m^\m{aper}_\m{UV, th}$} & \colhead{$M_\m{UV, th}$} & \colhead{$\left<M_\m{UV}\right>$} & \colhead{$\m{log}\m{SFR}_{\m{th}}$} & \colhead{$\m{log}M_{*,\m{th}}$} & \colhead{$N$} & \colhead{$n_\m{g}$} & \colhead{$A_\omega$}& \colhead{$r_0$} & \colhead{$b_\m{g}$}& \colhead{$\chi^2_{\nu}$}\\
\colhead{} & \colhead{} & \colhead{} & \colhead{} & \colhead{} & \colhead{}& \colhead{} & \colhead{$(10^{-4}\ \m{Mpc}^{-3})$} & \colhead{$(\m{arcsec^{0.8}})$} & \colhead{$\m{(Mpc)}$} & \colhead{} & \colhead{} \\
\colhead{(1)}& \colhead{(2)}& \colhead{(3)}& \colhead{(4)} &  \colhead{(5)}& \colhead{(6)}& \colhead{(7)} & \colhead{(8)}& \colhead{(9)}& \colhead{(10)}& \colhead{(11)}& \colhead{(12)}}

\startdata
3.8 & 27.2 & -19.6 & -20.2 & 0.97 & 9.1 & 1406 & $20.1\pm3.5$     & $1.2\pm0.2$ & $5.7_{-0.4}^{+0.4}$ & $2.9_{-0.2}^{+0.2}$ & 0.6\\
      & 27.6 & -19.2 & -19.8 & 0.77 & 8.9 & 2301 & $31.3\pm4.9$     & $1.0\pm0.1$ & $5.1_{-0.3}^{+0.3}$ & $2.6_{-0.1}^{+0.1}$ & 0.8\\
      & 28.2 & -18.4 & -19.3 & 0.35 & 8.4 & 2509 & $68.6\pm11.7$   & $0.8\pm0.1$ & $4.4_{-0.2}^{+0.2}$ & $2.4_{-0.1}^{+0.1}$ & 1.4\\
      & 29.2 & -17.3 & -18.4 & -0.15 & 7.9 & 161   & $154.7\pm53.8$ & $0.4\pm0.3$ & $3.0_{-1.5}^{+1.0}$ & $1.7_{-0.8}^{+0.5}$ & 0.4\\
      & 29.8 & -16.7 & -17.9 & -0.49 & 7.5 & 244   & $251.6\pm65.6$ & $0.2\pm0.1$ & $2.2_{-1.0}^{+0.8}$ & $1.2_{-0.5}^{+0.4}$ & 0.3\\
4.8 & 25.0 & -21.7 & -22.1 & 2.0 & 10.2 &   730 & $0.15\pm0.10$ & $8.8\pm3.4$ & $14.5_{-3.5}^{+2.9}$ & $8.2_{-1.8}^{+1.5}$ & 0.9\\
4.9 & 27.2 & -19.9 & -20.5 & 1.0 & 9.1 & 878   & $9.4\pm1.3$       & $2.0\pm0.4$ & $6.4_{-0.7}^{+0.7}$ & $4.0_{-0.4}^{+0.4}$ & 0.2\\
      & 27.6 & -19.5 & -20.0 & 0.84 & 8.9 &1467  & $15.2\pm1.7$      & $1.4\pm0.3$ & $5.2_{-0.6}^{+0.5}$ & $3.3_{-0.3}^{+0.3}$ & 0.4\\
      & 28.0 & -19.1 & -19.8 & 0.67 & 8.7 & 623   & $22.0\pm3.5$     & $0.8\pm0.3$ & $3.8_{-0.8}^{+0.7}$ & $2.5_{-0.5}^{+0.4}$ & 0.3\\
      & 29.2 & -17.9 & -18.7 & 0.011 & 7.9 & 120   & $72.7\pm30.4$   & $1.2\pm0.6$ & $4.9_{-1.4}^{+1.1}$ & $3.1_{-0.8}^{+0.6}$ & 0.4\\
5.9 & 27.4 & -20.0 & -20.5 & 1.1 & 9.1 & 285   & $3.8\pm0.6$       & $2.7\pm1.3$ & $6.4_{-1.9}^{+1.5}$ & $4.7_{-1.3}^{+1.0}$ & 0.6\\
      & 28.4 & -19.1 & -19.3 & 0.55 & 8.6 & 278   & $13.4\pm2.5$     & $1.1\pm0.7$ & $3.9_{-1.6}^{+1.2}$ & $3.0_{-1.2}^{+0.8}$ & 0.6\\
6.8 & 28.2 & -19.5 & -19.9 & 0.75 & 8.8 & 113   & $7.0\pm2.5$       & $4.0\pm1.2$ & $8.7_{-1.6}^{+1.4}$ & $7.1_{-1.2}^{+1.0}$ & 0.6\\
      & 28.4 & -19.3 & -19.8 & 0.65 & 8.7 & 150   & $9.0\pm2.2$       & $1.8\pm1.0$ & $5.5_{-2.2}^{+1.6}$ & $4.7_{-1.7}^{+1.2}$ & 0.7
\enddata

\tablecomments{Columns: (1) Mean redshift. (2) Threshold aperture magnitude in the rest-frame UV band. (3) Threshold absolute total magnitude in the rest-frame UV band. (4) Mean absolute total magnitude in the rest-frame UV band. (5) Threshold SFR in a unit of $\Msun\ \m{yr^{-1}}$ derived from the threshold total magnitude, $M_\m{UV, th}$. (6) Threshold stellar mass in a unit of $\Msun$ derived from $M_\m{UV, th}$ via equation (\ref{eq_MUVMs_z4}), (\ref{eq_MUVMs_z5}), (\ref{eq_MUVMs_z67}). (7) Number of galaxies in our subsample. (8) Number density of our subsample derived from a UV luminosity function of \citet{2015ApJ...803...34B}. (9) Power law amplitude (the power law index is fixed to $\beta=0.8$.). (10) Spatial correlation length. (11) Galaxy-dark matter bias estimated by the power law model. See column (6) in Table \ref{table_HOD} for the best estimate from the HOD modeling. (12) Reduced $\chi^2$ value.}
\label{table_PL}
\end{deluxetable*}

Contaminating sources in a galaxy sample reduce the value of $A_{\omega}$.
If contaminants have a homogeneous sky distribution,
the true $A_{\omega}$ is underestimated by a factor of $(1-f_\m{c})^2$,
where $f_\m{c}$ is a contamination fraction.
Because contaminants are more or less clustered, a clustering amplitude
multiplied by $1/(1-f_\m{c})^2$ provides the upper limit of the value of $A_{\omega}$,
\begin{equation}
A^\m{max}_{\omega}=\frac{A_{\omega}}{(1-f_\m{c})^2}.
\end{equation}

\begin{figure*}
\begin{center}
  \includegraphics[clip,bb=100 170 2000 2700,width=1\hsize]{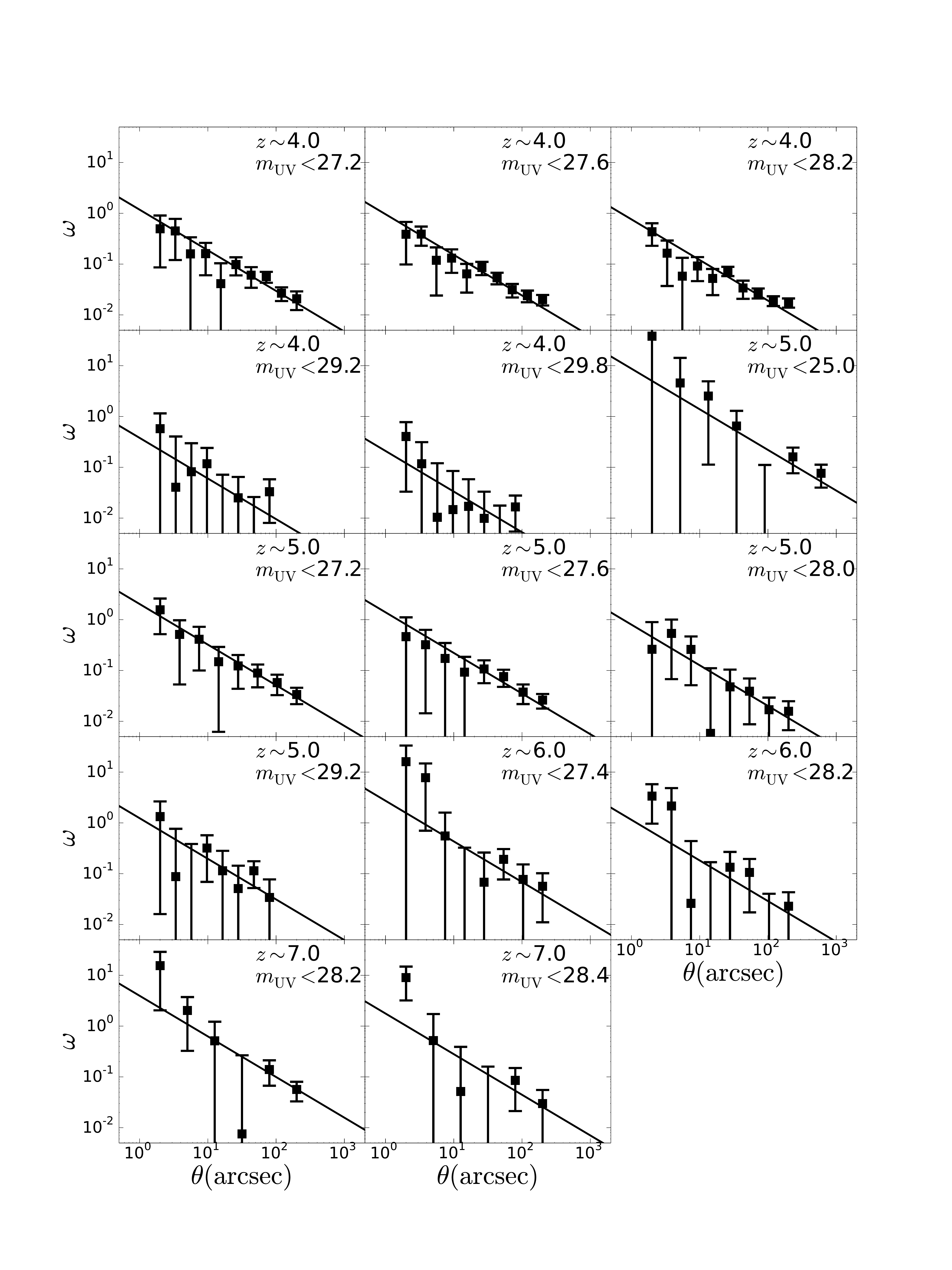}
\\
\end{center}
  \caption{ACF of each subsample.
    The solid lines indicate the best-fit power law function, $A_\omega\theta^{-\beta}$, where we fix $\beta=0.8$.
  The redshift and threshold magnitude are denoted in the upper right corner of each panel.
  \label{fig_ACF_PL}}
\end{figure*}

The contamination fractions are $f_\m{c}\sim2$, $5$, $7$, and $9\%$ for the $z\sim4$, $5$, $6$, and $7$ LBG 
samples, respectively \citep{2015ApJ...803...34B}. The corresponding $1/(1-f_\m{c})^2$ values are $\sim1.04$, $1.1$, $1.2$, and $1.2$
that are significantly smaller than the statistical errors.
Therefore, we do not apply these contamination corrections to our estimate of $A_{\omega}$.
Table \ref{table_PL} presents the best-fit $A_\omega$ values.
In Figure \ref{fig_ACF_PL}, we plot ACFs of our subsamples with the best-fit power law model.

\subsection{Correlation Length and Bias}
An ACF shows clustering properties of galaxies projected on the sky,
and depends upon a combination of a galaxy redshift distribution and a galaxy spatial correlation function $\xi_\m{g}(r)$.
The spatial correlation function is approximated by a single power law,
\begin{equation}
\xi_\m{g}(r)=\left(\frac{r}{r_0}\right)^{-\gamma},
\end{equation}
where $r_0$ is the correlation length.
We calculate correlation lengths from the amplitudes of the ACFs 
using the Limber equation \citep{1980lssu.book.....P,1991ApJ...380L..47E},
\begin{align}
\label{eq:limber}
A_\omega & =  Cr_0^\gamma\frac{\int^\infty_0F(z)D_\theta(z)^{1-\gamma}N(z)^2g(z)dz}{\left[\int^\infty_0N(z)dz\right]^2},\\
g(z) & =  \frac{H_0}{c}(1+z)^2\left\{1+\Omega_mz+\Omega_\Lambda\left[(1+z)^{-2}-1\right]\right\}^{1/2},\ \ \ \ \ \\
C & =  \frac{\sqrt{\pi}\Gamma[(\gamma-1)/2]}{\Gamma(\gamma/2)},
\end{align}
where $D_\theta(z)$ is the angular diameter distance and $N(z)$ is the redshift distribution of the sample.
$F(z)$ describes the redshift dependence of $\xi(r)$.
Assuming that the clustering pattern 
is fixed in comoving coordinates in the redshift range of our sample, we use the functional form 
$F(z)=[(1+z)/(1+z_c)]^{-(3+\epsilon)}$ for $\epsilon=-1.2$ \citep{1999MNRAS.307..703R}, where $z_c$ is the average redshift of the sample LBGs (Section 3.1).
The $r_0$ value does not significantly depend on $\epsilon$ over $-3<\epsilon<0$.
\redc{The slope of the spatial correlation function, $\gamma$, is related to that of the ACF, $\beta$, by
\begin{equation}
\gamma=\beta+1.
\end{equation}}
\redc{We adopt the redshift distribution of LBGs presented in \citet[][the left panel of Figure 1]{2015ApJ...803...34B}} and Y. Ono et al. (in preparation)
for our Hubble and Subaru samples, respectively.
These redshift distributions include the photometric uncertainties based on the Monte-Carlo simulations.
Y. Ono et al. obtain the redshift distribution by placing artificial objects randomly in the real images using a method similar to the one in  \citet{2015ApJ...803...34B}.
The object colors are calculated with redshifted model spectra \citep{2003MNRAS.344.1000B} and the HSC filter response curves.
We check the systematic errors on the halo mass estimates originating from the $N(z)$ uncertainties, and find that the errors change the mass estimates negligibly, only by $\lesssim0.05\ \m{dex}$, assuming $\lesssim15\%$ systematic shift of $N(z)$ that is found in the spectroscopic results of \citet{1999ApJ...519....1S}.

We calculate the galaxy-dark matter bias $b_\m{g}$ on scale of $r=8\ h^{-1}\m{Mpc}$, which is given by 
\begin{equation}\label{eq_bias}
b_\m{g}=\sqrt{\frac{\xi_\m{g}(r=8\ h^{-1}\m{Mpc})}{\xi_\m{DM}(r=8\ h^{-1}\m{Mpc},z)}}, 
\end{equation}
where $\xi_\m{DM}(r,z)$ is the spatial correlation function of the underlying dark matter calculated with the linear dark matter power spectrum, $P_\m{m}(k,z)$, which is defined by
\begin{equation}
\xi_\m{DM}(r,z)=\int\frac{k^2dk}{2\pi^2}\frac{\m{sin}\left(kr\right)}{kr}P_\m{m}(k,z).
\end{equation}
Table \ref{table_PL} presents the bias values thus obtained.

\subsection{HOD Model}

To connect observed galaxies to their host dark matter halos, we use 
a halo occupation distribution (HOD) model that is an analytic model of galaxy clustering
\citep[e.g.,][]{2000MNRAS.318..203S,2002ApJ...575..587B,2002PhR...372....1C,2003ApJ...593....1B,2004ApJ...609...35K,2005ApJ...633..791Z}.
The HOD model is adopted not only to low-redshift galaxies \citep[e.g.,][]{2005ApJ...630....1Z,2007ApJ...667..760Z,2013MNRAS.430..725V,2015ApJ...806....2M}, but also to high redshift galaxies \redc{\citep[e.g.,][]{2002MNRAS.329..246B,2004MNRAS.347..813H,2005ApJ...635L.117O,2006ApJ...642...63L,2009ApJ...695..368L,2007A&A...462..865H,2013ApJ...774...28B}}.
The key assumption of our HOD model is that the number of galaxy, $N$, in a given dark matter halo depends only on the halo mass, $M_\m{h}$.
We parameterize the mean number of galaxies in dark matter halos with a mass of $M_\m{h}$, $N(M_\m{h})$,
that is given by
\begin{equation}
N(M_\m{h})=DC(N_\m{c}(M_\m{h})+N_\m{s}(M_\m{h})),
\end{equation}
where $DC$ is the duty cycle of LBG activity (see Section \ref{ss_intro} for the definition).
$N_\m{c}(M_\m{h})$, and $N_\m{s}(M_\m{h})$ are the mean number of central and satellite galaxies, respectively.
Here the LBG activity for $DC$ is defined by the properties of galaxies
that are UV-bright star-forming galaxies selected as LBGs brighter than $m_{\rm UV, th}$.
We assume that DC does not depend on the halo mass in each subsample, because the present data are not large enough to investigate the mass dependence of DC that hides in the statistical errors.
We adopt functional forms of $N_\m{c}(M_\m{h})$ and $N_\m{s}(M_\m{h})$ that are motivated by N-body simulations, smoothed particle hydrodynamic simulations, and semi-analytic models for low-$z$ galaxies and LBGs \citep[e.g.,][]{2004ApJ...609...35K,2005ApJ...633..791Z,2015MNRAS.450.1279G}.
$N_\m{c}(M_\m{h})$ is approximated as a step function with a smooth transition, 
\begin{equation}
N_\m{c}(M_\m{h})=\frac{1}{2}\left[1+\m{erf}\left(\frac{\m{log}M_\m{h}-\logMmin}{\sigmalogM}\right)\right],
\end{equation}
where $\sigmalogM$ is a transition width reflecting the scatter in the luminosity-halo mass relation.
$M_\m{min}$ is the mass scale at which $50\%$ of halos host a central galaxy.
Similarly, the mean number of satellite galaxies, $N_\m{s}(M_\m{h})$, follows a power law with a mass cut,
\begin{equation}
\label{eq:n_s}
N_\m{s}(M_\m{h})=N_c(M_\m{h})\left(\frac{M_\m{h}-M_0}{M^\prime_1}\right)^\alpha,
\end{equation}
where $M_0$ is the cut off mass, and $M^\prime_1$ ($\alpha$) is the amplitude (slope) of the power law.

We calculate galaxy number densities from the HOD model with
\begin{equation}
n_\m{g}(z)=\int^\infty_0dM_\m{h}\frac{dn}{dM_\m{h}}(M_\m{h},z)N(M_\m{h}),
\label{eq:galaxy_number_density}
\end{equation}
where $\frac{dn}{dM_\m{h}}(M_\m{h},z)$ is the halo mass function.
We use the halo mass function derived in \citet{2013ApJ...770...57B}, 
which is a modification of the \citet{2008ApJ...688..709T} halo mass function 
for the high redshift universe ($z>2.5$) matching to the {\it Consuelo} simulation (\citealt{2009AAS...21342506M}; see also \citealt{2011ApJ...738...45L}; \citealt{2013ApJ...762..109B})\footnote{http://lss.phy.vanderbilt.edu/lasdamas/}.
The difference between the \citet{2013ApJ...770...57B} and \citet{2008ApJ...688..709T} halo mass functions is $\sim10\%$ in the number density of the $M_{\rm h}\sim 10^{10} M_\odot$ dark matter halo at $z\sim4-7$.
If we use the original mass function of \citet{2008ApJ...688..709T}, we find that none of our conclusions are changed.

In our HOD model, $\xi_\m{g}(r)$ is computed from the galaxy power spectrum, $P_\m{g}(k)$, through the Fourier transformation,
\begin{equation}
\xi_\m{g}(r)=\frac{1}{2\pi^2}\int^\infty_0dk\ k^2P_\m{g}(k)\frac{\m{sin}kr}{kr}.
\end{equation}
The galaxy power spectrum is described by 
\begin{equation}
P_\m{g}(k)=P_\m{g}^\m{1h}(k)+P_\m{g}^\m{2h}(k),
\end{equation}
where $P_\m{g}^\m{1h}(r)$ ($P_\m{g}^\m{2h}(r)$) is the one (two) halo term
for pairs of galaxies in one (two different) halo(s).

\begin{deluxetable*}{cccccccc}
\setlength{\tabcolsep}{0.35cm}
\tablecaption{Summary of the Clustering Measurements with our HOD Model}
\tablehead{
\colhead{$z_c$} & \colhead{$m^\m{aper}_\m{UV, th}$} & \colhead{$\logMmin$} & \colhead{$DC$}&\colhead{$\m{log}M_1^\prime$} & \colhead{$b_g^\m{eff}$} & \colhead{$\m{log}\left<M_\m{h}\right>$}&  \colhead{$\chi^2_{\nu}$}\\
\colhead{(1)}& \colhead{(2)}& \colhead{(3)}& \colhead{(4)} &  \colhead{(5)}& \colhead{(6)}& \colhead{(7)} & \colhead{(8)}}

\startdata
$3.8$ & $27.2$ & $11.42^{+0.07}_{-0.12}$ & $0.50^{+0.17}_{-0.16}$ &  $(12.19^{+0.08}_{-0.15})$ & $3.6^{+0.1}_{-0.2}$ &$11.89^{+0.04}_{-0.07}$ & 0.9\\
          & $27.6$ & $11.39^{+0.05}_{-0.09}$ & $0.99^{+0.01}_{-0.32}$ &  $12.35^{+0.20}_{-0.19}$ & $3.5^{+0.1}_{-0.1}$ &$11.80^{+0.04}_{-0.06}$ & 0.8\\
          & $28.2$ & $11.15^{+0.06}_{-0.13}$ & $0.80^{+0.12}_{-0.29}$ &  $12.07^{+0.21}_{-0.28}$ & $3.1^{+0.1}_{-0.1}$ &$11.67^{+0.04}_{-0.06}$ & 1.1\\
          & $29.2$ & $10.78^{+0.18}_{-0.25}$ & $0.95^{+0.03}_{-0.54}$ &  $(11.59^{+0.12}_{-0.35})$ & $2.8^{+0.2}_{-0.2}$ &$11.60^{+0.06}_{-0.07}$ & 0.8\\
          & $29.8$ & $10.55^{+0.14}_{-0.28}$ & $0.28^{+0.33}_{-0.13}$ &  $(11.17^{+0.17}_{-0.33})$ & $2.6^{+0.1}_{-0.1}$ &$11.54^{+0.04}_{-0.05}$ & 1.3\\
$4.8$ & $25.0$ & $12.25^{+0.05}_{-0.14}$ & $0.60$ (fix) &  $(13.18^{+0.06}_{-0.18})$ & $7.6^{+0.3}_{-0.7}$ &$12.35^{+0.05}_{-0.13}$ & 1.6\\
$4.9$ & $27.2$ & $11.35^{+0.05}_{-0.19}$ & $0.60$ (fix) &  $(12.12^{+0.24}_{-0.17})$ & $5.0^{+0.1}_{-0.7}$ &$11.66^{+0.02}_{-0.23}$ & 0.3\\
          & $27.6$ & $11.22^{+0.06}_{-0.18}$ & $0.60$ (fix) &  $(11.96^{+0.08}_{-0.20})$ & $4.7^{+0.2}_{-0.4}$ &$11.56^{+0.06}_{-0.10}$ & 0.9\\
          & $28.0$ & $11.11^{+0.10}_{-0.18}$ & $0.60$ (fix) &  $(11.80^{+0.11}_{-0.20})$ & $4.4^{+0.2}_{-0.3}$ &$11.48^{+0.06}_{-0.10}$ & 1.8\\
          & $29.2$ & $10.78^{+0.18}_{-0.24}$ & $0.60$ (fix) &  $(11.46^{+0.20}_{-0.26})$ & $3.8^{+0.3}_{-0.3}$ &$11.31^{+0.10}_{-0.10}$ & 0.5\\
$5.9$ & $27.4$ & $11.30^{+0.10}_{-0.13}$ & $0.60$ (fix) &  $(12.06^{+0.12}_{-0.16})$ & $6.3^{+0.4}_{-0.4}$ &$11.53^{+0.08}_{-0.10}$ & 0.5\\
          & $28.4$ & $11.03^{+0.05}_{-0.18}$ & $0.60$ (fix) &  $(11.75^{+0.07}_{-0.22})$ & $5.5^{+0.2}_{-0.4}$ &$11.30^{+0.07}_{-0.08}$ & 1.4\\
$6.8$ & $28.2$ & $11.04^{+0.08}_{-0.22}$ & $0.60$ (fix) &  $(11.77^{+0.07}_{-0.28})$ & $6.8^{+0.2}_{-0.8}$ &$11.28^{+0.04}_{-0.18}$ & 0.9\\
          & $28.4$ & $10.99^{+0.06}_{-0.20}$ & $0.60$ (fix) &  $(11.69^{+0.07}_{-0.27})$ & $6.3^{+0.4}_{-0.4}$ &$11.18^{+0.09}_{-0.11}$ & 0.6   
\enddata

\tablecomments{Columns: (1) Mean redshift. (2) Threshold magnitude in the rest-frame UV band.  (3) Best-fit value of $M_\m{min}$ in a unit of $\Msun$.  (4) Star formation duty cycle. (5) Best-fit value of $M_1^\prime$ in a unit of $\Msun$. The value in parentheses is derived from $\logMmin$ via equation (\ref{eq_M1p_Mmin}). (6) Effective bias. (7) Mean halo mass in a unit of $\Msun$. (8) Reduced $\chi^2$ value.}
\label{table_HOD}
\end{deluxetable*}

The one-halo term consists of a central-satellite part $P^\m{cs}_\m{g}(k)$ and a satellite-satellite part $P^\m{ss}_\m{g}(k)$,
\begin{eqnarray}\label{eq_1halo}
P_\m{g}^\m{1h}(k)=P^\m{cs}_\m{g}(k)+P^\m{ss}_\m{g}(k).
\end{eqnarray}
The quantities of $P^\m{cs}_\m{g}(k)$ and $P^\m{ss}_\m{g}(k)$ are given by
\begin{equation}
P^\m{cs}_\m{g}(k,z)=\frac{2}{n_\m{g}^2}\int dM_\m{h}\left<{N_\m{c}N_\m{s}}\right>(M_\m{h})\frac{dn}{dM_\m{h}}(M_\m{h},z)u(k,M_\m{h},z)
\end{equation}
and
\begin{align}
&P^\m{ss}_\m{g}(k,z)\notag\\
&=\frac{1}{n_\m{g}^2}\int dM_\m{h}\left<{N_\m{s}(N_\m{s}-1)}\right>(M_\m{h})\frac{dn}{dM_\m{h}}(M_\m{h},z)u^2(k,M_\m{h},z),
\end{align}
where $u(k,M_\m{h},z)$ is the Fourier transform of the dark matter halo density profile normalized by its mass \citep[e.g.,][]{2002PhR...372....1C}.
Here we assume that satellite galaxies in halos trace the density profile of the dark matter halo 
by NFW profile \citep{1996ApJ...462..563N,1997ApJ...490..493N}, and 
adopt the mass-concentration parameter relation by \citet{2001MNRAS.321..559B} with an appropriate correction \citep[see][]{2003ApJ...590..197S}.
If we assume the $z=4$ mass-concentration parameter relation for the halo mass estimate of the $z\sim7$ subsample, we find the negligible change of $0.05\ \m{dex}$ in $\logMmin$.
The values of $\left<{N_\m{c}N_\m{s}}\right>(M_\m{h})$ and $\left<{N_\m{s}(N_\m{s}-1)}\right>(M_\m{h})$ are 
the mean number of central-satellite and satellite-satellite galaxy pairs, respectively.
If we assume the independence of central and satellite galaxies
and a Poisson distribution of the satellite galaxy's distribution, 
these values are
\begin{eqnarray}
\left<{N_\m{c}N_\m{s}}\right>(M_\m{h})&=&N_\m{c}(M_\m{h})N_\m{s}(M_\m{h}),\\
\left<{N_\m{s}(N_\m{s}-1)}\right>(M_\m{h})&=&N_\m{s}^2(M_\m{h}).
\end{eqnarray}

The two-halo term is expressed as
\begin{align}\label{eq_2halo}
&P^\m{2h}_\m{g}(k,z)=P_\m{m}(k,z)\notag\\
&\left[\frac{1}{n_\m{g}}\int dM_\m{h}\ N(M_\m{h})\frac{dn}{dM_\m{h}}(M,z)b_\m{h}(M_\m{h},z)u(k,M_\m{h},z)\right]^2,
\end{align}
where $b_\m{h}(M_\m{h},z)$ is the halo bias factor \citep{2010ApJ...724..878T}.

To compare with the observational results,
we calculate the ACF from the galaxy power spectrum projecting on the redshift distribution using the Limber approximation \citep[see e.g. chapter 2 of][]{2001PhR...340..291B}, 
\begin{equation}\label{eq_limber}
\omega(\theta)=\int dzN^2(z)\left(\frac{dr}{dz}\right)^{-1}\int dk\frac{k}{2\pi}P_\m{g}(k,z)J_0[r(z)\theta k],
\end{equation}
where $N(z)$ is the normalized redshift distribution of galaxies and $J_0(x)$ is the zeroth-order Bessel function of the first kind.
Here we assume that $N_\m{c}(M)$ and $N_\m{s}(M)$ do not vary as a function of redshift
within the redshift ranges of the subsamples.
The quantity $r(z)$ is the radial comoving distance given by 
\begin{equation}
r(z)=\frac{c}{H_0}\int^z_0\frac{dz}{\sqrt{\Omega_\m{m,0}(1+z)^3+\Omega_{\Lambda,0}}},
\end{equation}
for a flat cosmology.
The mean galaxy number density with a redshift distribution $N(z)$ is calculated by
\begin{equation}
n_\m{g}=\frac{\int dz[{dV(r)}/{dz}]N(z)n_\m{g}(z)}{\int dz[{dV(r)}/{dz}]N(z)},
\end{equation}
where $n_\m{g}(z)$ is defined by Equation (\ref{eq:galaxy_number_density}),
and ${dV(r)}/{dz}$ is the comoving volume element per unit solid angle,
\begin{equation}
\frac{dV(z)}{dz}=r^2(z)\frac{dr}{dz}.
\end{equation}

\section{Dark matter halo Mass}\label{ss_result_mass}

\subsection{HOD Model Fitting}
We fit our HOD model to the ACF and the number density of each subsample, minimizing the $\chi^2$ value,
\begin{equation}
\chi^2=\sum_i \frac{\left[\omega_\m{obs}(\theta_i)-\omega_\m{model}(\theta_i)\right]^2}{\sigma_\omega^2(\theta_i)}+\frac{\left[\m{log}n_\m{g}^\m{obs}-\m{log}n_\m{g}^\m{model}\right]^2}{\sigma^2_{\m{log}n_\m{g}}}.
\end{equation}
We simply use the diagonal elements in the covariance matrix, in the same manner as \citet{2004MNRAS.347..813H} and \citet{2007ApJ...667..760Z}, because the errors of our correlation functions are dominated by the Poisson errors due to the small number statistics.
In fact, if we include off diagonal elements in the covariance matrix, changes in the best-fit value and $1\sigma$ errors of $\logMmin$ are small, $\sim0.06\ \m{dex}$, which does not change our conclusions.

We constrain the parameters of our HOD model using the Markov Chain Monte Carlo (MCMC) parameter estimation technique.
Our HOD model has a total of 6 parameters, $DC$, $M_\m{min}$, $\sigmalogM$, 
$M_0$, $M^\prime_1$, and $\alpha$. However, it is difficult to constrain all of these 6 parameters with our data
whose statistical accuracies are not high.
We thus fix $\sigmalogM=0.2$ and $\alpha=1.0$, following results of previous studies \citep[e.g.,][]{2004ApJ...609...35K,2005ApJ...633..791Z,2006ApJ...647..201C}.
To derive $M_0$ from $M^\prime_1$,
we use the relation
\begin{equation}\label{eq_M0_M1p}
\m{log}M_0=0.76\ \m{log}M_1^\prime+2.3,
\end{equation}
that is given by \citet{2006ApJ...647..201C} at $z\sim0-5$ based on their simulations.

Our HOD model with 3 parameters, $DC$, $M_\m{min}$, $M^\prime_1$,
is fitted to the subsamples of $z\sim 4$ LBGs with $m_{\rm UV, th}=27.6-28.2$
whose measurements have moderately high signal-to-noise ratios.
The best-fit parameters are summarized in Table \ref{table_HOD}.

\newcommand{\subsample}{z4_2720}
\begin{figure}
 \begin{center}
 \renewcommand{\subsample}{z4_2720}
 \begin{minipage}{0.55\hsize}
  \begin{center}
   \includegraphics[clip,bb=0 6 325 280,width=1\hsize]{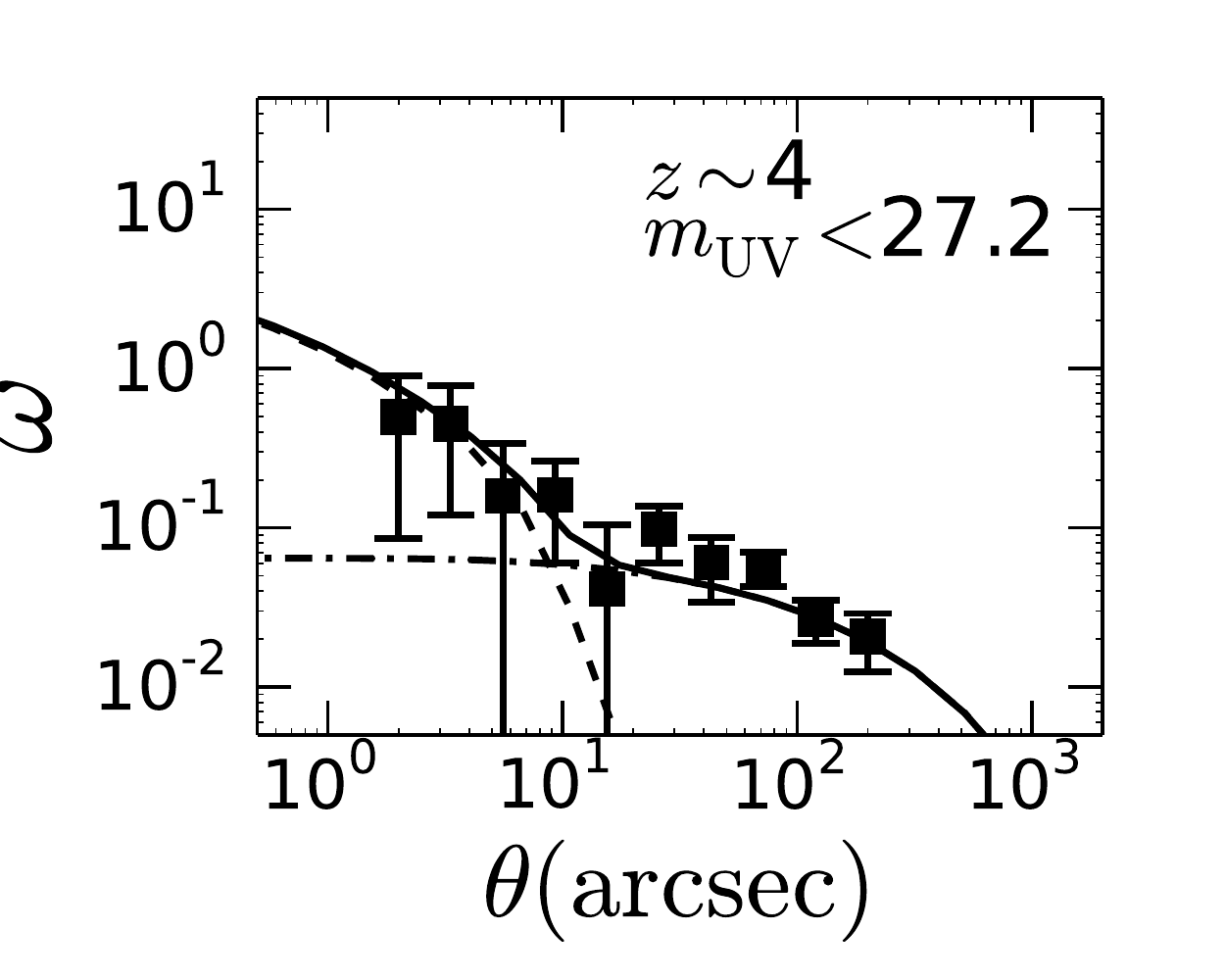}
     \end{center}
 \end{minipage}
 \begin{minipage}{0.42\hsize}
 \begin{center}
  \includegraphics[clip,bb=15 10 260 280,width=1\hsize]{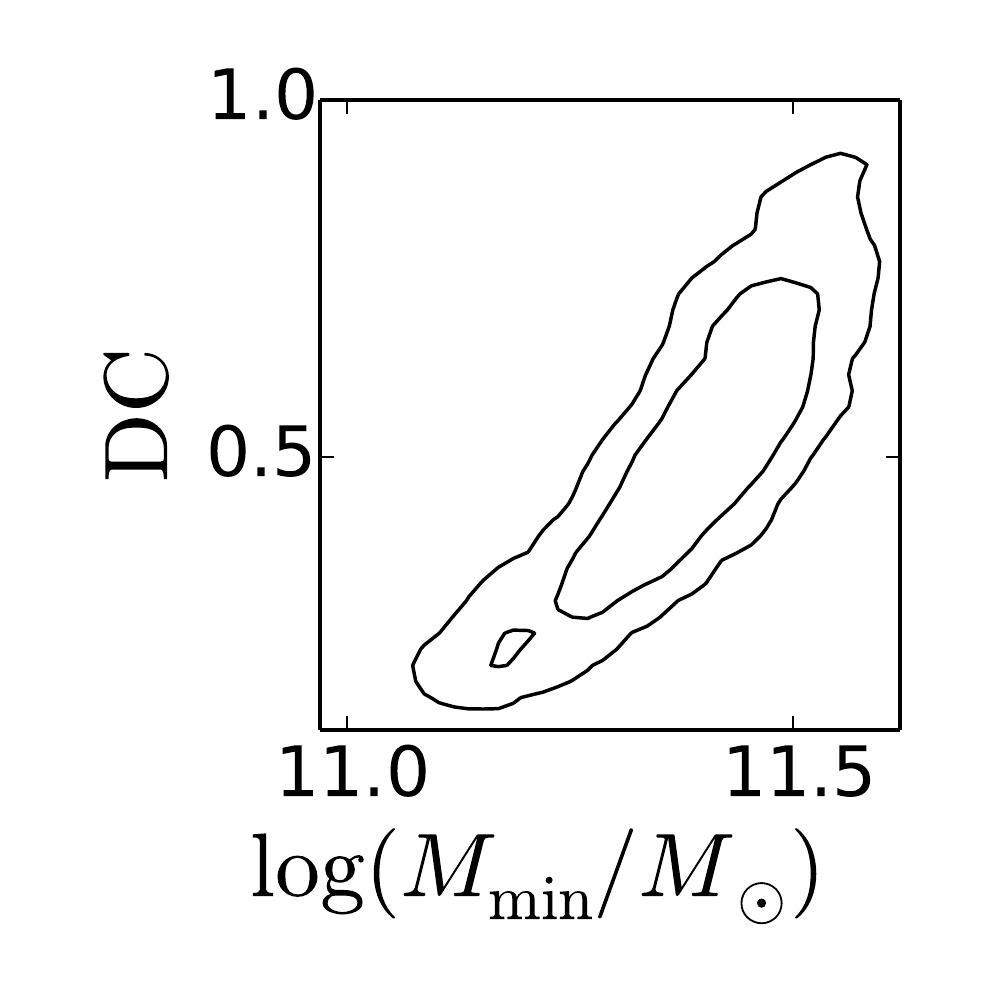}
 \end{center}
 \end{minipage}
 \\
  \renewcommand{\subsample}{z4_2760}
 \begin{minipage}{0.55\hsize}
  \begin{center}
   \includegraphics[clip,bb=0 6 325 280,width=1\hsize]{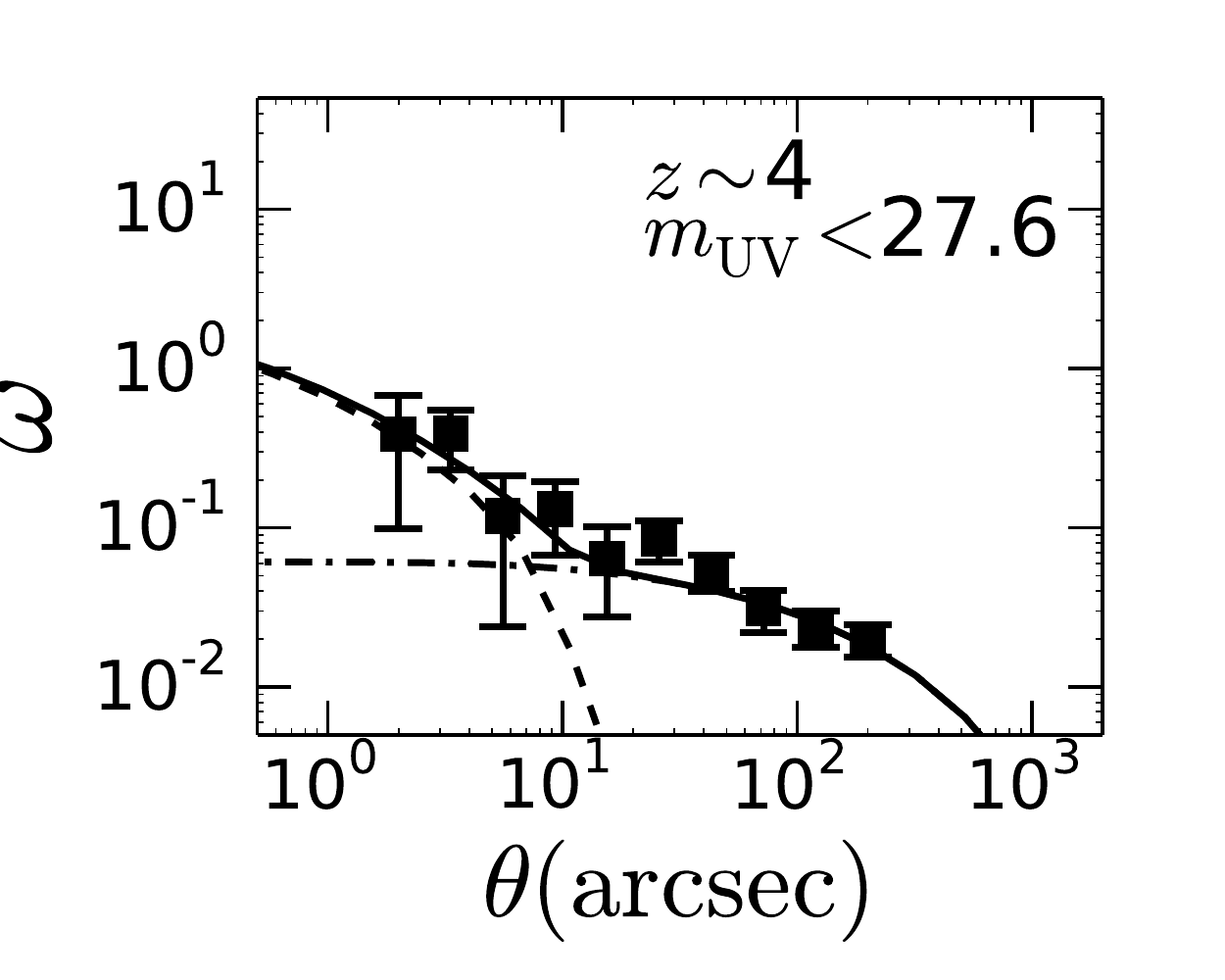}
     \end{center}
 \end{minipage}
 \begin{minipage}{0.42\hsize}
 \begin{center}
  \includegraphics[clip,bb=15 10 260 280,width=1\hsize]{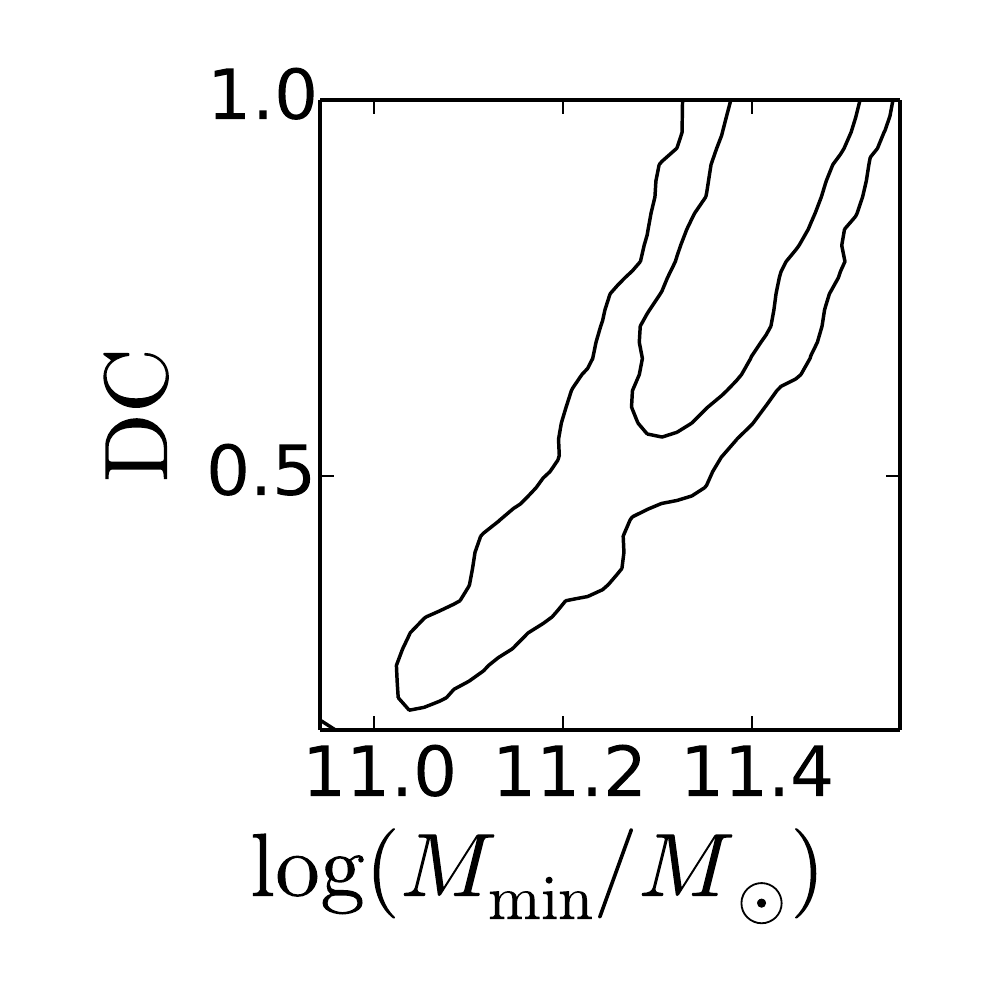}
 \end{center}
 \end{minipage}
 \\
  \renewcommand{\subsample}{z4_2820}
 \begin{minipage}{0.55\hsize}
  \begin{center}
   \includegraphics[clip,bb=0 6 325 280,width=1\hsize]{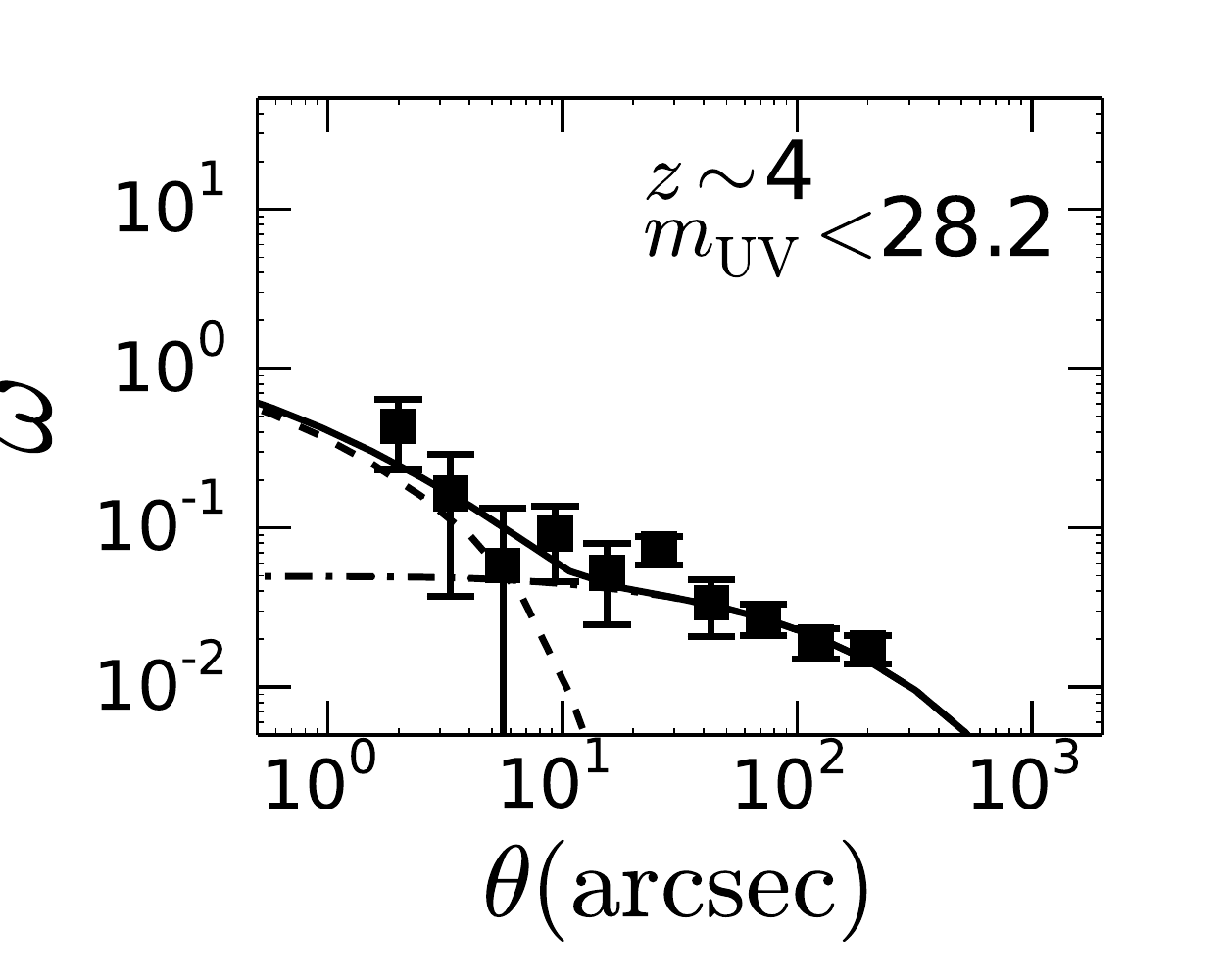}
     \end{center}
 \end{minipage}
 \begin{minipage}{0.42\hsize}
 \begin{center}
  \includegraphics[clip,bb=15 10 260 280,width=1\hsize]{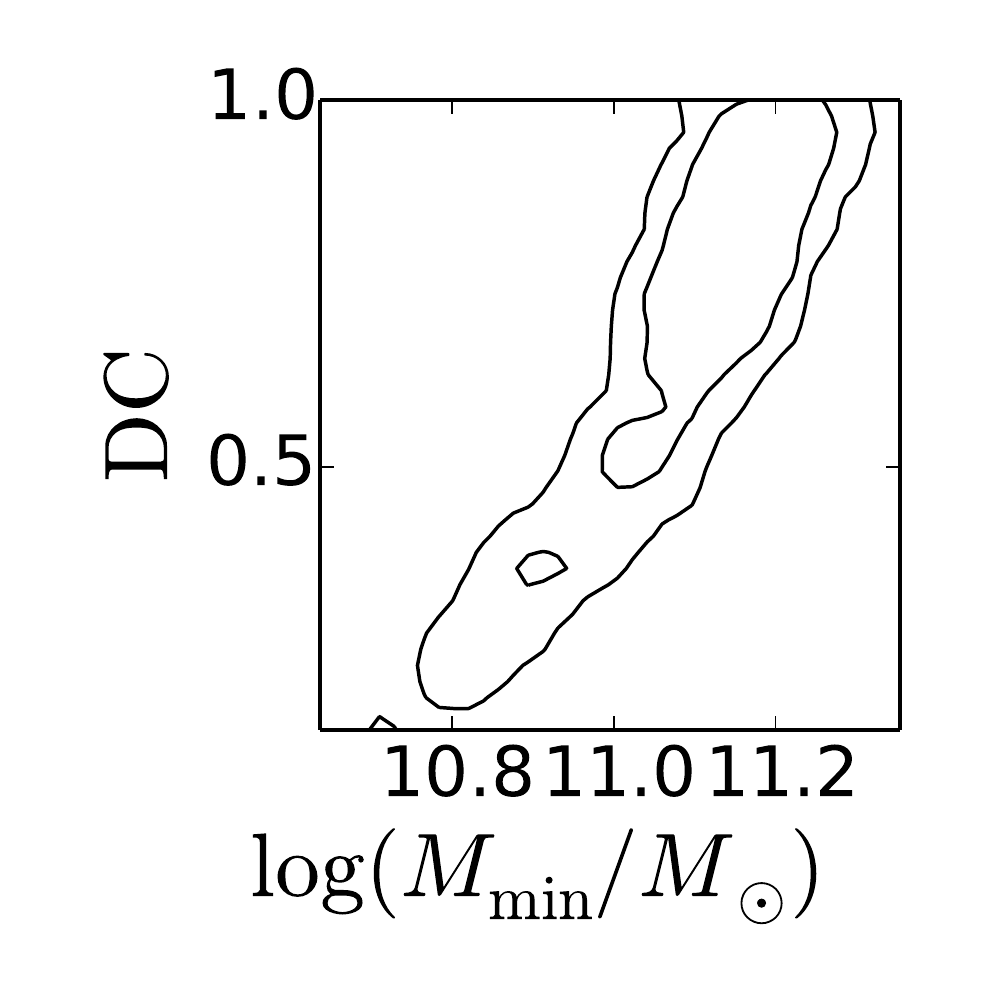}
 \end{center}
 \end{minipage}
 \\
  \renewcommand{\subsample}{z4_2920}
 \begin{minipage}{0.55\hsize}
  \begin{center}
   \includegraphics[clip,bb=0 6 325 280,width=1\hsize]{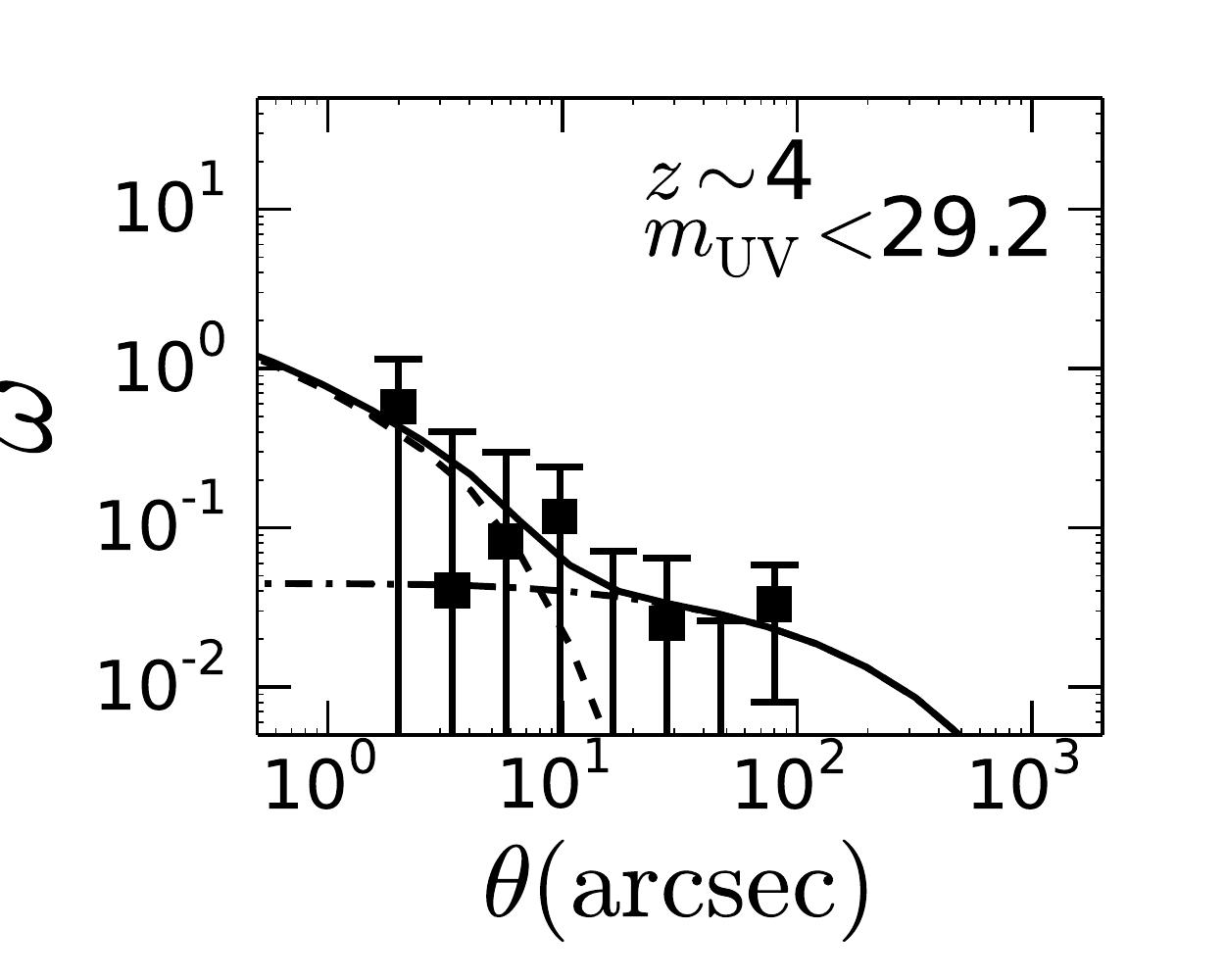}
     \end{center}
 \end{minipage}
 \begin{minipage}{0.42\hsize}
 \begin{center}
  \includegraphics[clip,bb=15 10 260 280,width=1\hsize]{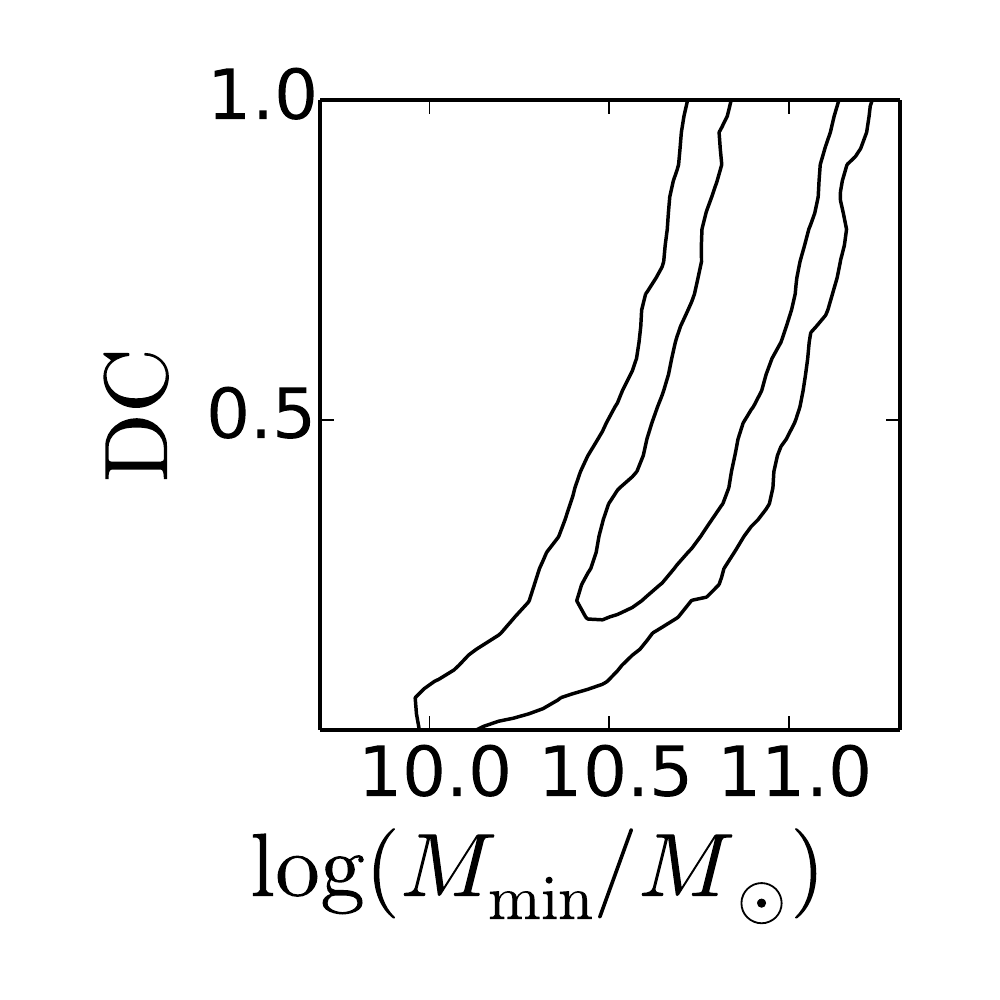}
 \end{center}
 \end{minipage}
 \\
  \renewcommand{\subsample}{z4_2980}
 \begin{minipage}{0.55\hsize}
  \begin{center}
   \includegraphics[clip,bb=0 6 325 280,width=1\hsize]{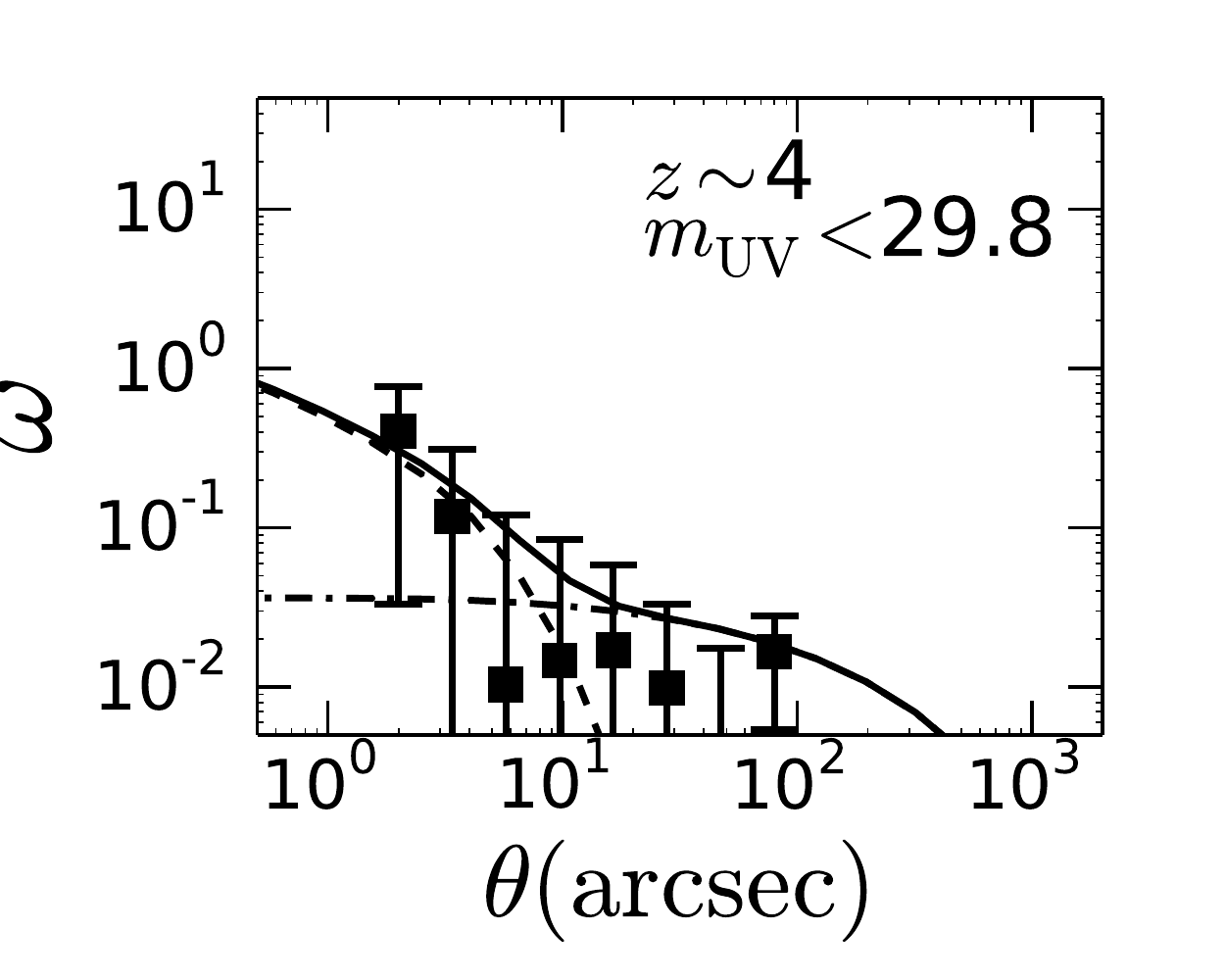}
     \end{center}
 \end{minipage}
 \begin{minipage}{0.42\hsize}
 \begin{center}
  \includegraphics[clip,bb=15 10 260 280,width=1\hsize]{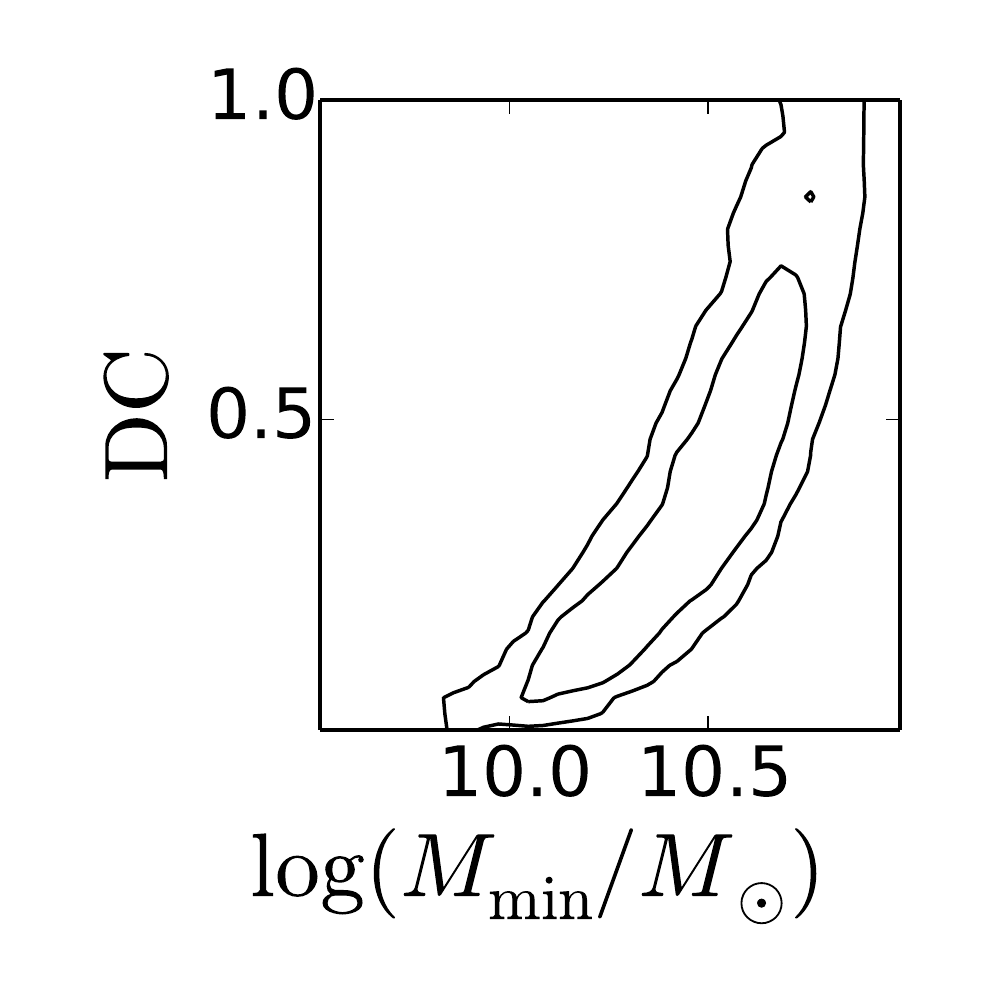}
 \end{center}
 \end{minipage}
 \end{center}
 \caption{ACF and the best-fit HOD model at $z\sim4$.
  Left panels: ACF with the prediction from our best-fit HOD model.
The dashed and dot-dashed curves denote the 1-halo and 2-halo 
terms (Equations \ref{eq_1halo} and \ref{eq_2halo}),
  respectively.
  Right panels: error contour obtained from our MCMC run.
The contours indicate the 68\% and 95\% confidence regions.
  \label{fig_hod_z4}}
\end{figure}

\begin{figure}
 \begin{center}
 \renewcommand{\subsample}{z5_2500}
 \begin{minipage}{0.55\hsize}
  \begin{center}
   \includegraphics[clip,bb=0 6 325 280,width=1\hsize]{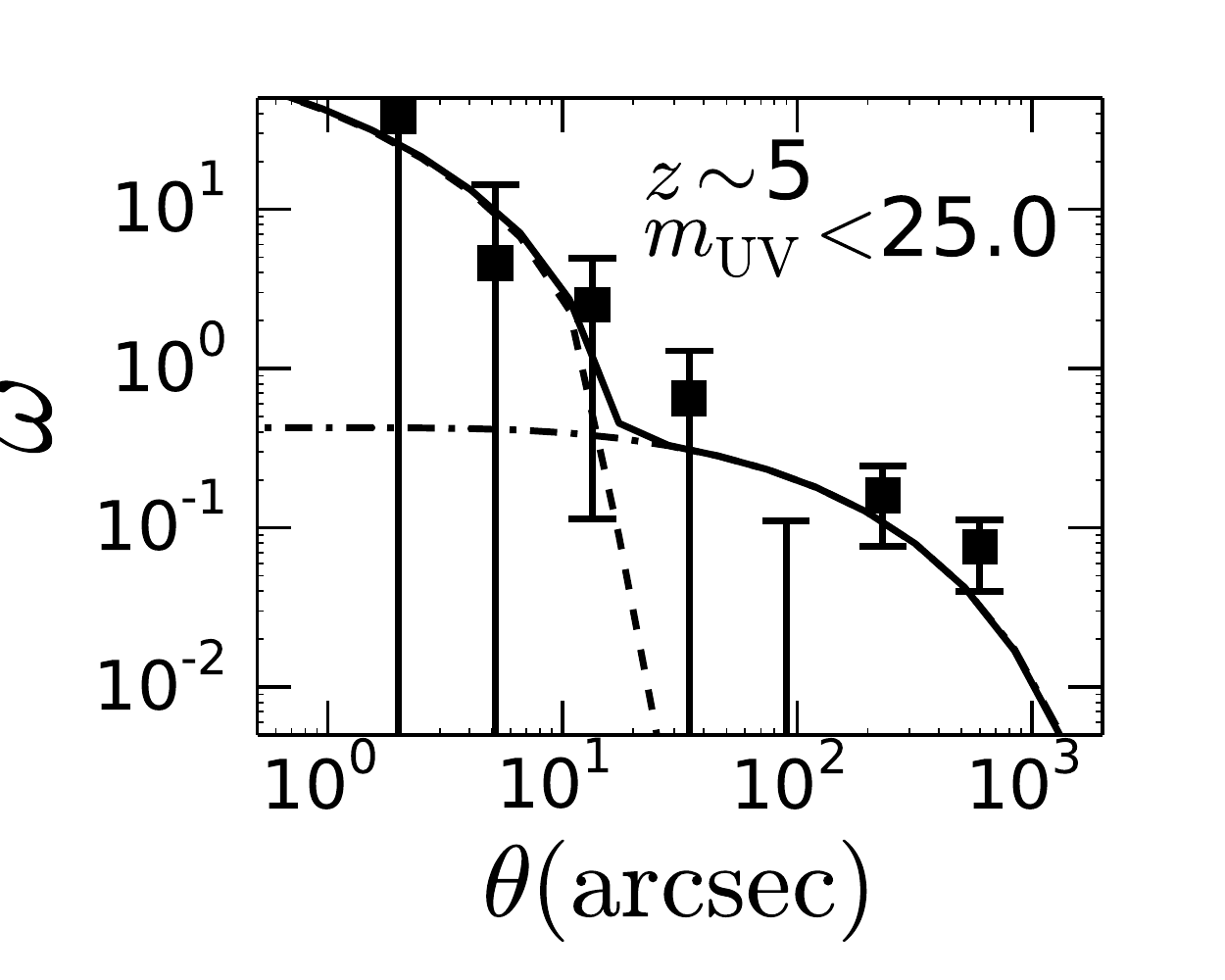}
     \end{center}
 \end{minipage}
 \begin{minipage}{0.42\hsize}
 \begin{center}
  \includegraphics[clip,bb=10 0 260 280,width=1\hsize]{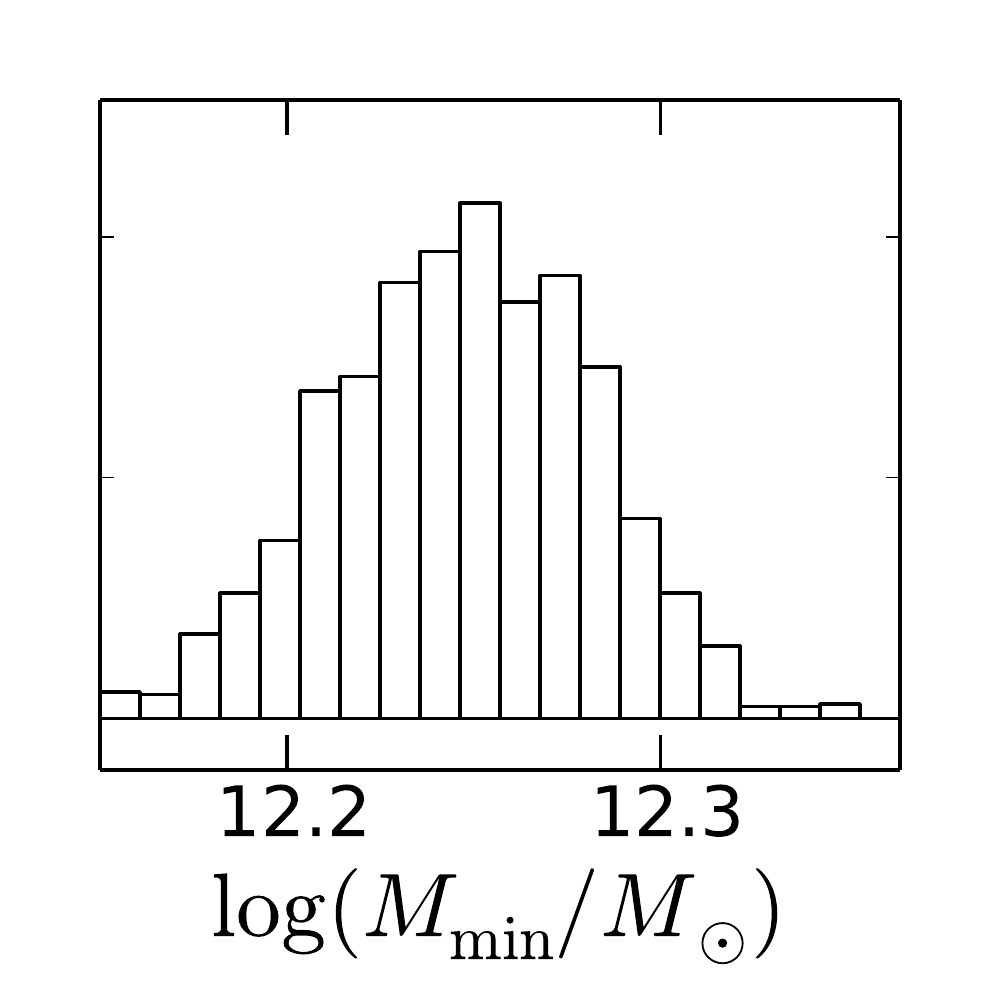}
 \end{center}
 \end{minipage}
 \\
 \renewcommand{\subsample}{z5_2720}
 \begin{minipage}{0.55\hsize}
  \begin{center}
   \includegraphics[clip,bb=0 6 325 280,width=1\hsize]{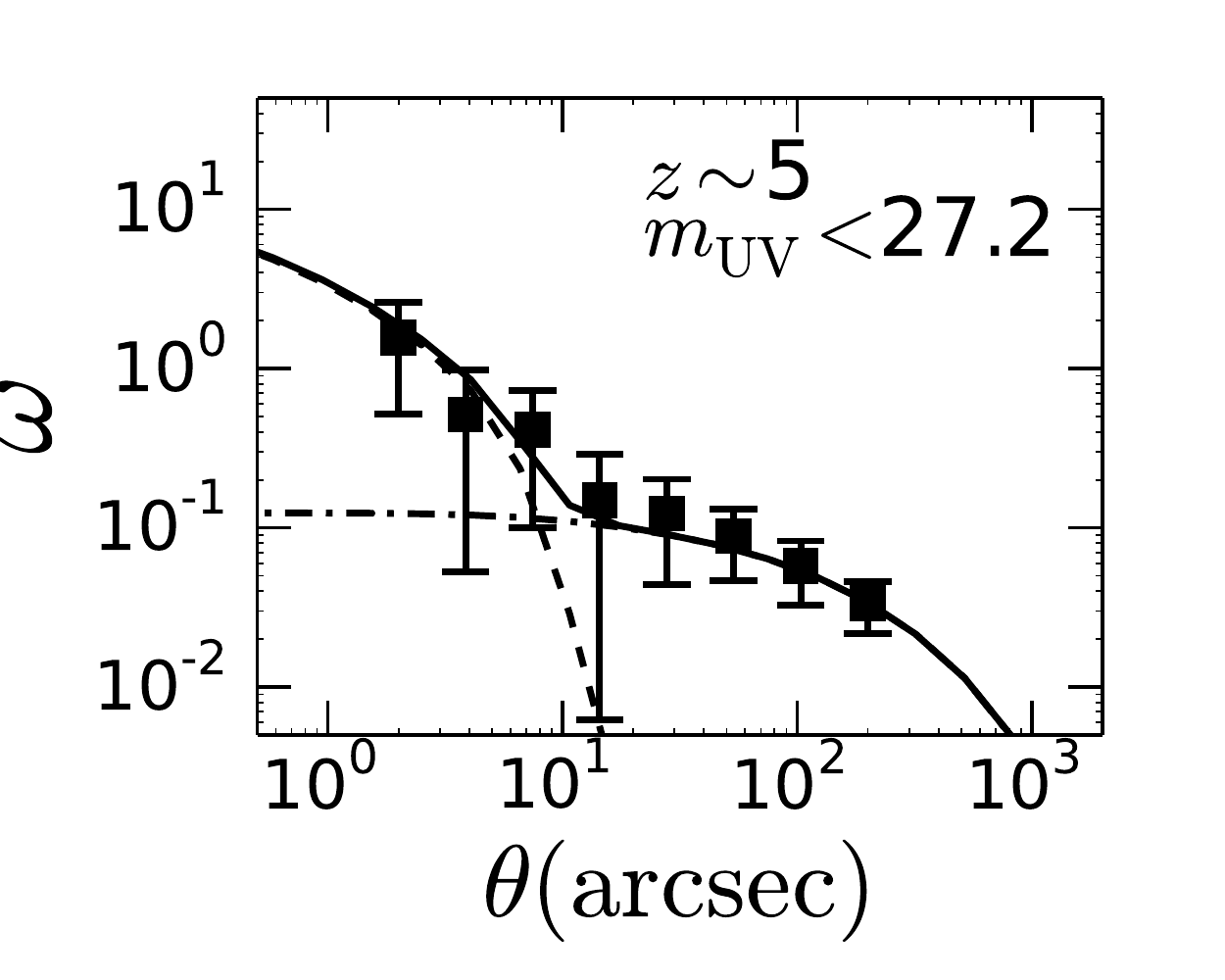}
     \end{center}
 \end{minipage}
 \begin{minipage}{0.42\hsize}
 \begin{center}
  \includegraphics[clip,bb=10 0 260 280,width=1\hsize]{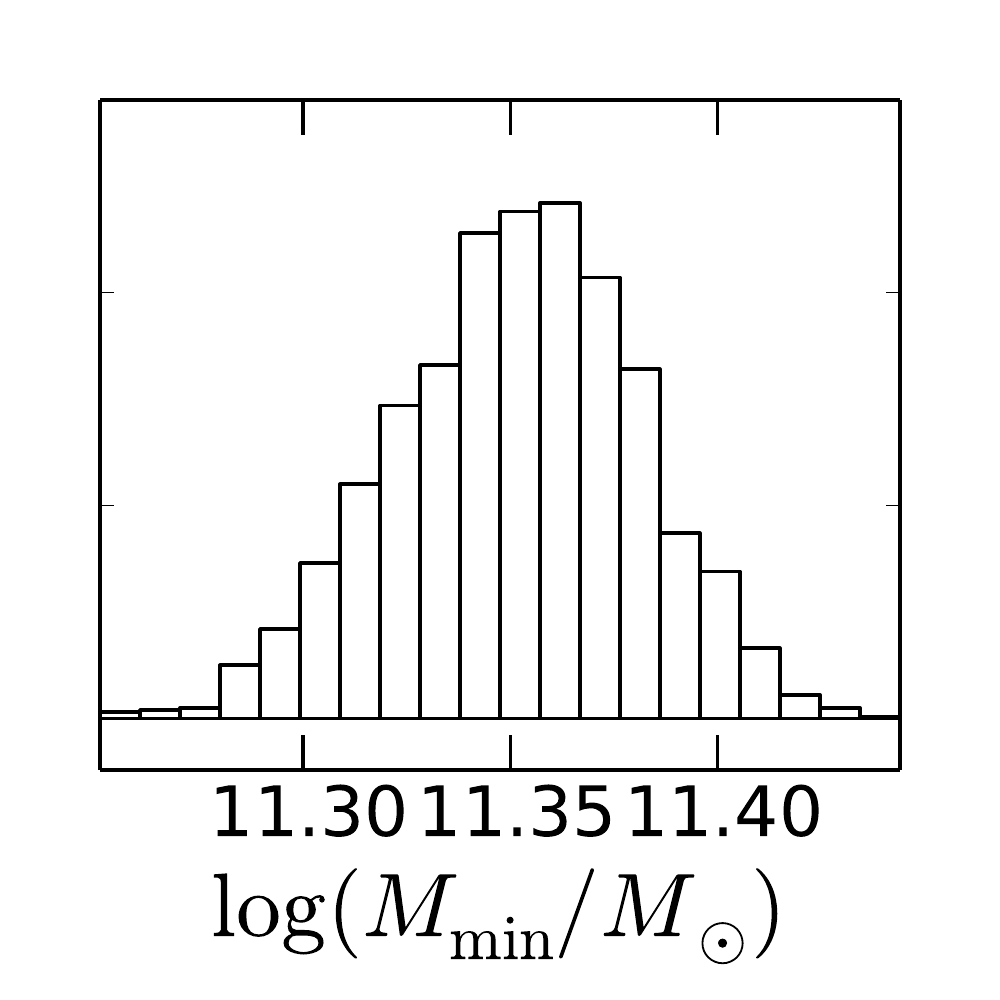}
 \end{center}
 \end{minipage}
 \\
 \renewcommand{\subsample}{z5_2760}
 \begin{minipage}{0.55\hsize}
  \begin{center}
   \includegraphics[clip,bb=0 6 325 280,width=1\hsize]{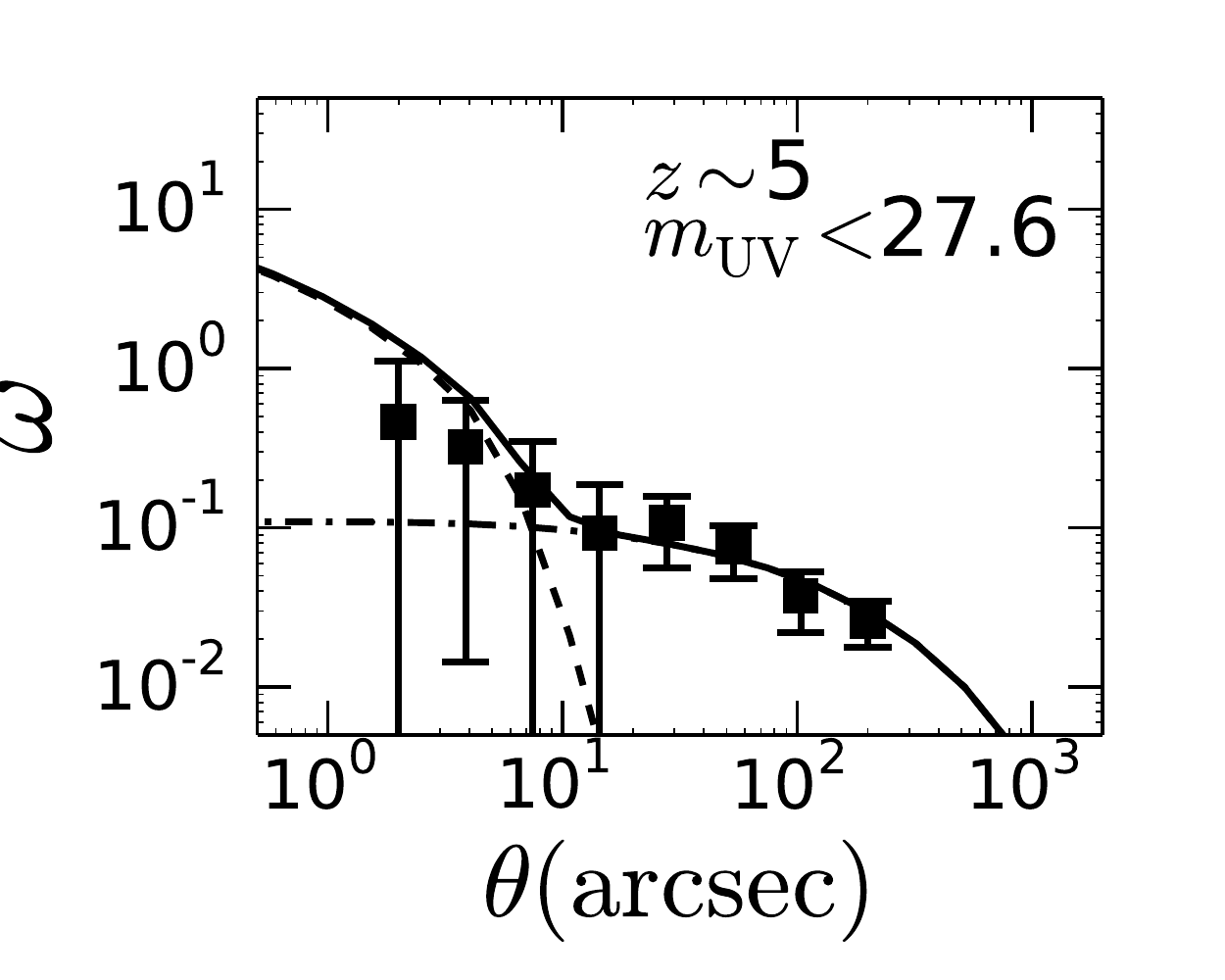}
     \end{center}
 \end{minipage}
 \begin{minipage}{0.42\hsize}
 \begin{center}
  \includegraphics[clip,bb=10 0 260 280,width=1\hsize]{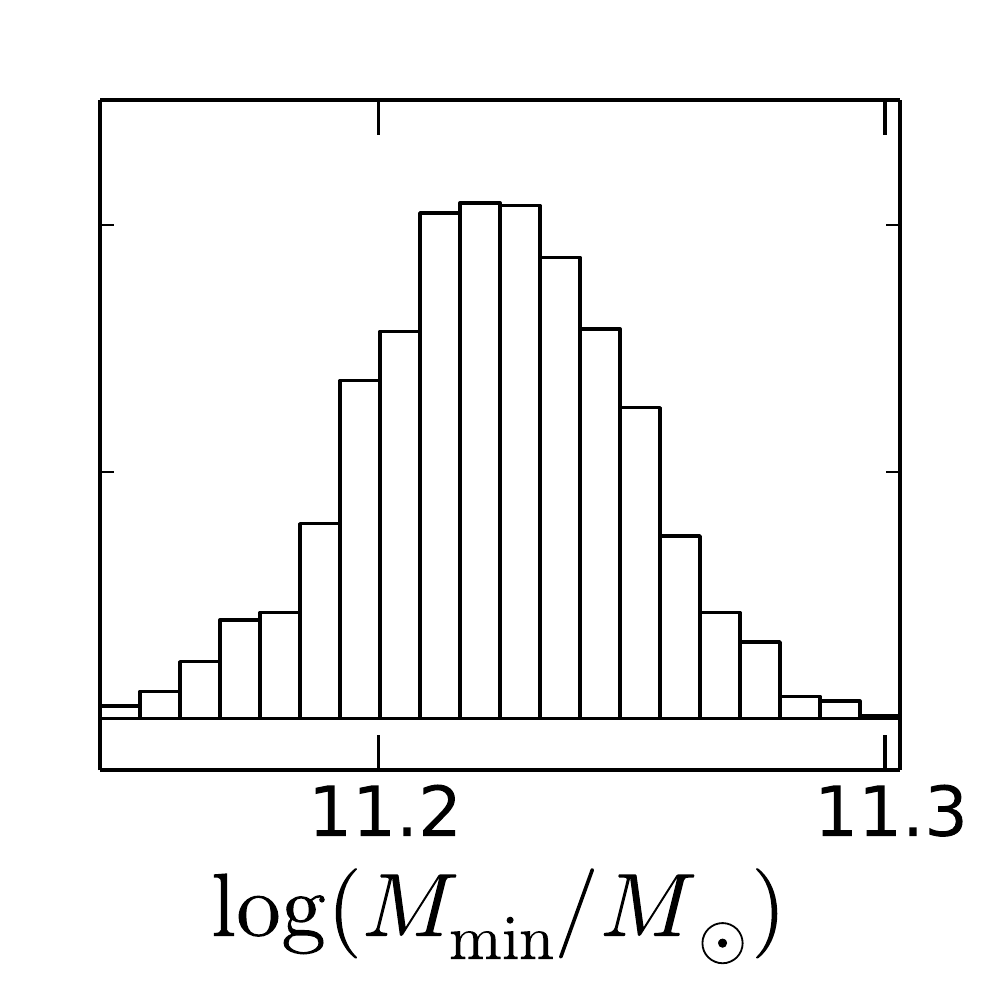}
 \end{center}
 \end{minipage}
 \\
 \renewcommand{\subsample}{z5_2800}
 \begin{minipage}{0.55\hsize}
  \begin{center}
   \includegraphics[clip,bb=0 6 325 280,width=1\hsize]{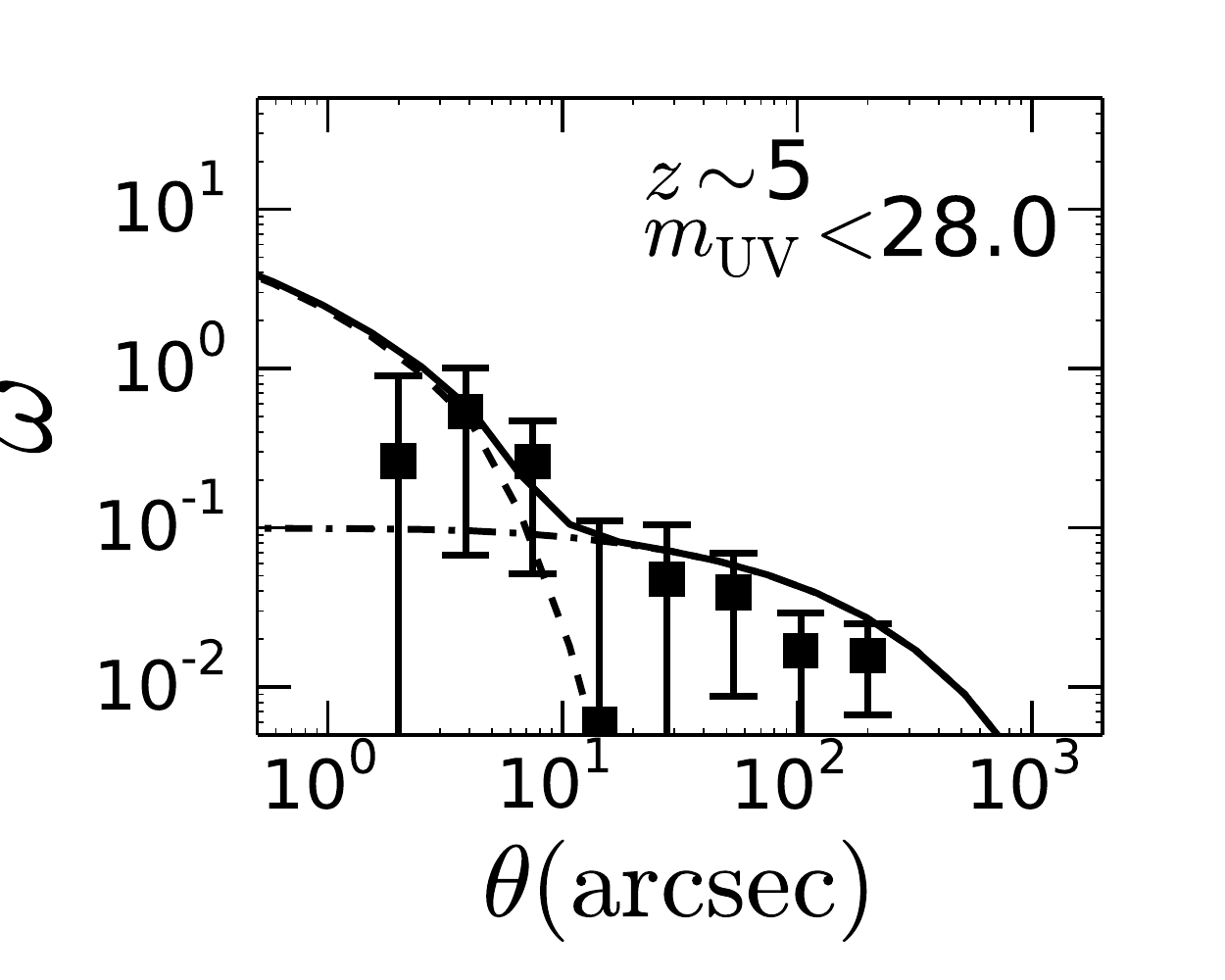}
     \end{center}
 \end{minipage}
 \begin{minipage}{0.42\hsize}
 \begin{center}
  \includegraphics[clip,bb=10 0 260 280,width=1\hsize]{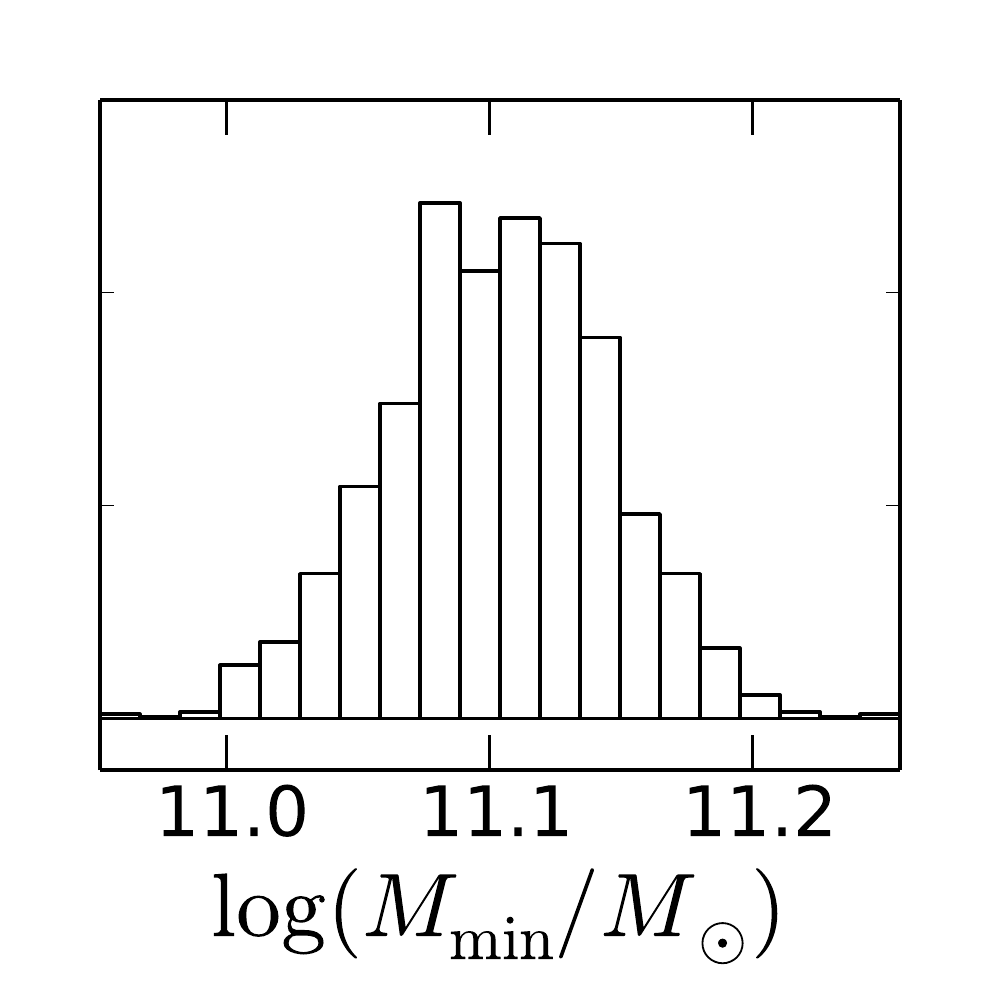}
 \end{center}
 \end{minipage}
 \\
 \renewcommand{\subsample}{z5_2920}
 \begin{minipage}{0.55\hsize}
  \begin{center}
   \includegraphics[clip,bb=0 6 325 280,width=1\hsize]{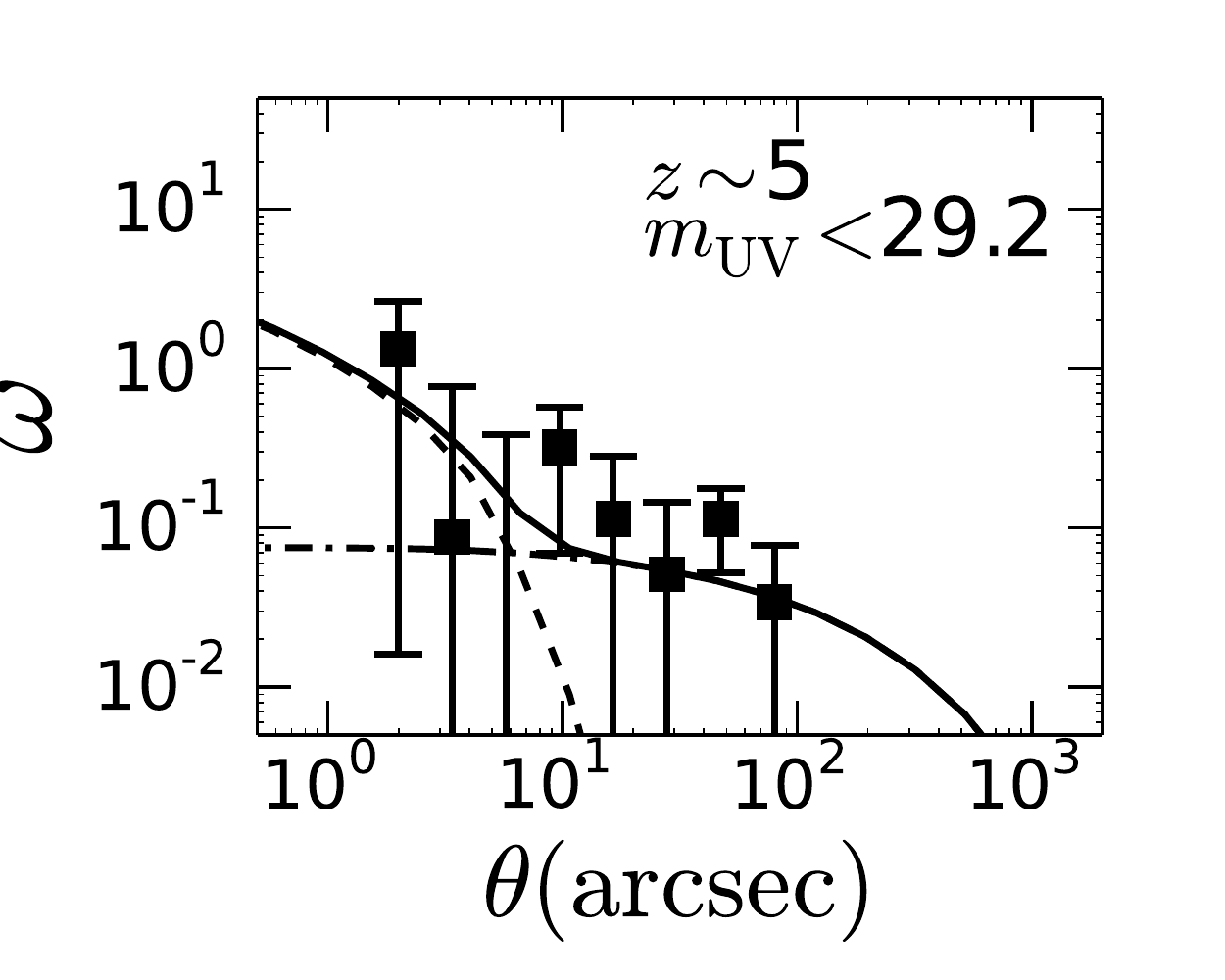}
     \end{center}
 \end{minipage}
 \begin{minipage}{0.42\hsize}
 \begin{center}
  \includegraphics[clip,bb=10 0 260 280,width=1\hsize]{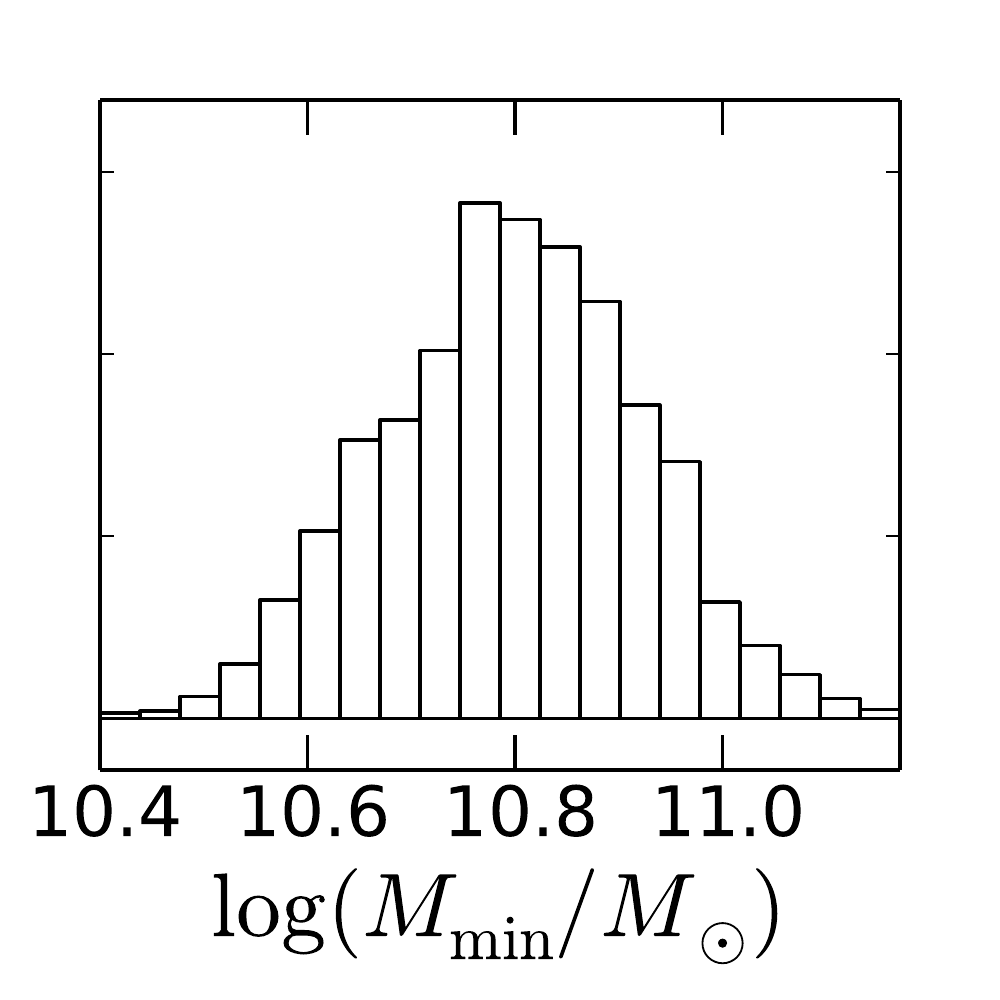}
 \end{center}
 \end{minipage}
 \end{center}
  \caption{ACF and the best-fit HOD model at $z\sim5$.
  Left panels: ACF with the prediction from our best-fit HOD model.
The dashed
  and dot-dashed curves denote the 1-halo and 2-halo 
terms (Equations \ref{eq_1halo} and \ref{eq_2halo}),
  respectively.
  Right panels: probability distribution for $\logMmin$
obtained
  from our MCMC run.
  \label{fig_hod_z5}}
 \end{figure}

Note that the rest of $z\sim 4$ LBG subsamples with $m_{\rm UV, th}=27.2$, $29.2$, and $29.8$
do not have statistical accuracies high enough to constrain
$M^\prime_1$ that describes the small scale clustering.\footnote{$M^\prime_1$ is sensitive to the 1-halo term (see Equation (\ref{eq:n_s})).}
We thus adopt another relation
\begin{equation}
\m{log}M_1^\prime=1.18\ \m{log}M_\m{min}-1.28, \label{eq_M1p_Mmin}
\end{equation}
that is calibrated with the results of \citet{2015MNRAS.446..169M}.
We obtain the best-fit parameters that are summarized in Table \ref{table_HOD}.
In the left panels of Figure \ref{fig_hod_z4}, we plot the ACFs 
with the best-fit HOD model curves. The models and the data agree well.
The right panels of Figure \ref{fig_hod_z4} are error contours 
in our HOD model obtained from the MCMC run.

\begin{figure}
 \begin{center}
 \renewcommand{\subsample}{z6_2740}
 \begin{minipage}{0.55\hsize}
  \begin{center}
   \includegraphics[clip,bb=0 6 325 280,width=1\hsize]{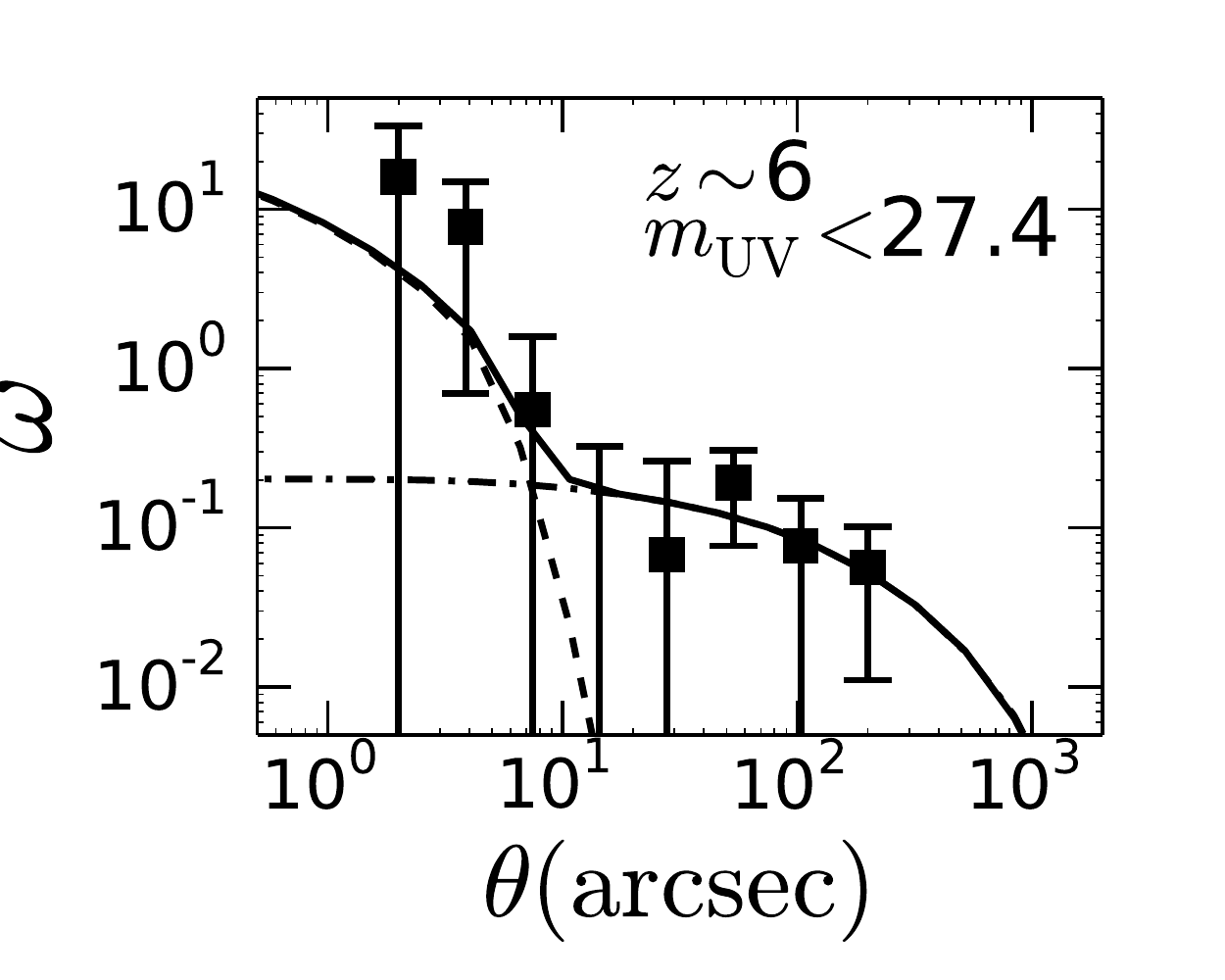}
     \end{center}
 \end{minipage}
 \begin{minipage}{0.42\hsize}
 \begin{center}
  \includegraphics[clip,bb=10 0 260 280,width=1\hsize]{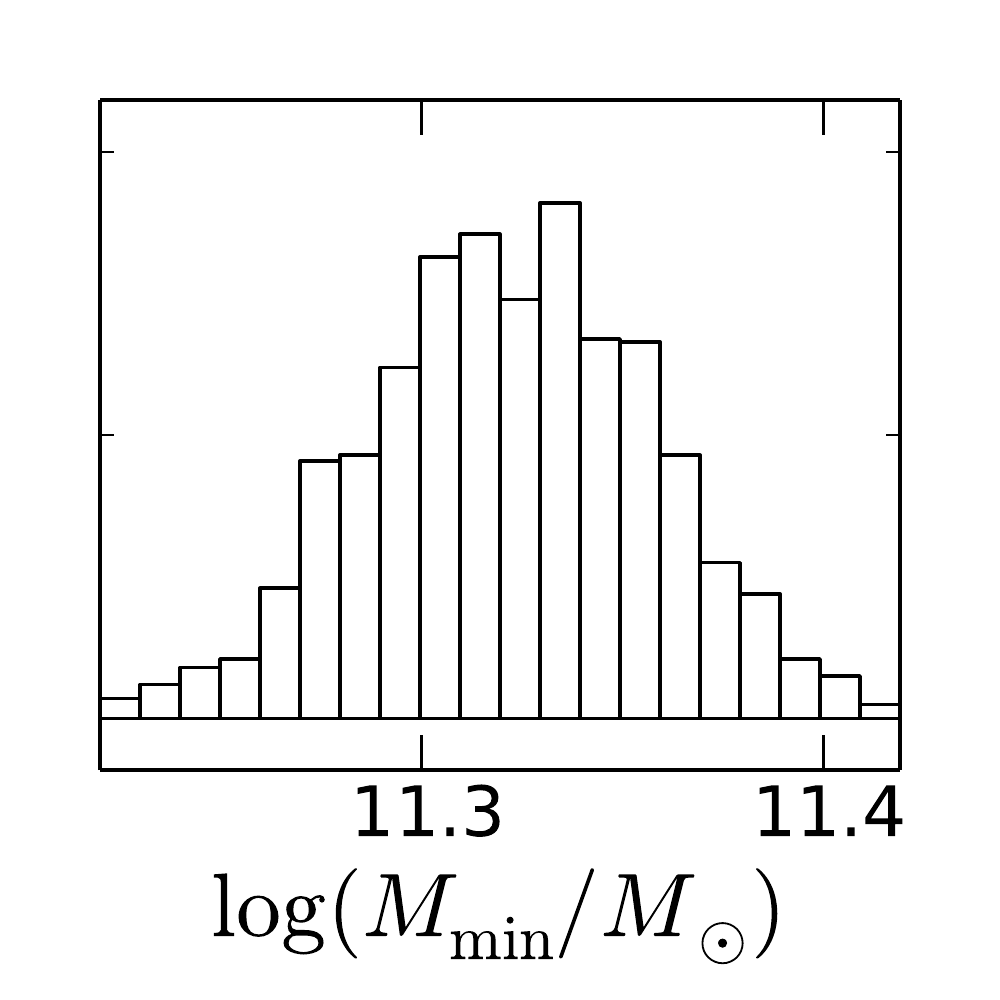}
 \end{center}
 \end{minipage}
 \\
 \renewcommand{\subsample}{z6_2840}
 \begin{minipage}{0.55\hsize}
  \begin{center}
   \includegraphics[clip,bb=0 6 325 280,width=1\hsize]{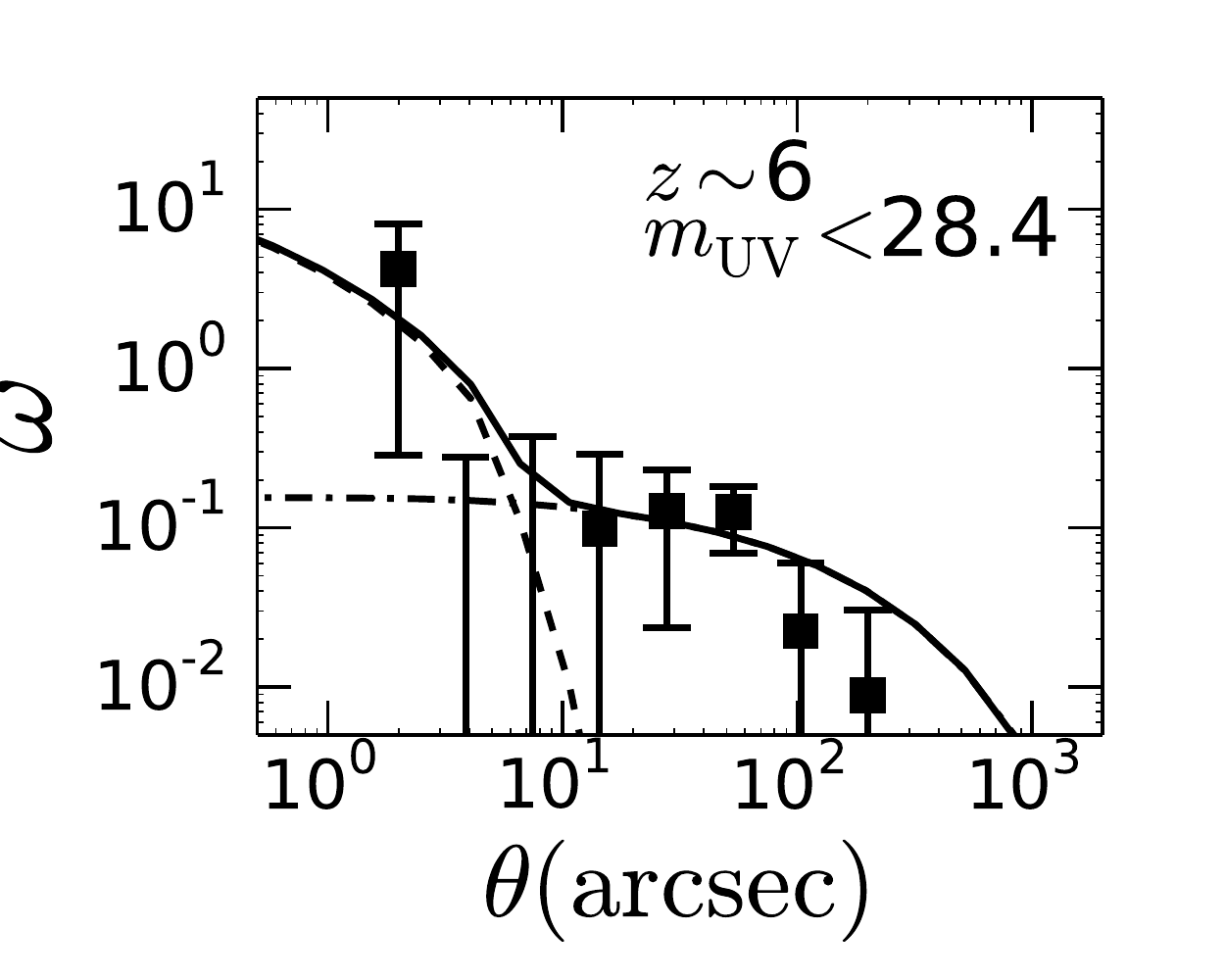}
     \end{center}
 \end{minipage}
 \begin{minipage}{0.42\hsize}
 \begin{center}
  \includegraphics[clip,bb=10 0 260 280,width=1\hsize]{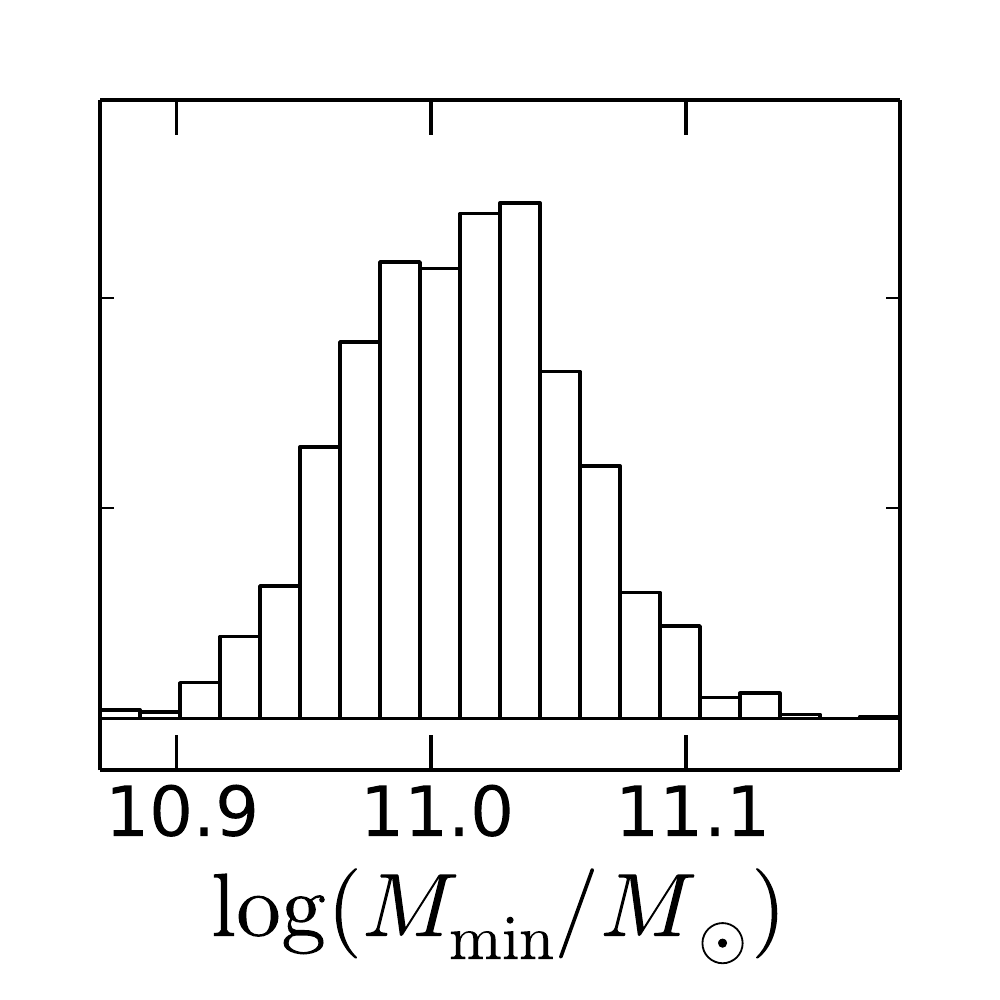}
 \end{center}
 \end{minipage}
 \\
 \renewcommand{\subsample}{z7_2820}
 \begin{minipage}{0.55\hsize}
  \begin{center}
   \includegraphics[clip,bb=0 6 325 280,width=1\hsize]{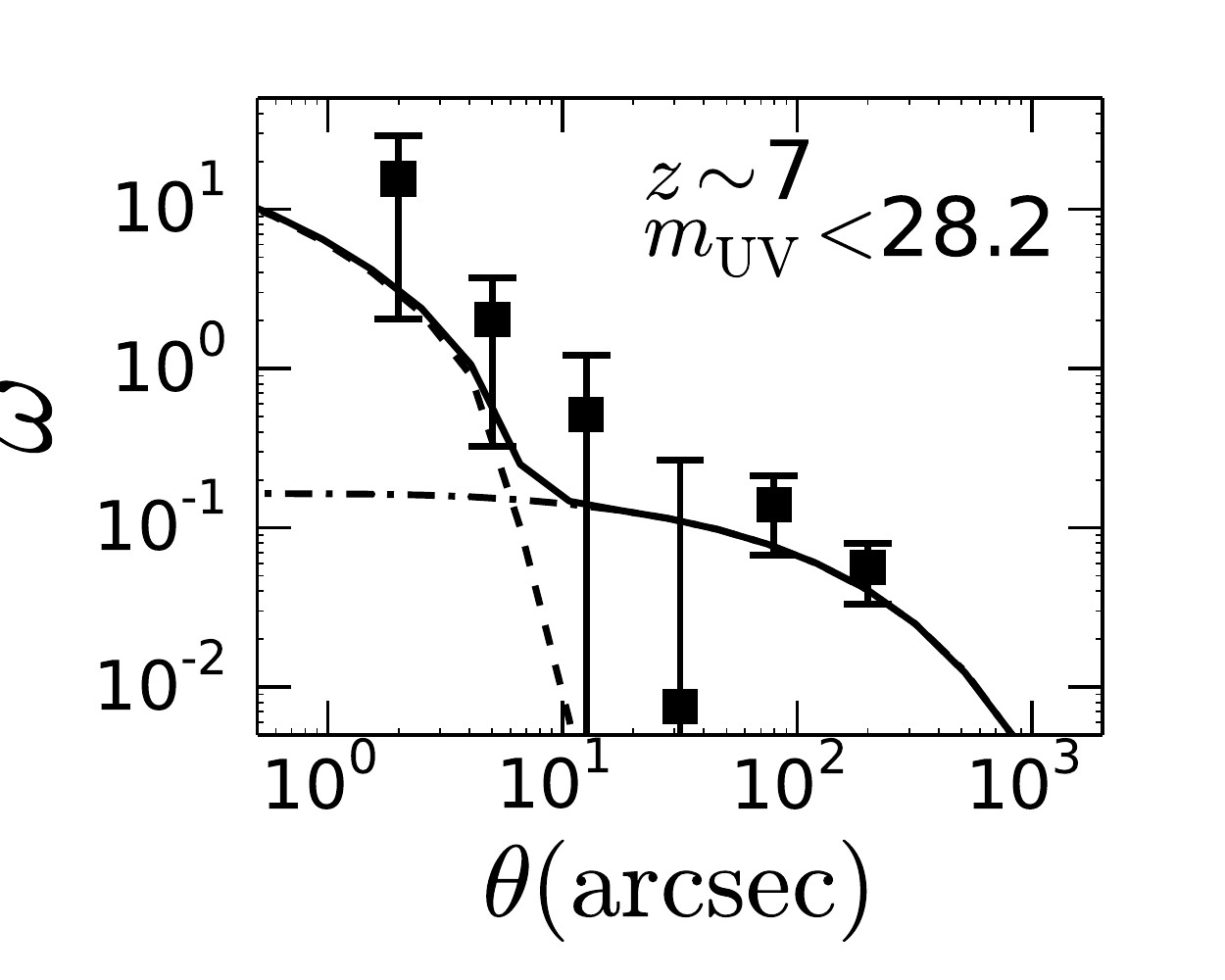}
     \end{center}
 \end{minipage}
 \begin{minipage}{0.42\hsize}
 \begin{center}
  \includegraphics[clip,bb=10 0 260 280,width=1\hsize]{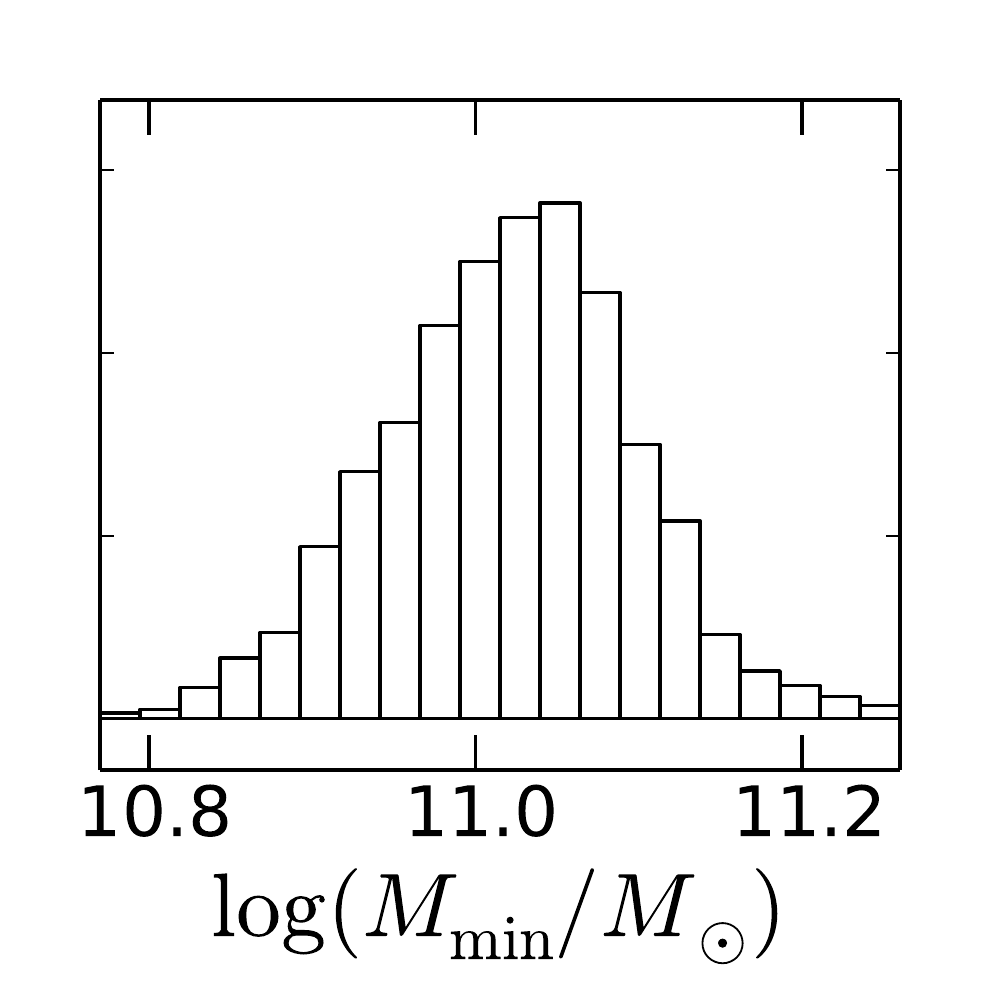}
 \end{center}
 \end{minipage}
 \\
 \renewcommand{\subsample}{z7_2840}
 \begin{minipage}{0.55\hsize}
  \begin{center}
   \includegraphics[clip,bb=0 6 325 280,width=1\hsize]{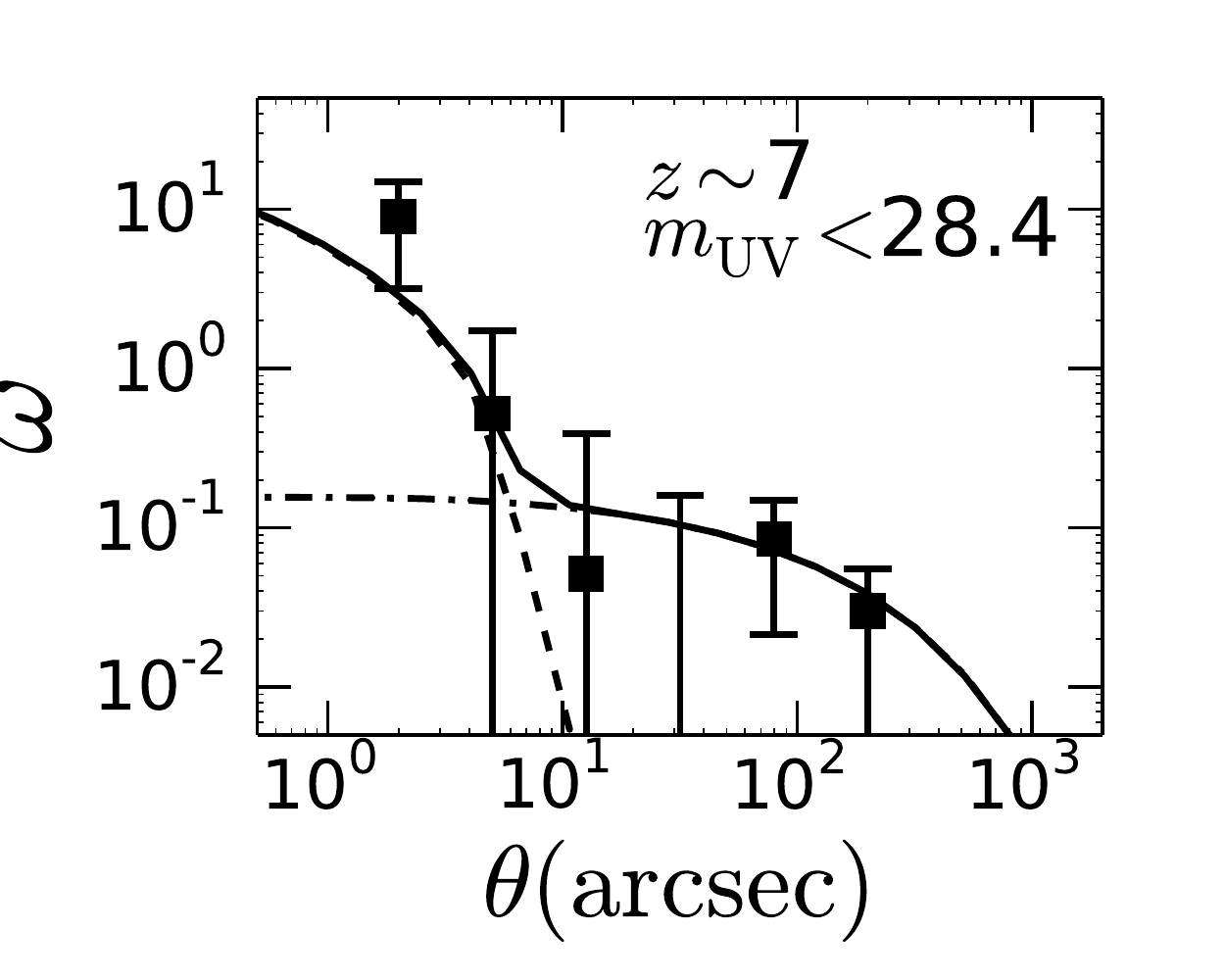}
     \end{center}
 \end{minipage}
 \begin{minipage}{0.42\hsize}
 \begin{center}
  \includegraphics[clip,bb=10 0 260 280,width=1\hsize]{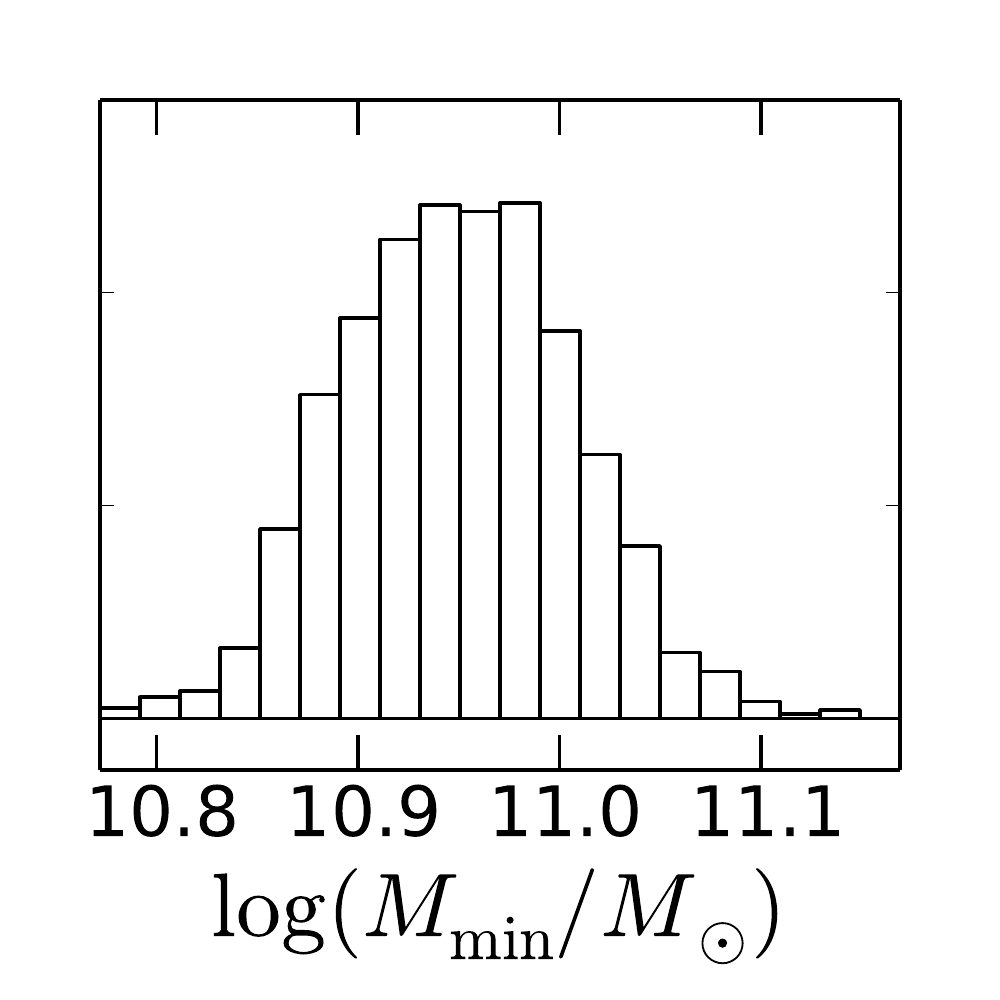}
 \end{center}
 \end{minipage}
   \end{center}
  \caption{
Same as Figure \ref{fig_hod_z5}, but for $z\sim6$ and $7$.
  \label{fig_hod_z67}}
\end{figure}

The $z\geq5$ subsamples have statistical uncertainties even higher than 
$z\sim 4$ subsamples. Here we calculate the mean $DC$ value
from the $z\sim 4$ subsample fitting results to be $DC=0.6^{+0.2}_{-0.3}$.
Assuming that $DC$ does not evolve by redshift, we use $DC=0.6$
in our model fitting for the $z\geq5$ subsamples. We thus obtain the $M_\m{min}$ estimates
as summarized in Table \ref{table_HOD}. 
The errors of $M_\m{min}$ include
statistical uncertainties in the fitting with $DC=0.6$ and 
uncertainties originating from the DC determination ($DC=0.6^{+0.2}_{-0.3}$).
Figures \ref{fig_hod_z5} and \ref{fig_hod_z67} are the same as Figure \ref{fig_hod_z4},
but for the $z\geq5$ subsamples. Because $DC$ is fixed to $DC=0.6$,
the right panels of Figures \ref{fig_hod_z5} and \ref{fig_hod_z67}
show the probability distributions of $M_\m{min}$.

\subsection{Dark matter halo Mass Estimates}

From the best-fit HOD model parameters, we calculate 
the effective galaxy bias
\begin{equation}
b_\m{g}^\m{eff} =\frac{1}{n_\m{g}}\int dM_\m{h}\frac{dn}{dM_\m{h}}(M_\m{h},z)N(M_\m{h})b_\m{h}(M_\m{h},z) \label{eq_bias_HOD}
\end{equation}
and the mean dark matter halo mass of central and satellite galaxies
\begin{equation}
\left<M_\m{h}\right> = \frac{1}{n_\m{g}}\int dM_\m{h}\frac{dn}{dM_\m{h}}(M_\m{h},z)N(M_\m{h})M_\m{h},
\end{equation}
and present the results in Table \ref{table_HOD}.
The most of the effective biases estimated by Equation (\ref{eq_bias_HOD}) are systematically larger than the biases estimated by Equation (\ref{eq_bias}), probably because the effective bias with Equation (\ref{eq_bias_HOD}) is the average bias including the satellite galaxies, majority of which reside in the massive halos.

\begin{figure}
 \begin{center}
  \includegraphics[clip,bb=10 10 340 340,width=1\hsize]{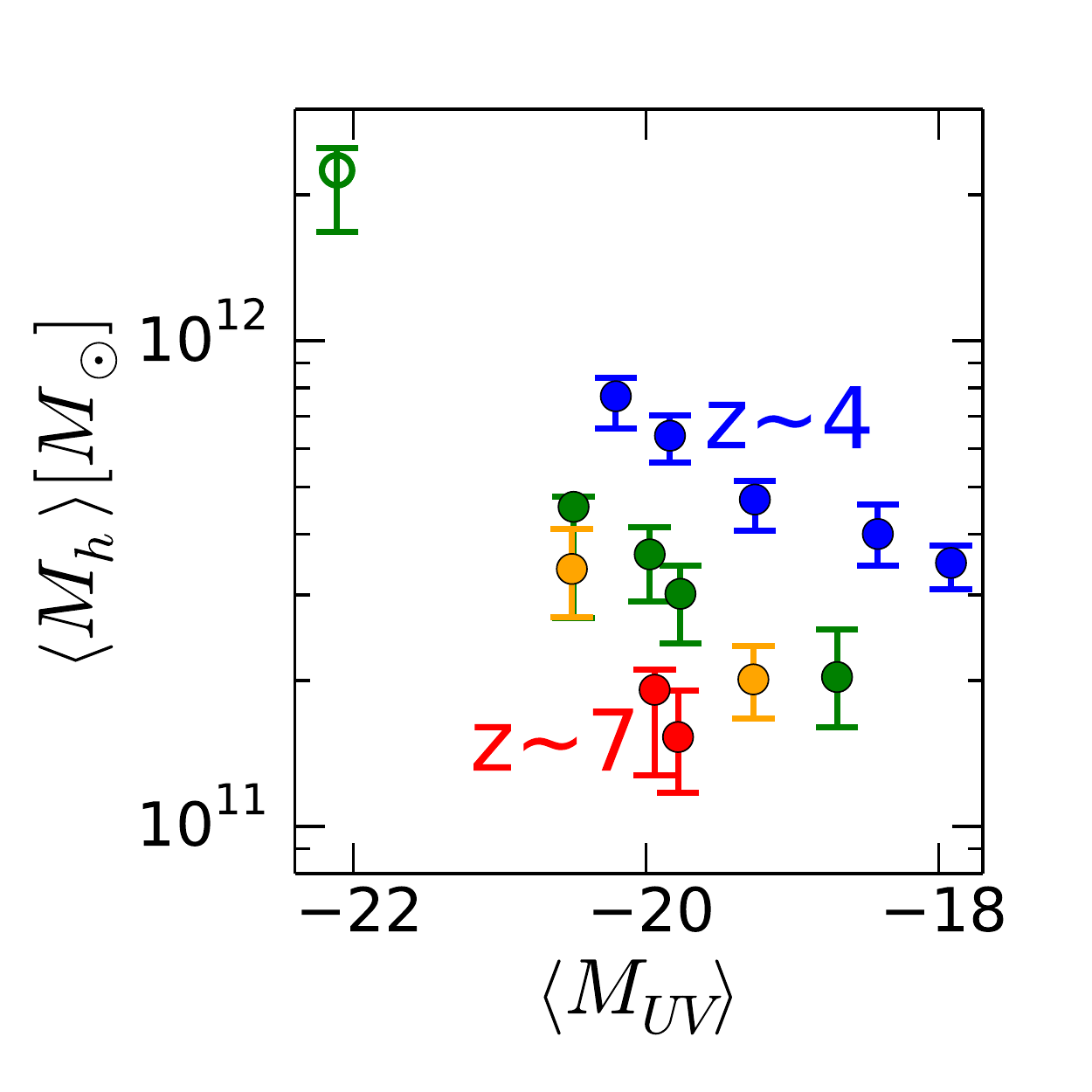}
 \end{center}
    \caption{Mean dark matter halo mass as a function of mean absolute UV magnitude.
The blue, 
    green, orange, and red circles represent the mean dark matter halo mass of our 
Hubble
    subsample 
at
    $z\sim4$, $5$, $6$, and $7$, respectively.
    The open green circle denotes the mean dark matter halo mass of our subsample constructed from the HSC data.
\label{fig_MUV_Mdh}}
\end{figure}

\begin{figure*}
\begin{center}
 \begin{minipage}{0.32\hsize}
 \begin{center}
  \includegraphics[clip,bb=10 16 332 330,width=1\hsize]{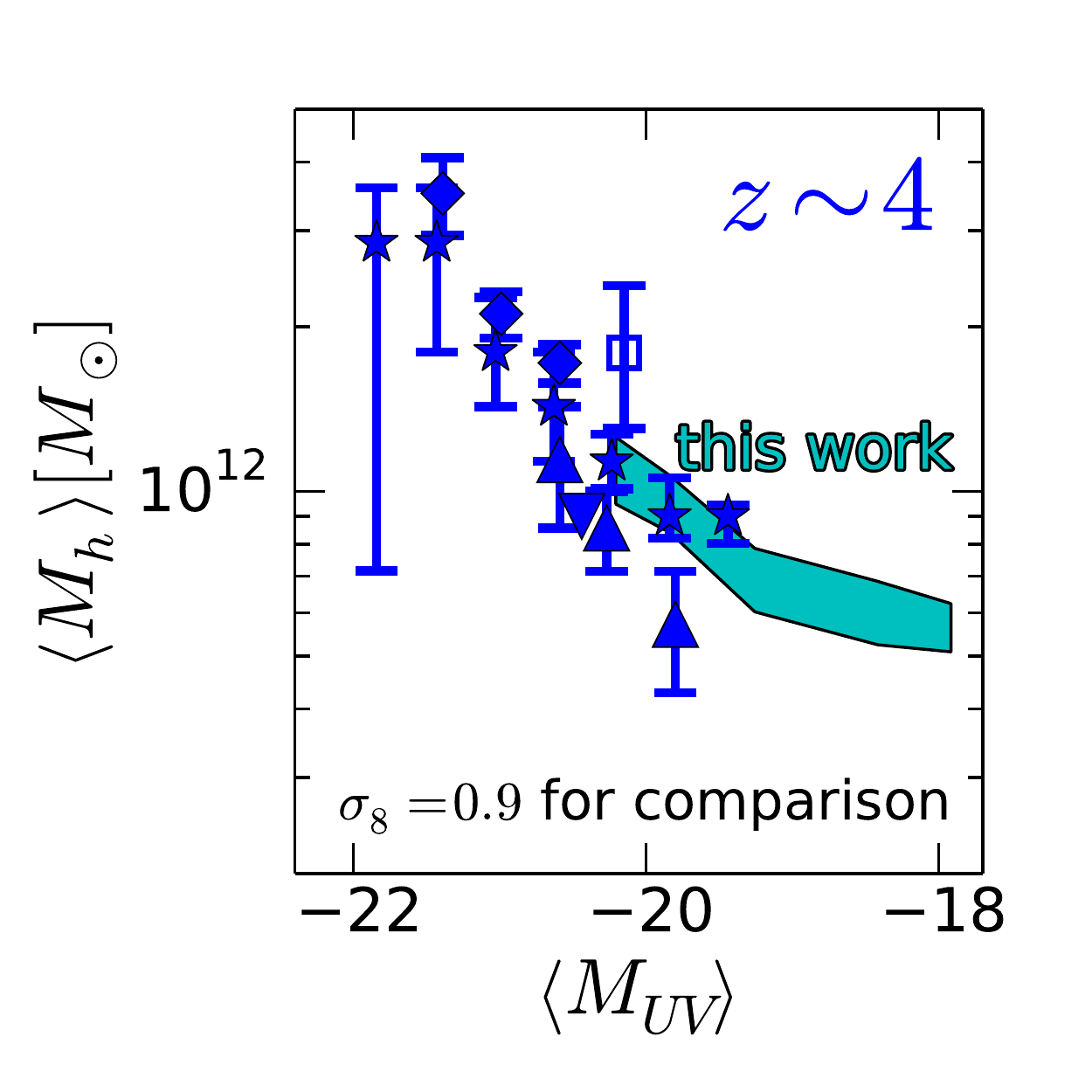}
 \end{center}
 \end{minipage}
  \begin{minipage}{0.32\hsize}
 \begin{center}
  \includegraphics[clip,bb=10 16 330 330,width=1\hsize]{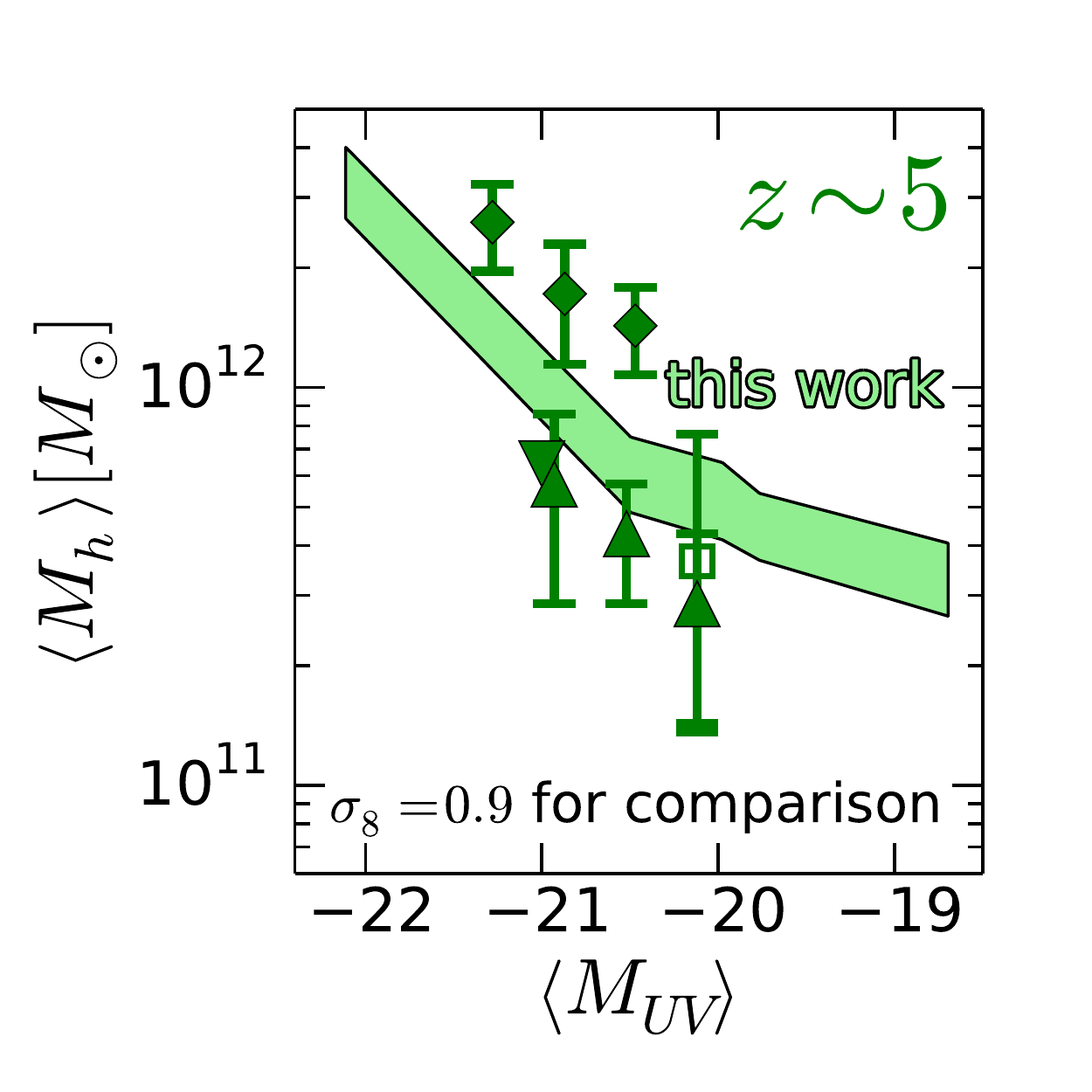}
  \end{center}
 \end{minipage}
  \begin{minipage}{0.32\hsize}
 \begin{center}
  \includegraphics[clip,bb=10 16 345 330,width=1\hsize]{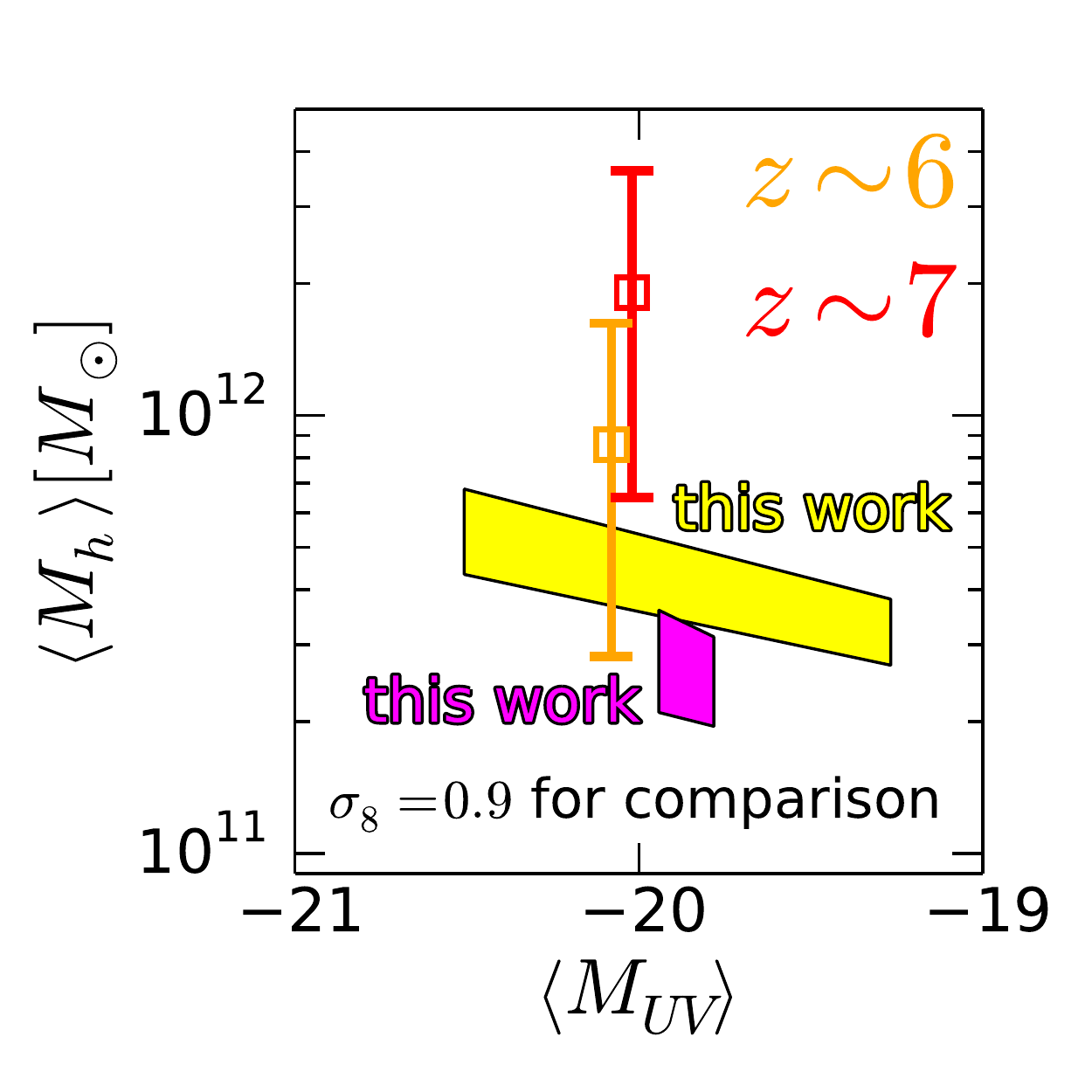}
 \end{center}
 \end{minipage}
 \end{center}
    \caption{Comparison with previous clustering studies under the same cosmology, $\sigma_8=0.9$.
    Our results are recalculated with $\sigma_8=0.9$.
    Left panel: comparison at $z\sim4$.
The cyan
shaded region represents the mean dark matter halo mass of our subsample 
at
$z\sim4$.
The blue symbols 
represent the results of the previous studies.
We plot the results of 
\citet[][downward triangle]{2004MNRAS.347..813H}, 
\citet[][stars]{2005ApJ...635L.117O}, \citet[][upward triangles]{2006ApJ...642...63L}, and \citet[][diamonds]{2009A&A...498..725H}. 
The downward triangle has
 no error bar, because \citet{2004MNRAS.347..813H} do not provide errors of the mean dark matter halo mass.
We also show the results of \citet{2014ApJ...793...17B} as a blue open square, 
who
use the simple power law model.
We compile the results of \citet{2014ApJ...793...17B} in the cosmology of $\Omega_\m{m}=0.3$, $\Omega_\Lambda=0.7$, $H_0=70\ \m{\kms Mpc^{-1}}$, and $\sigma_8=0.9$.
Center panel: comparison at $z\sim5$.
The light
green shaded region represents the mean dark matter halo mass of our subsample 
at
$z\sim5$.
The green
symbols represent the results of the previous studies of 
\citet[][downward triangle]{2004MNRAS.347..813H}, 
\citet[][upward triangles]{2006ApJ...642...63L}, \citet[][diamonds]{2009A&A...498..725H}, and 
\citet[][open square]{2014ApJ...793...17B}.
Right panel: comparison at $z\sim6,7$.
The yellow
and magenta shaded regions represent the mean dark matter halo masses of our subsample 
at
$z\sim6$, and $7$, respectively.
The orange
and red open squares represent the results of \citet{2014ApJ...793...17B} at $z\sim6$, and $7$, respectively.
    \label{fig_MUV_Mdh_09}}
    \end{figure*}

Figure \ref{fig_MUV_Mdh} shows $\left<M_\m{h}\right>$
as a function of mean absolute UV magnitude, $\left<M_\m{UV}\right>$ that is the mean of the absolute UV magnitude of the LBG subsample.
The dark matter halo masses from the Hubble data fall in the range of $\left<M_\m{h}\right>\sim(1-8)\times10^{11}\ \Msun$, 
while the one from the HSC data is at the massive regime of $\left<M_\m{h}\right>\sim2\times10^{12}\ \Msun$.
There is a trend of increasing the dark matter halo mass with increasing the UV luminosity at all redshifts.
Our results suggest that more UV luminous LBGs reside in more massive dark matter halos,
and agree with the conclusions of previous high-$z$ galaxy studies 
\redc{\citep{2001ApJ...558L..83O,2004ApJ...611..685O,2005ApJ...635L.117O,2003A&A...409..835F,2005ApJ...619..697A,2006ApJ...642...63L,2014ApJ...793...17B,2013ApJ...774...28B}}.

\subsection{Comparison with Previous Clustering Studies}
Figure \ref{fig_MUV_Mdh_09} compares our results with previous clustering studies 
\citep{2004MNRAS.347..813H,2005ApJ...635L.117O,2006ApJ...642...63L,2009A&A...498..725H,2014ApJ...793...17B}.
Because most of the previous studies assume $\sigma_8=0.9$ that is different from our assumption ($\sigma_8=0.8$),
we obtain our HOD model fitting results for our data with $\sigma_8=0.9$ for comparison.
Similarly, the results of the previous studies are re-calculated with the cosmological parameter sets
with $\Omega_\m{m}=0.3$, $\Omega_\Lambda=0.7$, $H_0=70\ \m{\kms Mpc^{-1}}$, and $\sigma_8=0.9$.
\redc{In this way, we conduct our comparisons using an equivalent set of cosmological parameters across all data sets.}
In Figure \ref{fig_MUV_Mdh_09}, we find that our $z\sim 4$ results are consistent with 
those of the previous studies within the uncertainties \citep[see also][]{2015arXiv151101983P}. While the previous results at $z\sim 5$ 
are largely scattered, our $z\sim 5$ results are placed near the center of
the distribution of the previous studies. At $z\sim 6$, 
our result agrees with that of \citet{2014ApJ...793...17B}.
Over the full redshift range ($z\sim4-6$) considered here, we confirm that our results are consistent with those of the previous studies.
However, there is a $1-2\sigma$ difference between
our results and those in \citet{2014ApJ...793...17B} at $z\sim 7$. This difference may be
simply explained by the measurement uncertainties or 
the difference of the LBG sample selections.
As shown in Figure \ref{fig_MUV_Mdh_09}, we estimate the dark matter halo mass at $z\sim6-7$ for the first time by the clustering analysis with the HOD model \cite[c.f.,][by no HOD modeling]{2014ApJ...793...17B}.
\redc{We also compare our results with the dark matter halo mass of $z\sim3$ LBGs.
\citet{2009A&A...498..725H} estimate the mean dark matter halo mass of the $z\sim3$ LBGs with $M_\m{UV}<-20.0$ to be $(1.6\pm0.6)\times10^{12}\ \Msun$, which is comparable to our result of $\left<M_\m{UV}\right>\sim-20$ at $z\sim4$ \citep[Figure \ref{fig_MUV_Mdh_09}; see also][]{2005ApJ...619..697A,2006ApJ...642...63L,2007A&A...462..865H,2013ApJ...774...28B}.
}

\section{SHMR}\label{ss_result_SHMR}
\subsection{Stellar Mass Estimates}
We estimate stellar masses $M_*$ of our LBGs with the \citet{2003PASP..115..763C} IMF 
from UV magnitudes $M_\m{UV}$, exploiting the star-formation main sequence, a tight correlation
between $M_*$ and star-formation rate (SFR) found at high-$z$ \citep[e.g.,][]{2007ApJ...670..156D,2014ApJS..214...15S}.
Because LBGs at $z\gtrsim 4$ are generally very dust poor,
SFR well correlates with $M_\m{UV}$ \citep[see also][]{2014ApJ...791L..25S}. Thus, $z\gtrsim 4$ LBGs have 
a correlation between $M_\m{UV}$ and $M_*$.
We use the $M_\m{UV}-M_*$ relations in \citet{2015arXiv150307481S}.
These are the empirical $M_\m{UV}-M_*$ relations at $z=0-6$ 
with photo-$z$ galaxies in the 3D-HST catalog in \citet{2014ApJS..214...24S}, who carry out the SED fitting for the stellar mass with the \citet{1955ApJ...121..161S} IMF.
In order to convert our stellar mass estimates of \citet{1955ApJ...121..161S} to \citet{2003PASP..115..763C} IMFs, our final estimates are divided by a factor of $1.8$.
We thus estimate $M_*$ from $M_\m{UV}$ with the relations,
\begin{eqnarray}
\m{log}M_*&=&-1.16-0.54\times M_\m{UV}\ (z\sim4) \label{eq_MUVMs_z4}\\
\m{log}M_*&=&-2.28-0.59\times M_\m{UV}\ (z\sim5)\label{eq_MUVMs_z5}\\
\m{log}M_*&=&-2.45-0.59\times M_\m{UV}\ (z\sim6,7).\label{eq_MUVMs_z67}
\end{eqnarray}
Because \citet{2015arXiv150307481S} show the $M_\m{UV}-M_*$ relations in the redshift ranges of 
$z=3-4$, $4-5$, and $5-6$, we interpolate the relations in \citet{2015arXiv150307481S} by redshift 
to derive Equations (\ref{eq_MUVMs_z4}) and (\ref{eq_MUVMs_z5}) for $z\sim 4$ and $5$ LBGs, respectively.
For $z\sim6$ and $7$ LBGs, we use the $M_\m{UV}-M_*$ relation of $z\sim5-6$ given in 
\citet{2015arXiv150307481S}.
If we extrapolate the $M_\m{UV}-M_*$ relation to $z\sim6$ and $7$, the derived stellar mass at $M_\m{UV}=-20$ 
becomes larger by $0.1\ \m{dex}$ at $z\sim6$ and by $0.3\ \m{dex}$ at $z\sim7$,
which do not change our conclusions (Section \ref{ss_SHMR_evol}). 
We also confirm that our conclusions do not change, if we use the $M_\m{UV}-M_*$ relations of \citet{2015arXiv150705636S}.
Note that the error on the mean relation between $M_*$ and $M_\m{UV}$ is $<0.01\ \m{dex}$ in the stellar mass while the dispersion is $0.5\ \m{dex}$ \citep{2015arXiv150307481S}.
In Table \ref{table_PL}, we present estimates of the stellar masses $M_{*,\m{th}}$ that correspond
to the threshold UV magnitudes, $M_\m{UV, th}$, of the subsamples.

\subsection{SHMRs and the Evolution}\label{ss_SHMR_evol}
\begin{figure}
\begin{center}
  \includegraphics[clip,bb=10 16 330 270,width=1\hsize]{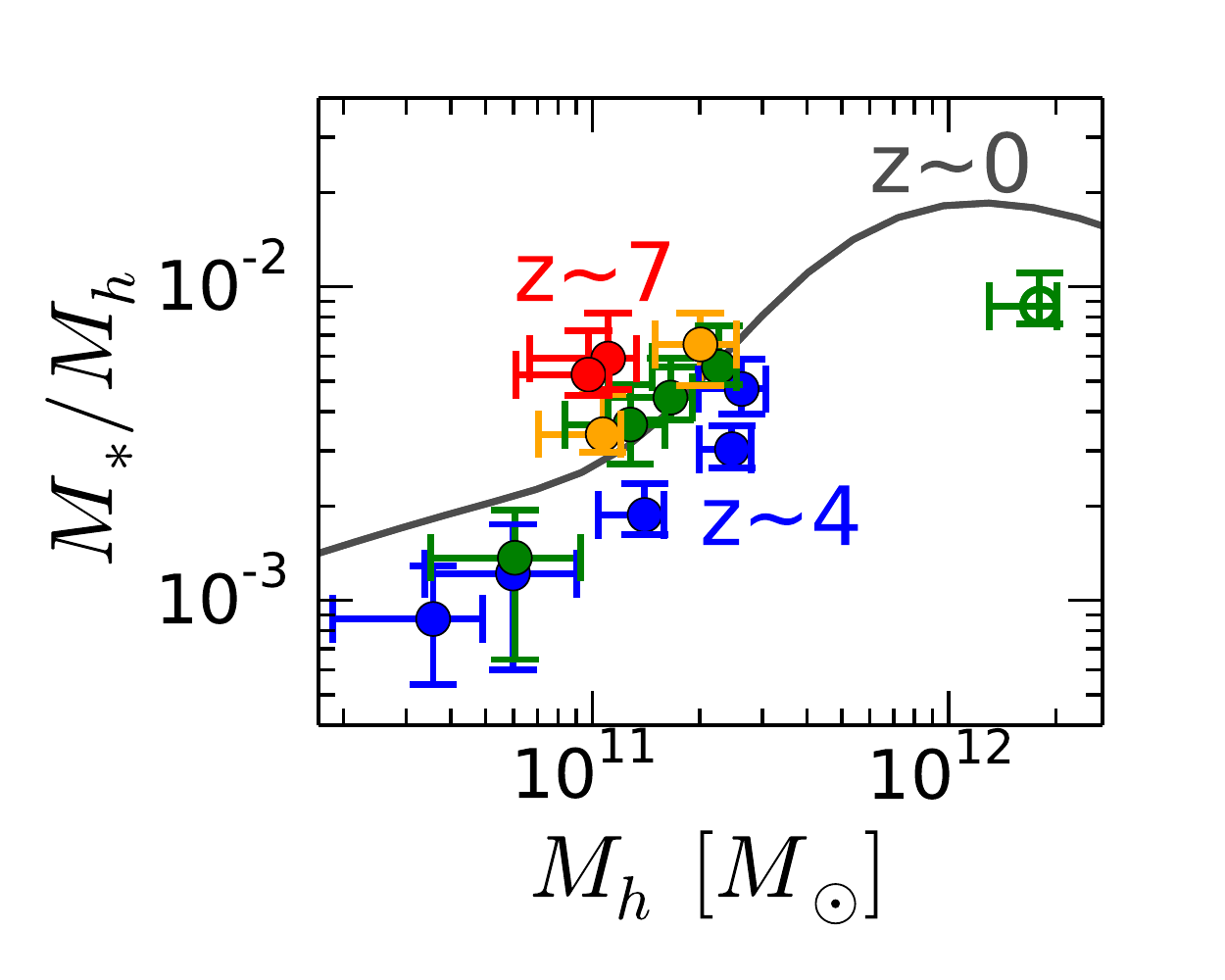}
 \end{center}
  \caption{SHMR of central galaxies as a function of 
dark
  matter halo mass at $z\sim4$, $5$, $6$, and $7$.
The blue, 
    green, orange, and red circles represent the SHMR of our Hubble subsample 
at
    $z\sim4$, $5$, $6$, and $7$, respectively.
    The open green circle denotes the SHMR of our subsample constructed from the HSC data.
The gray
    solid curve is the SHMR of \citet{2013ApJ...770...57B} at $z\sim0$, which is computed by P. Behroozi with the cosmological parameters and halo mass definition same as our analysis.
  \label{fig_SHMR_result}}
\end{figure}

Figure \ref{fig_SHMR_result} presents SHMRs of central galaxies 
for our LBG subsamples at $z\sim4$, $5$, $6$, and $7$. 
Hereafter, we use $M_{*,\m{th}}$ and $M_\m{min}$ values
for stellar masses $M_*$ and halo masses $M_{\rm h}$, respectively, 
because these quantities define our subsamples in a self-consistent manner.
The black curve in Figure \ref{fig_SHMR_result} represents the SHMR function at $z\sim 0$ 
that is the same as the one of \citet{2013ApJ...770...57B}, but for
our cosmological parameters and stellar-mass estimate assumptions
(P. Behroozi, private communication).

\begin{figure*}
\begin{center}
 \begin{minipage}{0.48\hsize}
 \begin{center}
  \includegraphics[clip,bb=10 10 330 270,width=1\hsize]{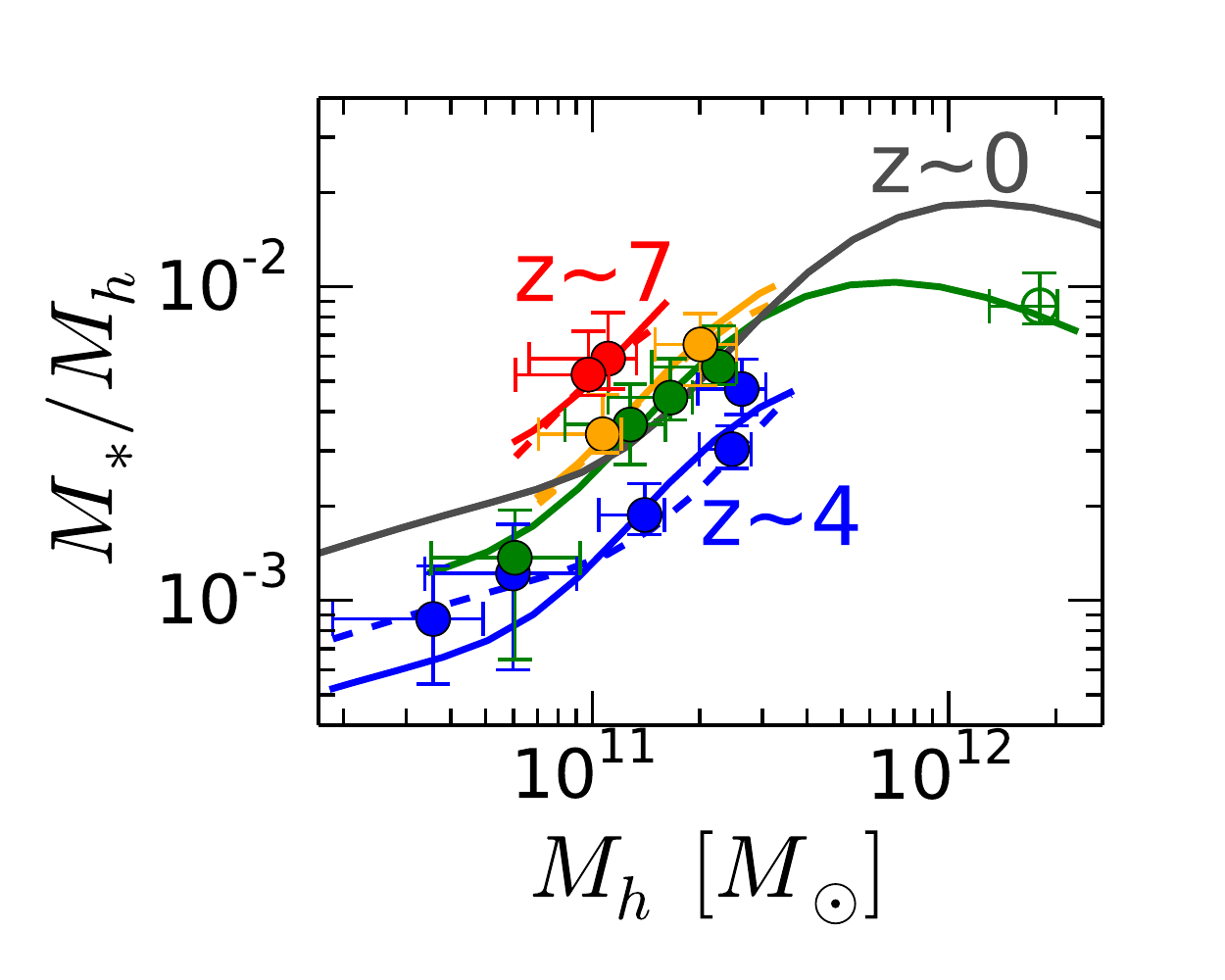}
 \end{center}
 \end{minipage}
  \begin{minipage}{0.48\hsize}
 \begin{center}
  \includegraphics[clip,bb=10 10 330 270,width=1\hsize]{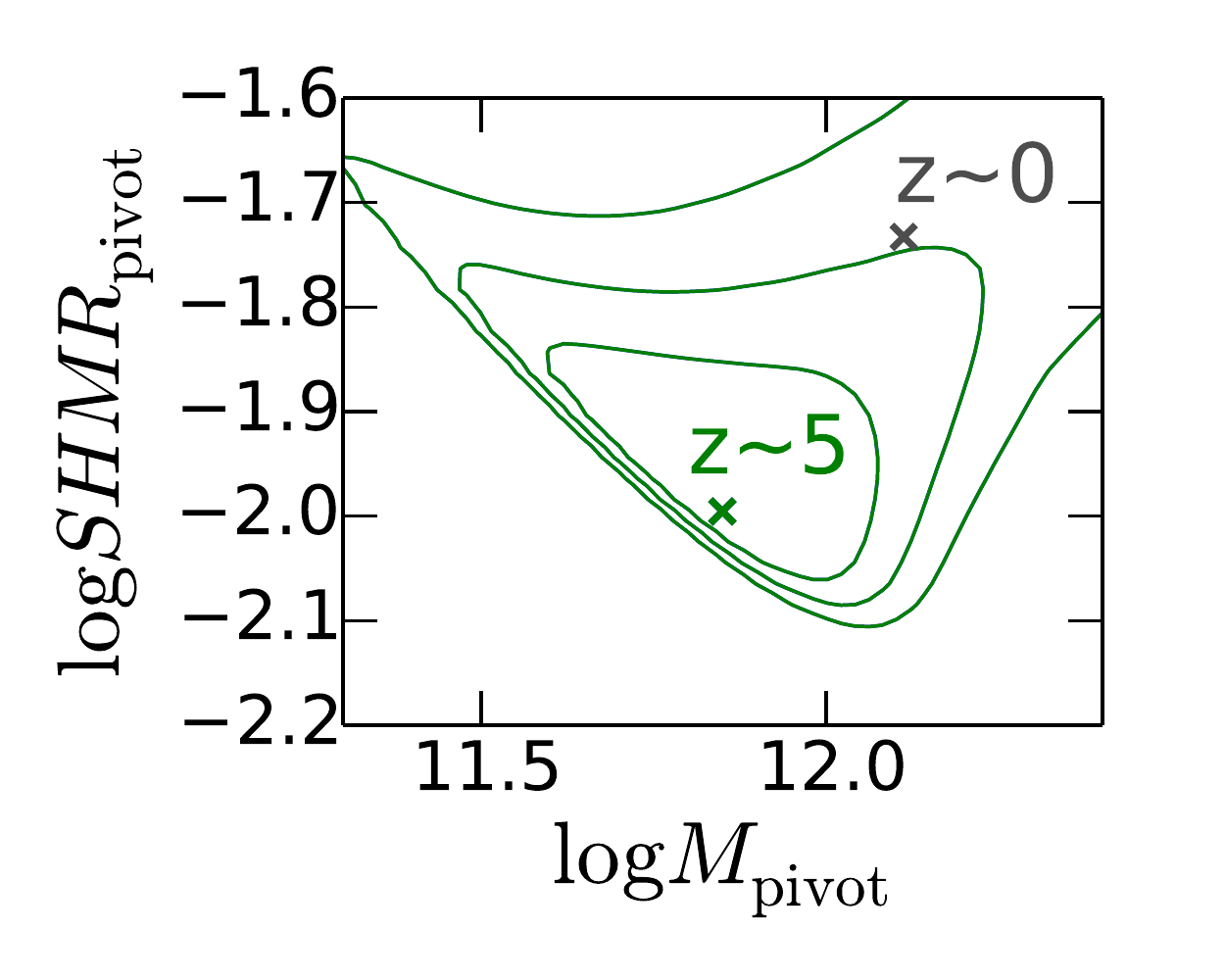}
 \end{center}
 \end{minipage}
\\
 \end{center}
   \caption{SHMR evolution.
The left panel shows the results of our SHMR function fittings.
The green
   curve represents the best-fit SHMR function 
at
   $z\sim5$.
    We fit 
SHMR functions to our $z \sim 5$ SHMR--$M_{\rm h}$ data (green circles)
    by parametrizing the $z\sim0$ SHMR function of \citet{2013ApJ...770...57B} 
    with a pivot mass, $M_\m{pivot}$, and an SHMR amplitude at the pivot mass, $SHMR_\m{pivot}$.
The blue, 
    orange, and red solid (dashed) curves describe the best-fit SHMR 
functions
    of $z\sim4$, $6$, and $7$, respectively, in the $M_\m{pivot}$-fixed
    ($SHMR_\m{pivot}$-fixed) case.
These curves are shown only in the range where measurements are available.
    The details of the fitting are presented in Section \ref{ss_SHMR_evol}.
In the right panel, the green contours represent the $1.0$, $1.5$ and $2.0\sigma$ confidence levels 
    of $M_\m{pivot}$ and $SHMR_\m{pivot}$ at $z\sim5$.
     The green cross in the contours 
corresponds to
     the best-fit values 
at
     $z\sim5$.
The gray
     cross shows the values of $M_\m{pivot}$ and $SHMR_\m{pivot}$ at $z\sim0$ \citep{2013ApJ...770...57B}.
      \label{fig_SHMR_fit}}
\end{figure*}

In Figure \ref{fig_SHMR_result}, the combination of the Hubble and HSC data covers the wide halo mass range
of $6\times 10^{10}-2\times 10^{12} M_\odot$ at $z\sim 5$. The SHMR increases from $\sim10^{-3}$ to $\sim10^{-2}$, 
with increasing $M_\m{h}$ from $\sim6\times10^{10}$ to $\sim 2\times10^{12}\ M_\odot$ at $z\sim 5$.
Similar positive correlations are found in the SHMR-$M_{\rm h}$ relations at $z\sim 4$ and $6$
as well as $z\sim 0$. 
The SHMRs at $z\sim7$ are consistent with the positive correlation.
At low redshift ($z\lesssim 1$), this correlation is claimed in various studies 
\citep[e.g.,][]{2012ApJ...744..159L,2015MNRAS.449.1352C}.
Our results newly show the positive correlations of SHMR-$M_{\rm h}$ near $M_{\rm h} \sim 10^{11} M_\odot$ 
at high redshift, $z\sim4-6$.

At $z=0-7$, the SHMR values at $M_{\rm h}\sim 10^{11}$ are obtained, 
which allow us to investigate evolution of SHMRs. From $z\sim 0$ to $z\sim4$, 
the SHMR decreases by a factor of \redc{$\sim 2$,
from $\sim2.7\times10^{-3}$ to $\sim1.3\times10^{-3}$}.
In contrast, 
the SHMR increases by a factor of \redc{$\sim 4$, 
from $\sim1.3\times10^{-3}$ at $z\sim 4$ to $\sim5.3\times10^{-3}$ at $z\sim 7$.}
To quantify the evolution of the SHMR-$M_{\rm h}$ relation,
we parameterize the $z\sim 0$ SHMR function of \citet{2013ApJ...770...57B} with a pivot halo mass $M_\m{pivot}$ and 
an SHMR amplitude at the pivot halo mass, $SHMR_\m{pivot}$,
\begin{align}
&\m{log}(SHMR-SHMR_\m{pivot})\notag\\
=&\m{log}\left[\frac{M_{*}(M_\m{h}-M_\m{pivot})}{M_\m{h}-M_\m{pivot}}\right]\notag\\
=&\m{log}\left(\frac{{\epsilon}M_1}{M_\m{h}-M_\m{pivot}}\right)+f\left[\m{log}\left(\frac{M_\m{h}-M_\m{pivot}}{M_1}\right)\right]-f(0),
\end{align}
\begin{equation}
f(x)=-\m{log}(10^{{\alpha}x}+1)+\delta\frac{\left\{\m{log}\left[1+\m{exp}(x)\right]\right\}^\gamma}{1+\m{exp}\left(10^{-x}\right)},
\end{equation}
where $\m{log}\epsilon=-1.85$, $\m{log}M_1=11.50$, $\alpha=-1.39$, $\delta=3.76$, and $\gamma=0.33$ at $z\sim0$ (P. Behroozi in private communication).
We fit this parameterized SHMR function to 
our SHMR-$M_{\rm h}$ data of the $z\sim 4-7$ LBGs.

Removing the dependent results of our $m_{\rm UV,th}$
subsamples whose bright LBGs 
are repeatedly included in the subsamples, except for the HSC and some HUDF data, we only use 
the independent SHMR data in our fitting.

We use the subsamples of $m^\m{aper}_\m{UV}<27.6\ \m{mag}$ and $m^\m{aper}_\m{UV}<29.8\ \m{mag}$
($m^\m{aper}_\m{UV}<25.0\ \m{mag}$, $m^\m{aper}_\m{UV}<28.0\ \m{mag}$, and $m^\m{aper}_\m{UV}<29.2\ \m{mag}$)
for $z\sim 4$ ($z\sim 5$). 
Similarly, the $m^\m{aper}_\m{UV}<28.4$ subsamples are fitted for $z\sim 6$, $7$.

Because our $z\sim 5$ SHMR estimates are obtained 
in the wide halo mass range 
that allows us to investigate the SHMR and $M_{\rm h}$ evolution
simultaneously, we perform fitting to the $z\sim 5$ SHMR estimates 
with the SHMR function varying $M_\m{pivot}$ and $SHMR_\m{pivot}$.
The best-fit function and the error contours are presented in the left and right panels of Figure \ref{fig_SHMR_fit},
respectively. We compare these results with those at $z\sim 0$ obtained by \citet{2013ApJ...770...57B}.
The left panel of Figure \ref{fig_SHMR_fit} indicates that the SHMRs of $z\sim 0$ and $5$
are similar at $M_{\rm h}\sim 10^{11} M_\odot$, but different at $M_{\rm h}\sim 10^{12} M_\odot$.
The massive end of our data makes a difference in the fitting result shown in the right panel of 
Figure \ref{fig_SHMR_fit}.

Although the mass ranges of our SHMR data are limited,
the SHMR results of $z\sim 4$ and $7$ show large differences from those of $z\sim 0$
at $M_{\rm h}\sim 10^{11} M_\odot$. We quantify the differences
by two extreme scenarios of $M_\m{pivot}$-fixed and
$SHMR_\m{pivot}$-fixed cases that bracket the realistic
scenario including both $M_\m{pivot}$ and $SHMR_\m{pivot}$ evolutions.
Adopting the best-fit $M_\m{pivot}$ or $SHMR_\m{pivot}$ value at $z\sim 5$,
we carry out the SHMR function fitting in these two cases.
The left panel of Figure \ref{fig_SHMR_fit} presents the best-fit
SHMR functions for $M_\m{pivot}$-fixed and 
$SHMR_\m{pivot}$-fixed cases with the solid and dashed lines,
respectively. These two cases show very similar best-fit SHMR functions
in the left panel of Figure \ref{fig_SHMR_fit}, because the $M_{\rm h}$ ranges
for the fitting are narrow and limited to $M_{\rm h}\sim 10^{11} M_\odot$.
Moreover, the notable differences between $z\sim 0$, $4$, and $7$ curves are identified,
suggesting the evolution of SHMR and/or $M_{\rm h}$ from $z\sim 0$ to $z\sim4$
($z\sim 0-4$) and $z\sim 4$ to $z\sim7$ ($z\sim 4-7$).
In the $M_\m{pivot}$-fixed ($SHMR_\m{pivot}$-fixed) case,
the differences of $z\sim 0-4$ and $z\sim4-7$ are found at the $\redc{5.6}$ and $3.1\sigma$ ($\redc{3.3}$ and $2.5\sigma$)
levels, respectively.
\redc{In the calculations of these significance levels, we use the statistical error presented in \citet{2010ApJ...717..379B} as the error of the $z\sim0$ SHMR, because \citet{2013ApJ...770...57B} do not provide the statistical errors. 
\citet{2010ApJ...717..379B} use the similar dataset to that of \citet{2013ApJ...770...57B}}.
We also investigate $M_\m{*,pivot}$-fixed case with varying $M_\m{pivot}$, where $M_\m{*,pivot}$ is the pivot {\it stellar} mass
that is not the pivot {\it halo} mass of $M_\m{pivot}$. We find the differences at the redshift ranges of
$z\sim0-4$ and $z\sim4-7$ are $\redc{4.8}$ and $2.7\sigma$ significance levels, respectively.
In addition, we adopt the best-fit $M_\m{pivot}$ or $SHMR_\m{pivot}$ value of $z\sim 0$, instead of $z\sim 5$,
and confirm that the arguments above are unchanged.
In any cases of these scenarios, we find the SHMR evolutions at the redshift ranges of $z\sim 0-4$ and $z\sim4-7$
at the $>99$\% and $>98$\% confidence levels, respectively.
These SHMR evolutionary trends 
at $z\sim 0-4$ and $z\sim4-7$
are identified, for the first time, based on the clustering analyses.

We examine whether
these results are produced by systematic biases in
our HOD model fitting, where we fixed some parameters and the analytic relations.
Firstly, we have assumed 
the fixed parameter of $\sigmalogM=0.2$
over $z=4-7$ in Section \ref{ss_result_mass}, although it is known that $\sigmalogM$ could vary with the redshift and the halo mass.
According to the formulation of \citet{2013ApJ...770...57B}, $\sigmalogM$ values of $z\sim 4$ and $7$ galaxies of $M_\m{h}\sim10^{11}\ \Msun$
are $0.3$ and $0.5$, respectively.
Adopting $\sigmalogM=0.3$ for our $z\sim 4$ LBGs, we find negligible differences from 
the original $\sigmalogM=0.2$ results in the SHMR evolution from $z\sim 0$ to $4$.
We also estimate SHMR and $M_{\rm h}$ values with $\sigmalogM=0.5$ for our $z\sim 7$ LBGs.
Although the estimated $M_{\rm h}$ values are larger than those of the original $\sigmalogM=0.2$ results by a factor of 1.5,
the SHMR evolution from $z\sim 4$ to $7$ is still found 
at the $\sim2\sigma$ significance level.
Secondly, we have adopted the analytic relations of Equations (\ref{eq_M0_M1p}) and (\ref{eq_M1p_Mmin})
to derive $M_0$ and $M^\prime_1$ in Section \ref{ss_result_mass}.
Here we fit our ACFs with varying $M_0$ and $M^\prime_1$ as free parameters
in the ranges of $9<\m{log}M_1^\prime<14$ and $8<\m{log}M_0<14$, respectively,
to evaluate the impacts of the $M_0$ and $M^\prime_1$ values on our results.
For all subsamples that we use in the SHMR evolution discussion,
we find that the new $M_\m{min}$ values from these fitting analyses
agree with our best-estimate values (Table \ref{table_HOD}) within
the uncertainties.
For example, the subsample of $z\sim4$ $m_{\rm UV}<27.6\ \m{mag}$
gives the new $M_\m{min}$ value of $\logMmin=11.4^{+0.1}_{-0.1}$
that is consistent with our best-estimate value of $\logMmin=11.4^{+0.1}_{-0.1}$ (Table \ref{table_HOD}).
Although some errors of the new $M_\m{min}$ value are larger than those of
the best-estimate value by a factor of $\sim1.5$,
the SHMR evolution
at $z\sim 0-4$
is still found at the $>3\sigma$ level
due to the scatter of the new $M_\m{min}$ value 
that separates the SHMRs of $z\sim 0$ and $z\sim4$.
The evolutions of the SHMR we find are more significant than the systematic biases.

\redc{We discuss systematic uncertainties in our SHMRs at $z\sim4-7$, by a comparison with that of \citet{2013ApJ...770...57B} at $z\sim0$. 
We derive the stellar masses assuming the \cite{2003PASP..115..763C} IMF and the \citet{2003MNRAS.344.1000B} stellar population synthesis models, which are also used in \citet{2013ApJ...770...57B}.
Although the dust attenuation model in our analysis \citep{2000ApJ...533..682C} is different from the one \citep{2007AJ....133..734B} adopted by \citet{2013ApJ...770...57B}, this difference only changes the stellar mass estimate by 0.02 dex \citep{2010ApJ...717..379B}. 
The halo mass function in our analysis is the same as the one \citet{2013ApJ...770...57B} use. 
The $z\sim0$ SHMRs shown in our paper are re-computed by P. Behroozi with the cosmological parameter set and halo mass definition same as ours. 
Thus, the evolution of the SHMR at $z\sim0-4$ is significant beyond these systematic uncertainties.
}

\section{Discussion}\label{ss_discussion}
\subsection{Interpretations of the SHMR evolution}
Figure 10 shows the redshift evolution of the SHMR at $z\sim 0-7$ (Section 6.2).
At the dark matter halo mass of $\sim10^{11}\ \m{\Msun}$, 
SHMRs decrease from $z\sim 7$ to $4$ by \redc{$0.6\ \m{dex}$,}
and increase from $z\sim 4$ to $0$ by \redc{$0.3\ \m{dex}$.}
These two evolutionary trends of $\sim10^{11}\ \m{\Msun}$ dark matter halos suggest that the stellar mass assembly is 
very (moderately) efficient at $z\gtrsim 7$ ($z\sim 0-4$), while it is inefficient at $z\sim 4-7$.
There are three possible physical origins to explain the increase and decrease of SHMRs.

The first is the evolution of the gas cooling efficiency.
The gas cooling efficiency is expected to be high if the number density or the chemical abundance of gas are high.
At high redshift, the average density of the collapsed dark matter halos is very high due to the high cosmic matter density.
Towards low redshift, the inter-stellar medium in galaxies becomes chemically enriched.
If one tries to explain the SHMR evolution at $M_\m{h}\sim10^{11}\ \Msun$ by the change of the gas cooling efficiency,
the high SHMR at $z\sim7$ would be reduced by the decreasing gas cooling efficiency due to the small number density at $z\sim4$.
In fact, Figure \ref{fig_MUV_Mdh} indicates that low-redshift galaxies tend to have a high $\left < M_{\rm UV} \right>$ (i.e. low SFR)
for a given $\left < M_{\rm h} \right >$.
From $z\sim4$ to $z\sim0$, the SHMR may increase due to the increase of gas cooling efficiency originated from the chemical enrichment.

The second is the evolution of feedback strengths. Because the mass of $\sim10^{11}\ \m{\Msun}$ falls
in a moderately low mass regime,
the SN and radiation pressure feedback are probably more important than AGN feedback \cite[e.g.,][]{2014PASJ...66...70O}.
Typical LBGs at $z\gtrsim 7$ are very compact, having a half-light radius of $\lesssim 1$ kpc
(e.g. Shibuya et al. 2015). The gravitational potential at the dark matter halo center dominated by
baryonic mass \citep{2006ApJ...646..107E} is deeper at $z>7$ than $z\sim4-7$. 
At $z\sim4-7$, due to the moderately shallow gravitational potential, 
hot/warm gas is expelled from the center to the outer halo, and it takes time for the gas
to come back to the central region by radiative cooling in the low-density regime of the outer halo.
Although these dense and compact galaxies would allow the high SHMR values,
the energy production rates corresponding to SFRs are also high (see Figure 8 and the
discussion above). Thus, it is unclear whether feedback is responsible for
the SHMR evolution at $z>4$.
Below $z\sim4$, the decrease of specific SFRs (sSFR; SFR divided by stellar mass) is accelerated
towards low-$z$ \citep{2014ApJ...795..104W}.
This fast decrease of sSFR as well as the increase of the galaxy half-light radius towards low-$z$
do not raise SHMRs at $z=0-4$. Thus, the feedback may not be major physical origins of the SHMR evolution of $z\sim0-4$.

The third is the evolution of merger rates and merger-induced star-formation activities.
In Section 6.2, we find that the SHMR positively correlates with $M_{\rm h}$ at $\sim10^{11}\ \m{\Msun}$
over $z\sim 0-7$ (Figure 9). 
The shape of the SHMR-$M_{\rm h}$ function at $z\sim 0$
suggests that these positive correlations at $z\sim 0-7$ continue to a halo mass range 
much lower than $M_{\rm h}$ at $\sim10^{11}\ \m{\Msun}$.
If a merger is not associated with the star-formation activity (i.e. dry merger), such a merger does not change the SHMR, but increase
the halo mass.
For example, if a galaxy with $\sim10^{11}\ \m{\Msun}$ forms by multiple mergers of 
$\sim10^{10}\ \m{\Msun}$ galaxies with no merger induced star-formation activity,
the SHMR of the $\sim10^{11}\ \m{\Msun}$ galaxy is as low as those of 
the $\sim10^{10}\ \m{\Msun}$ galaxies. Under the condition of the positive correlation
between SHMR and $M_{\rm h}$, this mechanism of mergers suppresses SHMRs.
On the other hand, intensive star-formation is followed by merger events.
Such a merger-induced star-formation would build up stellar masses, and
boost SHMRs. Thus, there are two effects of mergers: increasing and decreasing SHMRs.
It is known that the merger rates decrease from high-$z$ to low-$z$ in simulations \citep{2010MNRAS.406.2267F}. 
If one tries to explain the SHMR evolution at $\sim10^{11}\ \m{\Msun}$ by merger alone,
the high SHMR raised by merger-induced star formation may be reduced by the decrease of the merger rate.
At $z\sim 0-4$, the merger induced star-formation should increase SHMRs faster than the
merger suppression.
Here we have discussed the three possible physical origins of the SHMR evolution at $z\sim 0-7$. 
Although our observational results alone cannot determine what is the major physical origins,
it is likely that the mixture of these effects give the evolutionary trends of the SHMRs. 

\begin{figure}
 \begin{center}
  \includegraphics[clip,bb=10 10 330 270,width=1\hsize]{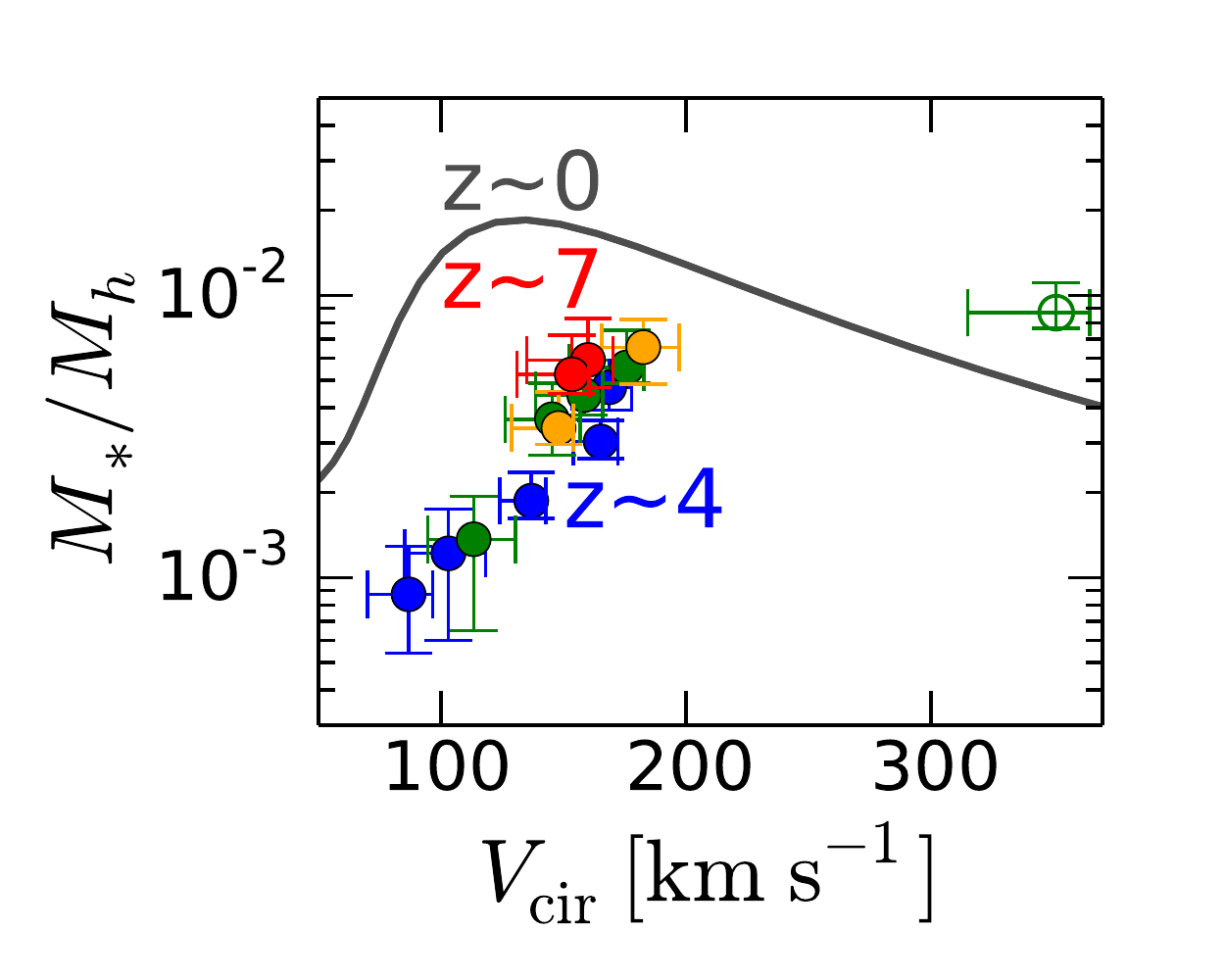}
 \end{center}
    \caption{SHMR as a function of circular velocity.
The blue, 
    green, orange, and red circles represent the SHMRs of our subsamples at $z\sim4$, $5$, $6$, and $7$, respectively.
The gray
    solid curve is the SHMR of \citet{2013ApJ...770...57B} at $z\sim0$.
     \label{fig_Vc_SHMR}}
\end{figure}

We discuss this SHMR evolution with
the circular velocity,
\begin{equation}
V_\m{cir}=\sqrt{\frac{GM_\m{h}}{r_{200}}}.
\end{equation}
Since the circular velocity is determined by
the halo mass and radius,
$V_\m{cir}$ provides the condition for gas
escaping from the halo by outflow.
Figure \ref{fig_Vc_SHMR} presents the SHMRs
as a function of circular velocity
for the $z\sim 4-7$ LBGs.
In Figure \ref{fig_Vc_SHMR}, the SHMRs of $z\sim7$ are higher than those
of $z\sim4$ by $0.3\ \m{dex}$ at $V_\m{cir}\sim160\ \m{\kms}$.
This increase is smaller than the $0.7\ \m{dex}$ increase of the SHMR
from $z\sim4$ to $7$ at $M_\m{h}\sim10^{11}\ \Msun$ (Figure \ref{fig_SHMR_fit}).
Because the circular velocity for a given halo mass increases by redshift,
the $0.7\ \m{dex}$ SHMR evolution in the SHMR-$M_{\rm h}$ plane (Figure \ref{fig_SHMR_fit})
is narrowed in the SHMR-$V_\m{cir}$ plane (Figure \ref{fig_Vc_SHMR}).
The small SHMR evolution in the SHMR-$V_\m{cir}$ plane
would indicate that the ($V_\m{cir}$-dependent) gas outflow conditions are similar
over the redshift range of $z\sim 4-7$.
In the SHMR-$V_\m{cir}$ plane (Figure \ref{fig_Vc_SHMR}), there is a significant SHMR evolution
of 0.5-1 dex at $z\sim 0-4$ in contrast with the small SHMR evolution at $z\sim 4-7$.
This significant SHMR evolution at $z\sim 0-4$ suggests that the early galaxies
at $z\sim 4-7$ have the gas outflow conditions clearly different from those of
matured galaxies at $z\sim 0$.

\begin{figure}
 \begin{center}
  \includegraphics[clip,bb=10 30 330 270,width=1\hsize]{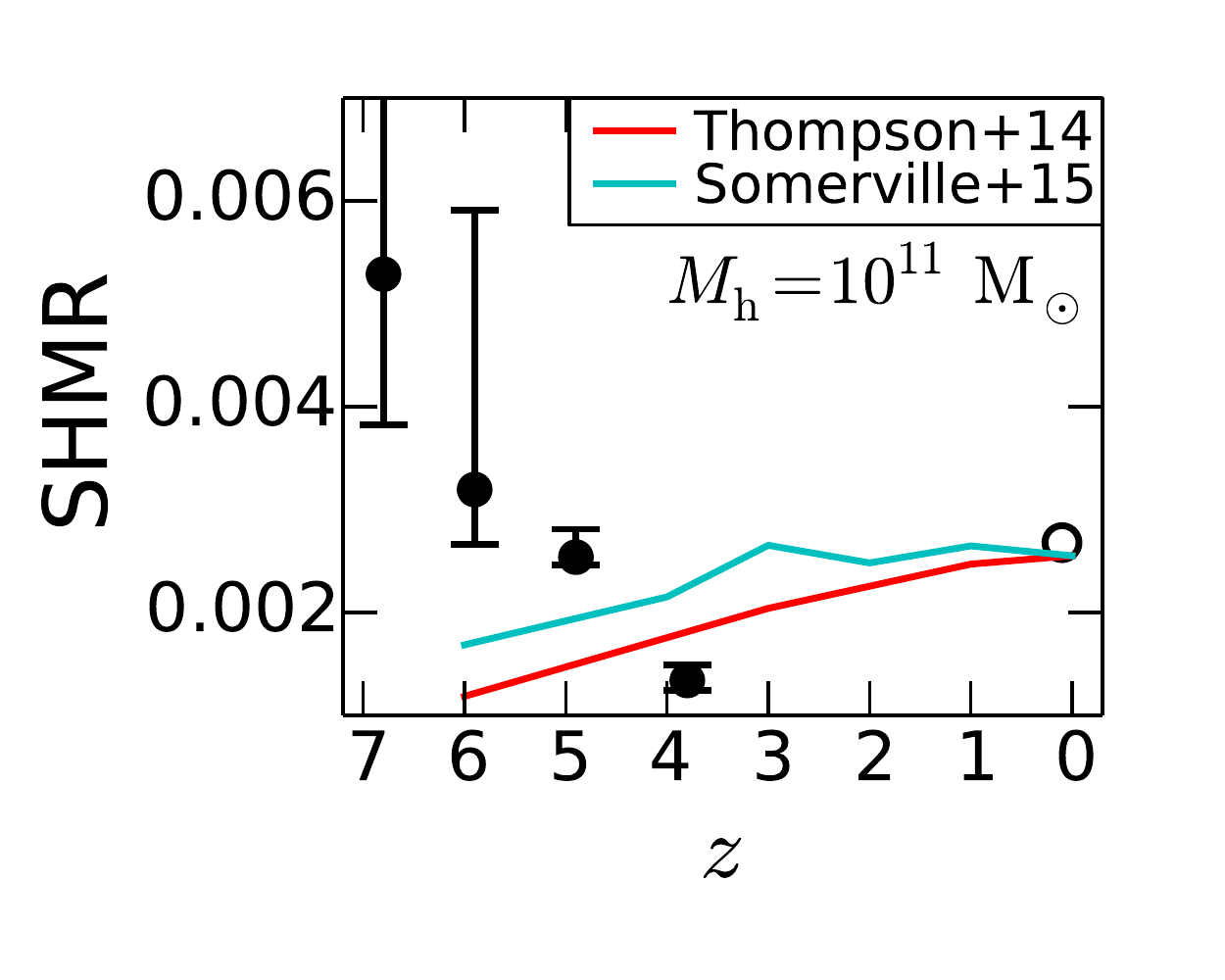}
 \end{center}
    \caption{\redc{SHMRs predicted by theoretical studies.
   The red and cyan lines are SHMRs at $M_\m{h}=10^{11}\ \Msun$ predicted by \citet{2014ApJ...780..145T} and \citet{2015MNRAS.453.4337S}, respectively.
   The black filled and open circles are the SHMRs at $M_\m{h}=10^{11}\ \Msun$ obtained in this study and in \citet{2013ApJ...770...57B}, respectively.
   To compare the trend with the obtained SHMR evolution, the amplitudes of the red and cyan lines are normalized to the $z\sim0$ SHMR.}
     \label{fig_theo_SHMR}}
\end{figure}

\redc{We compare our SHMRs with results of the theoretical studies, and investigate whether the theoretical models explain the SHMR evolution at $z\sim0-7$. 
In Figure \ref{fig_theo_SHMR}, we plot the SHMRs at $M_\m{h}=10^{11}\ \Msun$ predicted by the hydrodynamic simulation \citep{2014ApJ...780..145T} and the semi-analytic model \citep{2015MNRAS.453.4337S}.
These theoretical studies predict evolutionary trends of the SHMR decrease from $z\sim0$ to $4$ that are similar to our observational results. 
On the other hand, the theoretical studies can not reproduce the SHMR increase from $z\sim4$ to $7$ found in our observational study. 
This discrepancy may pose a challenge in the current theoretical study of galaxy formation.
}

\subsection{Baryon Conversion Efficiency}
In Section 6.2, we find that SHMR and $M_{\rm h}$ at $z\sim 0-7$ have the positive correlations 
in the mass range around $M_{\rm h}\sim 10^{11}\ \m{\Msun}$ (Figure 9).
To understand more details of these positive correlations,
we calculate the baryon conversion efficiency (BCE) of the $z\sim4$ subsamples
that have high statistical accuracies.
BCE is the ratio of the SFR to the baryon accretion rate, $\dot{M}_\m{b}$:
\begin{equation}
BCE=\frac{SFR}{\dot{M}_\m{b}}.
\label{eq:bce}
\end{equation}
Because most of the accreting baryon have a form of gas \citep[e.g.,][]{2015arXiv150502159S}, we adopt
$\dot{M}_\m{g}\simeq\dot{M}_\m{b}$. Thus, Equation (\ref{eq:bce}) can be 
written as follows, $BCE\simeq SFR/\dot{M}_\m{g}$, indicating
that BCE is the conversion rate from gas to stars.

We define the cosmic baryon fraction, $f_b\equiv \Omega_\m{b}/\Omega_\m{m}=0.15$.
The baryon accretion rate is computed with $f_b$ by
\begin{equation}
\dot{M}_\m{b}=f_\m{b}\times\dot{M}_\m{h},
\end{equation}
where $\dot{M}_\m{h}$ is the median halo mass accretion rate that is
a function of halo mass and redshift. 
We estimate $\dot{M}_\m{h}$ with the analytic formula obtained
from the N-body simulation results \citep{2013ApJ...770...57B}.
\redc{The SFRs are derived from the threshold total absolute magnitude in the rest-frame UV band, $M_{\m{UV,th}}$. 
We correct for the dust extinction with two empirical relations. 
One is the attenuation-UV slope, $\beta_\m{UV}$, relation \citep{1999ApJ...521...64M}, and the other is the $\beta_\m{UV}-M_\m{UV}$ relation \citep{2014ApJ...793..115B}.
}

\begin{figure}
 \begin{center}
  \includegraphics[clip,bb=10 20 330 270,width=1\hsize]{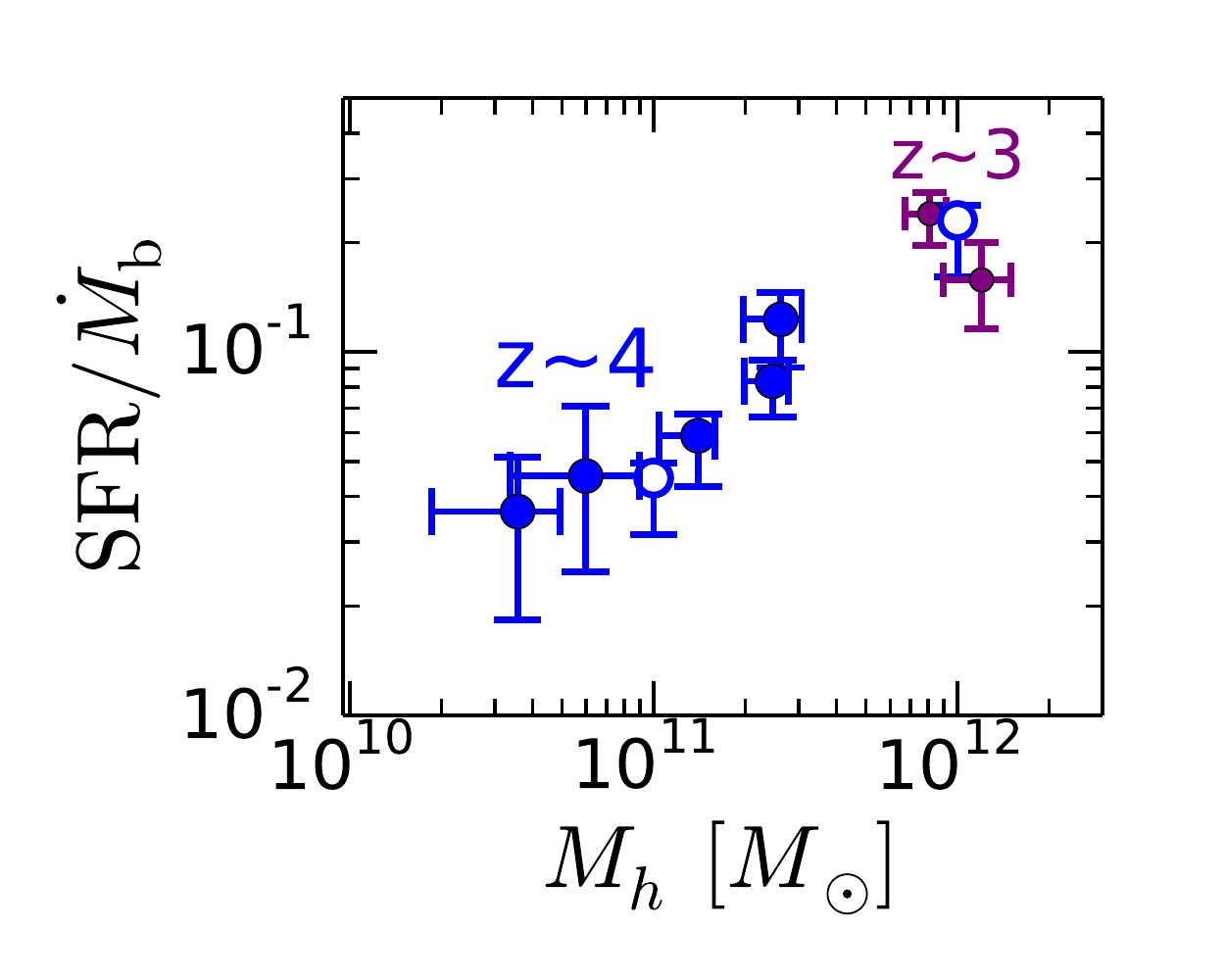}
 \end{center}
    \caption{BCE as a function of dark matter halo mass.
The blue circles represent the 
BCEs of our subsample 
at $z\sim4$.
The open blue circles at $M_\m{h}=10^{11}$ and $10^{12}\ \Msun$ describe the 
BCEs of \citet{2013ApJ...770...57B} at $z\sim4$.
\redc{BCEs at $z\sim3$ \citep{2013ApJ...774...28B} are shown with the purple circles.}
     \label{fig_BCE}}
\end{figure}

Figure \ref{fig_BCE} presents BCEs of our $z\sim4$ subsamples as a function of the dark matter halo mass.
The errors of these BCE estimates do not include the halo mass accretion rate scatters, but the halo mass estimates.
Although there exist moderately large uncertainties in the results of our $z\sim4$ subsamples in Figure \ref{fig_BCE},
there is a signature of positive correlation between BCE and $M_\m{h}$.
We compare BCE estimates of \citet{2013ApJ...770...57B} in Figure \ref{fig_BCE}, and confirm
that our results including the positive correlation signature are consistent with those of \citet{2013ApJ...770...57B}.
This consistency would indicate that the abundance matching technique provides results similar to our clustering analysis (see Section \ref{ss_com_AM}).

This positive correlation signature indicates the low conversion efficiency from gas to stars in low-mass halos, suggesting the inefficient star formation 
in the low-mass halos. The inefficient star-formation probably originates from the mass dependence of 
feedback and/or gas cooling. In low-mass halos, star-formation activities associated with
supernovae, stellar wind, and radiation produce outflowing gas that
suppress next generation star formation as the feedback process.
Moreover, in low-mass halos, the gas cooling is slow \citep{1993PhR...231..293S}.
The combination of these effects would make the positive correlation between BCE and $M_\m{h}$.

\redc{
We compare the BCEs at $z\sim4$ with those at $z\sim3$ given by \citet{2013ApJ...774...28B}. 
Because \citet{2013ApJ...774...28B} use the mean halo mass to derive BCEs in their paper, we re-calculate BCEs with $M_\m{min}$ values presented in \citet{2013ApJ...774...28B}.
We estimate the SFR from the upper limit of the magnitude for each subsample in \citet{2013ApJ...774...28B}. 
We find that the re-calculated BCEs of $M_\m{h}\sim10^{12}\ \m{\Msun}$ at $z\sim3$ are comparable to the one at $z\sim4$ within the error bars (Figure \ref{fig_BCE}).
}

\subsection{Comparisons with Abundance Matching Studies}\label{ss_com_AM}
\subsubsection{Impact of DC and HOD Systematics on the $M_{\rm h}$ Estimates}\label{ss_impact_DC_HOD}
We discuss differences between our clustering analysis and the abundance matching results.
As explained in Section \ref{ss_intro}, these differences should originate from
DC and the subhalo-galaxy relation that are implemented in our HOD model.
We investigate the physical impacts of DC and HOD (which affects the subhalo-galaxy relation) on 
the $M_\m{h}$ estimates, comparing those obtained from a simple abundance matching (SiAM) method
based on halo mass functions of \citet{2013ApJ...770...57B} and UV luminosity functions of \citet{2015ApJ...803...34B}
at $z\sim 4$.
We calculate $M_\m{h}$ as a function of $M_\m{UV}$ by the abundance matching technique
with and without DC and HOD terms, using
\begin{align}
&\int^\infty_0dM^\prime_h\frac{dn}{dM_h}(M^\prime_\m{h},z)\times DC\times HOD(M^\prime_\m{h},M_\m{h})\notag\\
&=\int_{M_\m{UV}}^{-\infty}dM^\prime_\m{UV}\Phi(M^\prime_{\m{UV}},z),
\end{align}
where $\frac{dn}{dM_h}(M^\prime_\m{h},z)$ and $\Phi(M^\prime_\m{UV},z)$ are the halo mass function 
and the UV luminosity function at $z\sim4$, respectively.
The term of $HOD(M^\prime_\m{h},M_\m{h})$ is changed on a case-by-case basis,
as detailed below.

\begin{figure}
 \begin{center}
  \includegraphics[clip,bb=10 16 350 340,width=1\hsize]{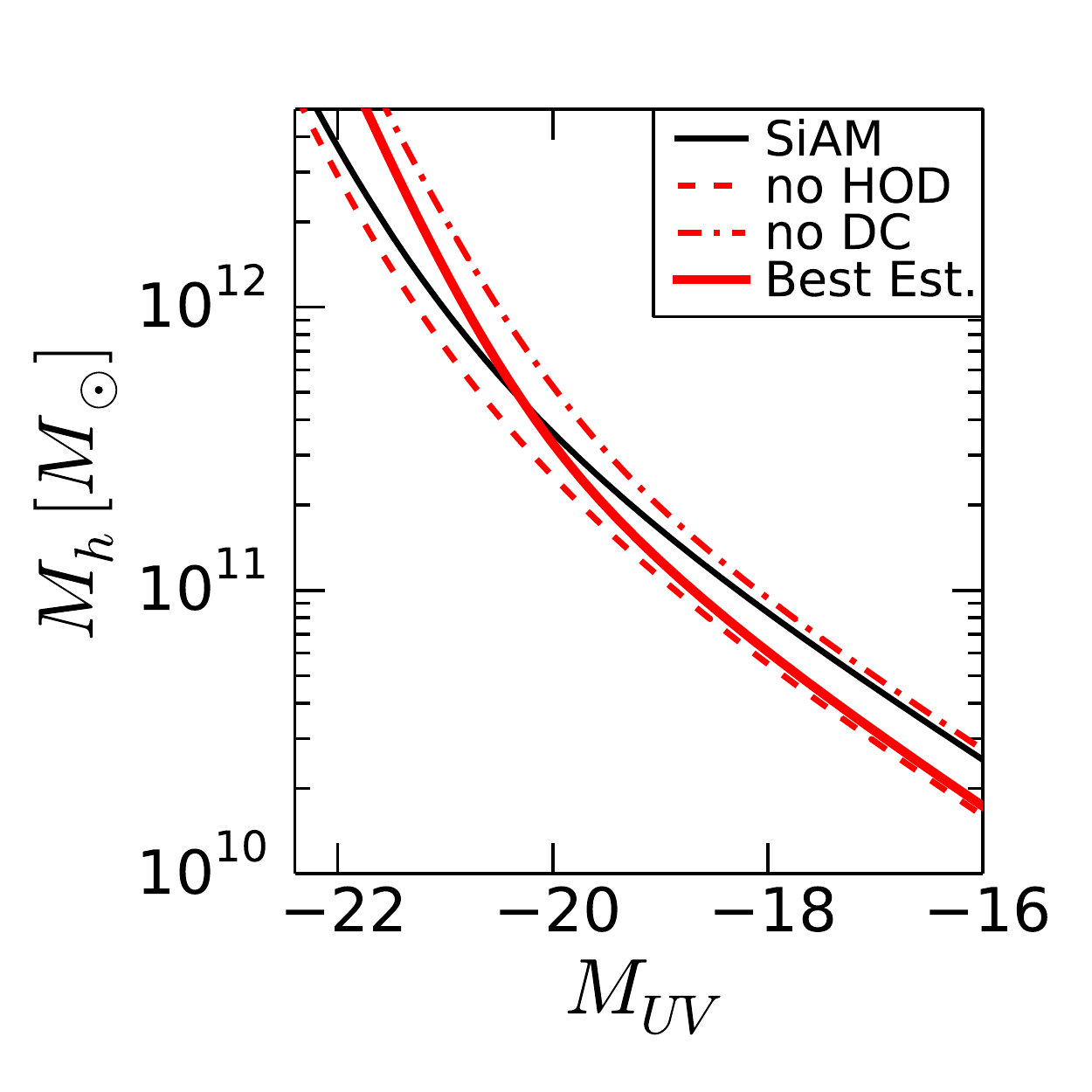}
 \end{center}
    \caption{
Impact
    of DC and HOD 
on the dark matter halo mass estimates.
    We plot the dark matter halo mass 
estimates for our $z \sim 4$ subsample
    as a function of 
absolute
magnitude.     
The black
    curve is the result of the simple abundance matching (SiAM). 
The dashed
    curve represents the result of the abundance matching with $DC=0.6$.
The dot-dashed
    curve 
denotes
    the result of the abundance matching considering the HOD ($\sigmalogM=0.2$, $\m{log}M_0=11.5$, $\m{log}M_1^\prime=12.1$, and $\alpha=1.0$) with $DC=1.0$.
The red
    solid curve 
{is}
    the result of the abundance matching considering the HOD with $DC=0.6$, 
mimicking our clustering analysis results.
     See Section \ref{ss_impact_DC_HOD} for more details.
     \label{fig_onoff}}
     \end{figure}

The calculation results are shown in the $M_\m{h}$ vs. $M_\m{UV}$ plot of Figure \ref{fig_onoff}.
The black solid curve represents the SiAM case that includes neither
DC nor HOD effect, i.e., $DC=1$ and $HOD(M^\prime_\m{h},M_\m{h})=H(M^\prime_\m{h}-M_\m{h})$,
where $H(x)$ is a step function.
The red dashed curve corresponds the no-HOD case with the DC effects,
$DC=0.6$ and $HOD(M^\prime_\m{h},M_\m{h})=H(M^\prime_\m{h}-M_\m{h})$.
 The red dot-dashed curve denotes the no-DC case with the HOD effects,
 $DC=1.0$ and $HOD(M^\prime_\m{h},M_\m{h})=N_\m{c}(M^\prime_\m{h})+N_\m{s}(M^\prime_\m{h})$,
 where we use our best-fit HOD parameter set of the $z\sim4$ LBG subsample of $m^\m{aper}_\m{UV}<28.2\ \m{mag}$ that is well determined,
 ($\sigmalogM$, $\m{log}M_0$, $\m{log}M_1^\prime$, $\alpha$)=(0.2, 11.5, 12.1, 1.0),
 and $M_\m{min}=M_\m{h}$ by definition.
The red solid curve indicates the best-estimate case with the DC and HOD effects, mimicking our clustering analysis results,
$DC=0.6$ and $HOD(M^\prime_\m{h},M_\m{h})=N_\m{c}(M^\prime_\m{h})+N_\m{s}(M^\prime_\m{h})$.

\begin{figure*}
\begin{center}
 \begin{minipage}{0.45\hsize}
 \begin{center}
  \includegraphics[clip,bb=10 16 345 340,width=1\hsize]{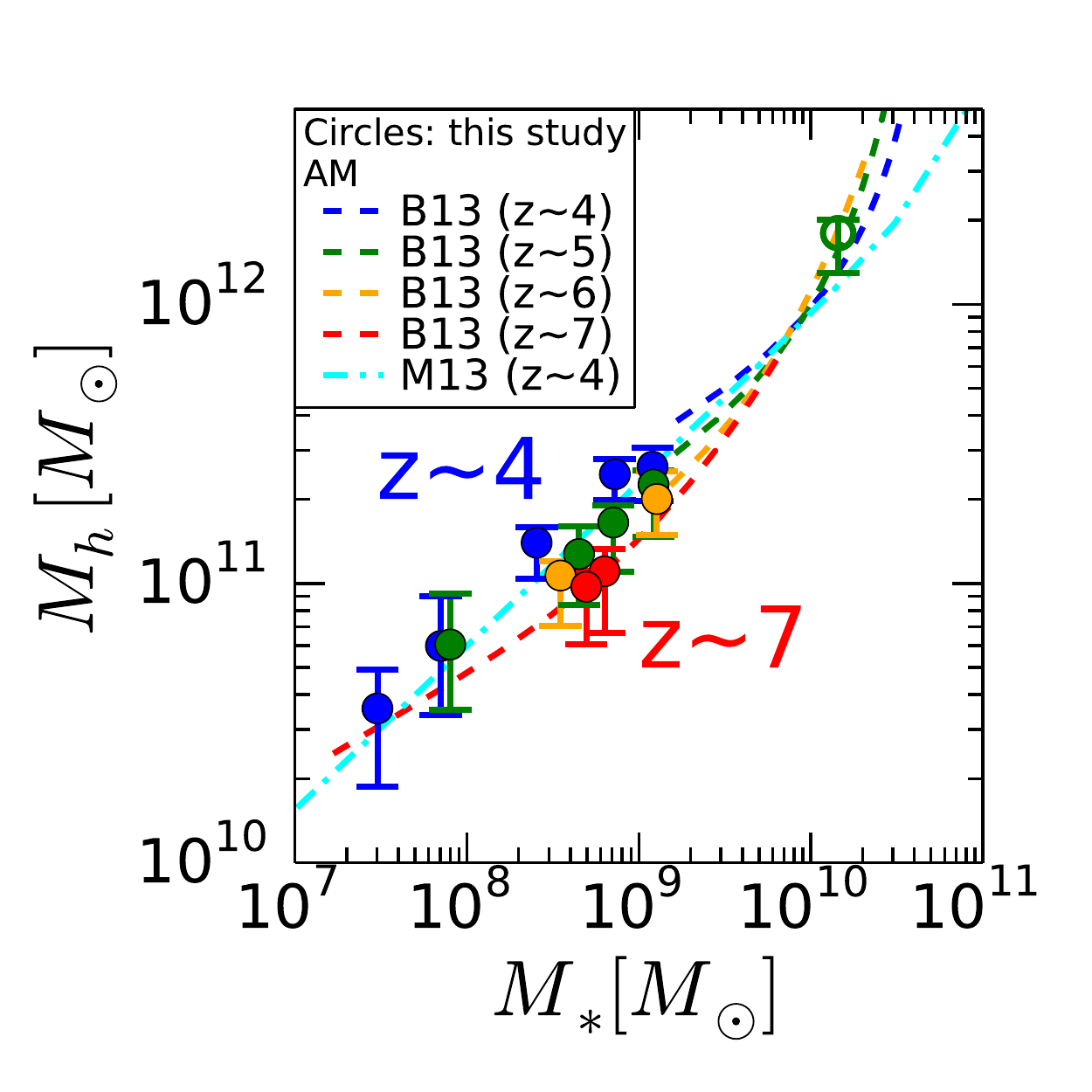}
 \end{center}
 \end{minipage}
  \begin{minipage}{0.53\hsize}
 \begin{center}
  \includegraphics[clip,bb=10 16 340 270,width=1\hsize]{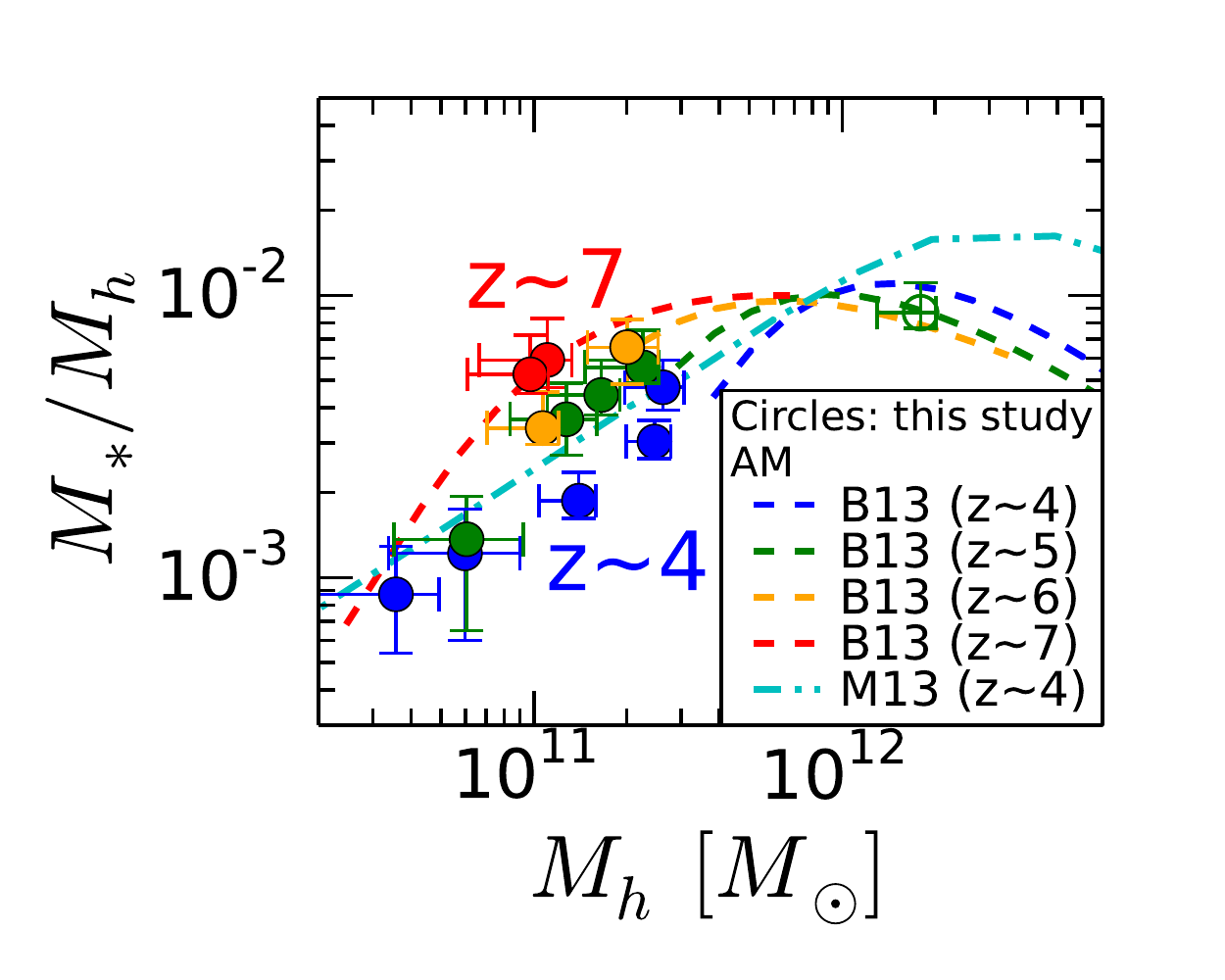}
   \end{center}
 \end{minipage}
\\
 \end{center}
 \caption{Comparison with \citet{2013ApJ...770...57B} and \citet{2013MNRAS.428.3121M}.
 Left panel: 
comparison
 of the dark matter halo mass as a function of 
stellar
 mass.
 \redc{The blue, green, orange, and red dashed curves are the results of \citet{2013ApJ...770...57B}. 
 These results are re-computed by P. Behroozi with the cosmological parameters and halo mass definition same as ours.}
The cyan
        dot-dashed curve is the result of \citet{2013MNRAS.428.3121M}.
The blue, 
    green, orange, and red circles represent the dark matter halo mass of our 
Hubble subsamples at
    $z\sim4$, $5$, $6$, and $7$, respectively.
    The open green circle denotes the dark matter halo mass of our subsample constructed from the HSC data.
   Right panel: 
comparison
   of the SHMR.
   Same as the left panel but the horizontal and vertical axes are the dark matter halo mass and the SHMR.
  \label{fig_Com_w_B13}}\end{figure*}

\begin{figure}
 \begin{center}
  \includegraphics[clip,bb=10 16 350 340,width=1\hsize]{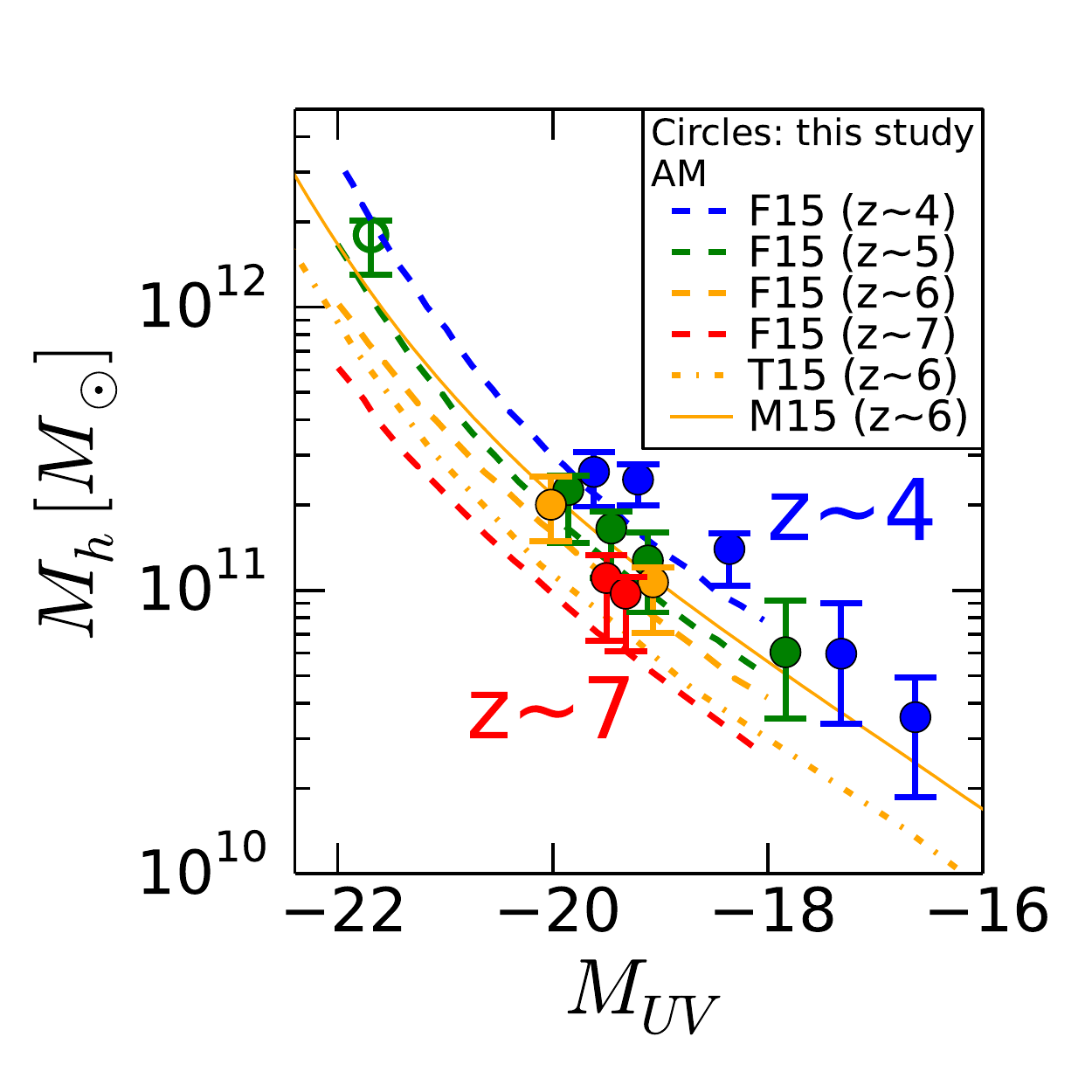}
 \end{center}
    \caption{Comparison with \citet{2015arXiv150400005F}, \citet{2015arXiv150702685T} and \citet{2015arXiv150801204M}.
The blue, 
       green, orange, and red dashed curves are the results of \citet{2015arXiv150400005F} at $z\sim4, 5, 6,$ and $7$, respectively.
The orange
    dot-dashed and 
the
    solid curve shows the $z\sim6$ 
results
    of \citet{2015arXiv150702685T} and \citet{2015arXiv150801204M}, respectively.
The blue
        green, orange, and red circles represent the dark matter halo 
masses
of our 
subsamples at
        $z\sim4$, $5$, $6$, and $7$, respectively.
\label{fig_Com_w_F15}}
\end{figure}

\begin{figure}
 \begin{center}
  \includegraphics[clip,bb=10 16 340 340,width=1\hsize]{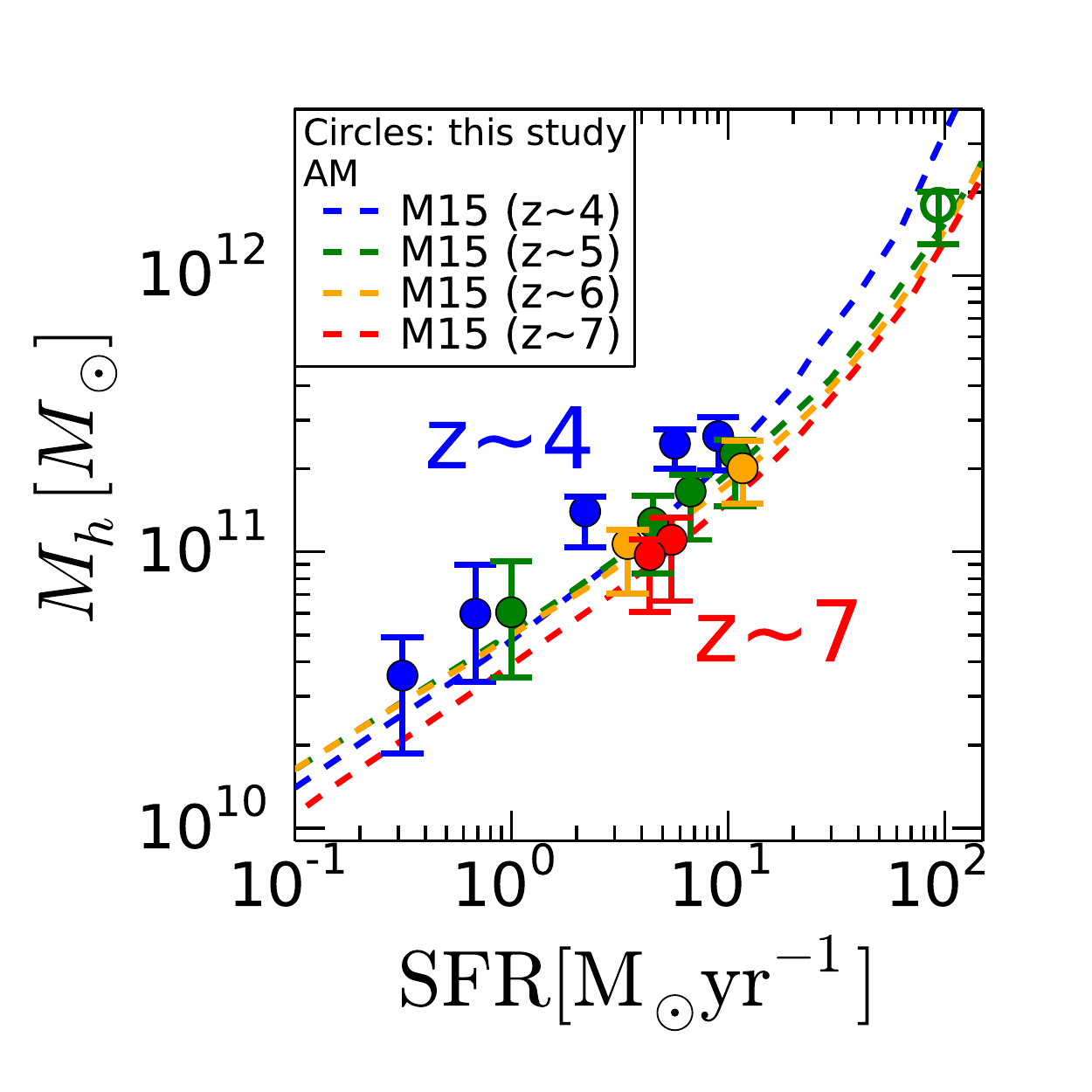}
 \end{center}
    \caption{Comparison with \citet{2015arXiv150700999M}.
The blue, 
        green, orange, and red dashed curves are the results of \citet{2015arXiv150700999M} at $z\sim4, 5, 6,$ and $7$, respectively.
The blue, 
    green, orange, and red circles represent the dark matter halo 
masses
    of our 
subsamples at
    $z\sim4$, $5$, $6$, and $7$, respectively.
\label{fig_Com_w_M15}}
\end{figure}

Comparing the best-estimate case with the DC and HOD effects (red solid curve) in Figure \ref{fig_onoff},
we find that the no-DC case (red dot-dashed curve) overestimates $M_{\rm h}$ by 
$\sim0.2\ \m{dex}$ at the faint magnitude of $M_\m{UV}\sim-16\ \m{mag}$.
However, this overestimate becomes small at the bright magnitude, suggesting that
the DC effect is more important at the faint magnitude.
In contrast to the no-DC case, the no-HOD case (red dashed curve) agrees with the best-estimate case (red solid curve)
at the faint magnitude, while the no-HOD case underestimates $M_{\rm h}$ by $0.4\ \m{dex}$ at the bright magnitude of $M_\m{UV}=-21.5\ \m{mag}$.
Note that the SiAM case (black solid curve) is bracketed by the no-DC case (red dot-dashed curve)
and the no-HOD case (red dashed curve). The SiAM case overestimates (underestimates) $M_{\rm h}$
at the faint (bright) magnitudes by $\sim0.2$ ($\sim 0.4$) $\m{dex}$, following the no-DC (no-HOD) case. 
Because the DC and HOD effects cancel out, the SiAM case provides $M_{\rm h}$ values comparable to
those of the best-estimate case at the intermediate magnitude of $M_\m{UV}\sim-21 - (-20)\ \m{mag}$ 
that corresponds to $L^*$.

\subsubsection{Comparisons with the Abundance Matching Results}
\label{sec:comparison_AM}

We compare results of our clustering analyses with those of recent abundance matching studies.
Figure \ref{fig_Com_w_B13} presents halo masses estimated by abundance matching techniques
of \citet{2013ApJ...770...57B} and \citet{2013MNRAS.428.3121M}, together with our halo mass estimates
from the clustering analyses. Here, again, we use the modified results of \citet{2013ApJ...770...57B}
whose cosmological parameters and halo mass definition
are the same as ours (P. Behroozi in private communication).
Figure \ref{fig_Com_w_F15} (Figure \ref{fig_Com_w_M15}) is the same as Figure \ref{fig_Com_w_B13},
but for \citet{2015arXiv150400005F}, \citet{2015arXiv150702685T}, and \citet{2015arXiv150801204M}
(\citealt{2015arXiv150700999M}). Although the comparisons in the $M_{\rm h}$ vs. $M_*$ or $SHMR$ vs. $M_{\rm h}$
plot (Figure \ref{fig_Com_w_B13}) are straightforward, we use $M_{\rm h}$ vs. $M_{\rm UV}$ and $M_{\rm h}$ vs. SFR
in Figures \ref{fig_Com_w_F15} and \ref{fig_Com_w_M15}, respectively. This is because
the abundance matching studies shown in Figures \ref{fig_Com_w_F15} and \ref{fig_Com_w_M15} do not
present $M_*$, but $M_{\rm UV}$ or SFR.

In Figure \ref{fig_Com_w_B13}, we find that the abundance matching results of \citet{2013ApJ...770...57B} 
agree with our clustering results at $z\sim 5-7$ very well within $1\sigma$ errors.
At $z\sim4 $, the stellar mass range of \citet{2013ApJ...770...57B} does not cover the one of ours,
and secure comparisons cannot be made.

Figure \ref{fig_Com_w_F15} indicates that the abundance matching results in \citet{2015arXiv150400005F}
are consistent with our clustering results within $1\sigma$ errors at $z\sim 5-7$, although all of the data 
of \citet{2015arXiv150400005F} appear to fall below our clustering results.
At $z\sim 4$, the $M_{\rm h}$ values of \citet{2015arXiv150400005F}
are lower than ours by $0.13$ dex at $z\sim 4$ at the $\sim 2\sigma$ levels. 
There are similar $\sim 2\sigma$ level $M_{\rm h}$ value offsets of $0.1$ ($\sim 0.2$) dex
to \citet{2013MNRAS.428.3121M} (\citealt{2015arXiv150702685T} and \citealt{2015arXiv150700999M})
at $z\sim 4$ ($z\sim 6$ and $4$). 
\redc{The results of \citet{2015arXiv150801204M} agree well with ours within the error bars. Note that all of these studies adopt the cosmological parameters different from ours. The cosmological parameter sets used in \citet{2013MNRAS.428.3121M}, \citet{2015arXiv150702685T}, \cite{2015arXiv150700999M}, and \citet{2015arXiv150801204M} are $(H_0, \Omega_\m{m}, \Omega_\m{\Lambda}, \sigma_8)=(70.4, 0.272, 0.728, 0.810)$, $(70, 0.27, 0.73, 0.8)$, $(70, 0.3, 0.7, 0.82)$, and $(67.31, 0.315, 0.685, 0.829)$, respectively.
\citet{2015arXiv150400005F} use the parameter sets of $(H_0, \Omega_\m{m}, \Omega_\m{\Lambda})=(70.2, 0.275, 0.725)$.}
If we correct these study results for the effects of the different cosmological parameters, the duty cycle, and the halo mass function, these study results
agree with our clustering analysis estimates within $\sim 0.1$ dex.
In other words, all of these abundance matching results agree with our clustering results
very well. In Section 7.3.1, we compare the results of SiAM with those of the best estimates
having clustering analysis parameter constraints, and conclude that the SiAM results
differ from the best estimates up to $0.4$ dex in $M_{\rm h}$. However, we find very good agreements 
between the recent abundance matching studies and the clustering analysis study,
within $\sim 0.1$ dex in $M_{\rm h}$. These good agreements are probably explained by the facts that
these abundance matching studies are different from the classical technique of SiAM, but
with the subhalos, incompleteness \citep[similar to DC;][]{2013ApJ...770...57B}, and/or SFR+stellar mass function evolution that reduce the systematics.


\subsubsection{Conclusions of the Comparisons}
In Section \ref{sec:comparison_AM}, we find that $M_{\rm h}$ estimates from our clustering analyses 
agree with most of the abundance matching results within the $\sim 0.1$ dex level at $z\sim4-7$, 
which are corrected for the differences of cosmological parameters, the duty cycle values, and the halo mass functions.
Although there exist the systematic $M_{\rm h}$ differences originating from the DC and HOD uncertainties
up to by $0.4\ \m{dex}$ or a factor of 3 in the SiAM at the dark matter halo mass of $10^{10}-10^{12}\ \Msun$ (Section 7.3.1), 
the recent abundance matching techniques including the subhalos, incompleteness, and/or SFR+stellar-mass evolution, appear to reduce the systematics down to 
the $\sim 0.1$ dex level. Thus, the abundance matching techniques are useful to estimate
$M_{\rm h}$ of high-$z$ galaxy halos, if one allows the systematic uncertainties up to a factor of 3.
The good agreements between the clustering and abundance matching techniques are found, 
probably because the systematics of the subhalo-galaxy relation is small due to the small satellite fraction at high-$z$
 (\citealt{2007ApJ...667..760Z}; \citealt{2012A&A...542A...5C}; see also \citealt{2015arXiv150700713J}).

\section{Summary}\label{ss_summary} 
We obtain clustering measurements of $z\sim 4-7$ galaxies
from the data set of the legacy deep Hubble/\redc{Advanced Camera for Surveys (ACS)} and \redc{Wide Field Camera 3 (WFC3)} images
and the complimentary large-area Subaru/\redc{Hyper-Suprime Cam (HSC)} images that are newly available.
Via our \redc{halo occupation distribution (HOD)} modeling, we investigate \redc{stellar-to-halo mass ratios (SHMRs)} at $z\sim 4-7$
for the first time by galaxy clustering analyses.
We compare our clustering analysis results with the abundance matching results
that are actively being obtained by recent studies.
Our major findings are summarized below.

\begin{enumerate}

\item
The mean dark matter halo masses of our \redc{Lyman break galaxies (LBGs)} at $z\sim 4-7$ are $\left < M_{\rm h}\right > \sim (1-20)\times10^{11}\ \Msun$.
There is an increasing trend in the dark matter halo mass with increasing the UV luminosity of LBGs at $z\sim 4-7$.
Our estimated dark matter halo masses are consistent with the previous clustering studies in the $\left < M_{\rm h}\right > -\left < M_{\rm UV}\right >$ plane,
if we use the same cosmological parameter set for comparison.

\item
We estimate SHMR for our LBGs.
We identify the SHMR evolutions in $z\sim0-4$ and $z\sim4-7$ at the $>98\%$ confidence level for the first time
based on clustering analyses.
We find that, at the dark matter halo mass of $M_\m{h}\sim10^{11}\ \Msun$, SHMRs decrease 
by a factor of \redc{$\sim 2$, from $\sim2.7\times10^{-3}$ ($z\sim0$) to $\sim1.3\times10^{-3}$} ($z\sim 4$),
and increase by a factor of \redc{$\sim4$, from $\sim1.3\times10^{-3}$ ($z\sim4$) to $\sim5.3\times10^{-3}$} ($z\sim7$).

\item
\redc{We compare our SHMRs with results of the hydrodynamic simulation and the semi-analytic model at $M_\m{h}=10^{11}\ \Msun$. 
These theoretical studies predict evolutionary trends of the SHMR decrease from $z\sim0$ to $4$ that are similar to our observational results. 
On the other hand, the theoretical studies can not reproduce the SHMR increase from $z\sim4$ to $7$ found in our observational study.
}

\item
We calculate \redc{baryon conversion efficiency (BCE)}, that is the ratio of the \redc{star formation rate (SFR)} to the baryon accretion rate corresponding to rates of the conversion from gas to stars.
The BCEs at $z\sim4$ increase from $\sim3\times10^{-2}$ to $\sim1\times10^{-1}$ with increasing dark matter halo mass up to $\sim10^{12}\ \Msun$.
The low mass halos form stars inefficiently, probably due to feedback effects and/or slow gas cooling.

\item
We compare our clustering+HOD results with abundance matching estimates.
We find that the $M_{\rm h}$ estimates of clustering+HOD analyses 
agree with those of the simple abundance matching within a factor of 3.
Moreover, the results of the recent studies' sophisticated abundance matching techniques including the subhalos, incompleteness, and/or SFR+stellar-mass evolution
are even better than those of the simple abundance matching technique,
some of which agree with our clustering results within $0.1$ dex
at $z\sim 4-7$. Due to the small galaxy occupation in one-halo at high-$z$,
abundance matching techniques are useful to estimate $M_{\rm h}$ for high-$z$ galaxies,
if one allows these reasonably small uncertainties raised by the assumptions of
the abundance matching techniques.
\end{enumerate}

\acknowledgments
We thank the anonymous referee for a careful reading and valuable comments that improved the clarity of the paper.
We are grateful to Peter Behroozi for useful discussions and providing his SHMR results that are recalculated with the cosmological parameters 
and halo mass definition same as ours.
We appreciate helpful comments on an early draft version of this paper from Michael A. Strauss and Masahiro Takada.
We thank Jim Bosch, Hisanori Furusawa, Robert H. Lupton, Sogo Mineo, and Naoki Yasuda for their helpful comments and discussions on the treatment of the HSC data.
We are also grateful to Masayuki Tanaka for kindly providing a masking code and discussing our results.
This work has benefited from many interesting conversations with Renyue Cen, Motohiro Enoki, Thibault Garel, Ryosuke Goto, Masao Hayashi, Chiaki Hikage, Yutaka Hirai, Soh Ikarashi, Akio Inoue, Masaru Kajisawa, Nobunari Kashikawa, Daichi Kashino, Charlotte A. Mason, Masakazu A.R. Kobayashi, Masato I.N. Kobayashi, Andrei Mesinger, Kentaro Nagamine, Tohru Nagao, Masahiro Nagashima, Atsushi J. Nishizawa, Takashi Okamoto, Kouji Ohta, Taku Okamura, Jaehong Park, Yoshiaki Sofue, Guochao Sun, Tsutomu Takeuchi, and Masayuki Umemura.

This work is based on observations taken by the 3D-HST Treasury Program (GO 12177 and 12328)  and CANDELS Multi-Cycle Treasury Program with the NASA/ESA HST, which is operated by the Association of Universities for Research in Astronomy, Inc., under NASA contract NAS5-26555.
This paper makes use of software developed for the Large Synoptic Survey Telescope.
We thank the LSST Project for making their code available as free software at http://dm.lsstcorp.org.

The Pan-STARRS1 Surveys (PS1) have been made possible through contributions of the Institute for Astronomy, the University of Hawaii, the Pan-STARRS Project Office, the Max-Planck Society and its participating institutes, the Max Planck Institute for Astronomy, Heidelberg and the Max Planck Institute for Extraterrestrial Physics, Garching, The Johns Hopkins University, Durham University, the University of Edinburgh, Queen's University Belfast, the Harvard-Smithsonian Center for Astrophysics, the Las Cumbres Observatory Global Telescope Network Incorporated, the National Central University of Taiwan, the Space Telescope Science Institute, the National Aeronautics and Space Administration under Grant No. NNX08AR22G issued through the Planetary Science Division of the NASA Science Mission Directorate, the National Science Foundation under Grant No. AST-1238877, the University of Maryland, and Eotvos Lorand University (ELTE).

This work is supported by World Premier International Research Center Initiative (WPI Initiative), MEXT, Japan, KAKENHI (23244025) and (15H02064) Grant-in-Aid for Scientific Research (A) through Japan Society for the Promotion of Science (JSPS), and a grant from
the Hayakawa Satio Fund awarded by the Astronomical Society of Japan.
Y.H. is supported by an Advanced Leading Graduate Course for Photon Science grant (ALPS).
S.M. is supported by KAKENHI (15K17600).

{\it Facilities:} \facility{HST (ACS, WFC3), Subaru (HSC, Suprime-Cam), CFHT (MEGACAM).}

\bibliographystyle{apj}
\bibliography{apj-jour,reference}

\end{document}